\documentclass[sigconf]{acmart}

\usepackage{enumitem}
\usepackage{multirow}
\usepackage{booktabs}
\usepackage{caption}
\usepackage{array}
\usepackage{makecell}
\usepackage{color}
\usepackage{xcolor}

\usepackage{subcaption}
\usepackage{algorithm}
\usepackage{algpseudocode}
\usepackage{threeparttable}
\usepackage{pifont}
\usepackage{pdfpages}
\usepackage{pdfcomment}
\usepackage{tikz}
\usepackage{amsmath}
\usepackage{pifont}
\usepackage{float}
\usepackage{bm}
\usepackage{listings} % 代码展示
\usepackage{xcolor} % 使用颜色
\usepackage{stfloats}
\usepackage{adjustbox}

\definecolor{deepgreen}{rgb}{0,0.5,0} % 深绿色

\lstdefinestyle{lfonts}{
    basicstyle = \footnotesize\ttfamily,
    stringstyle = \color{deepgreen},
    keywordstyle = \color{blue!60!black}\bfseries,
    commentstyle = \color{gray}\scshape,
}
\lstdefinestyle{lnumbers}{
    numbers = left,
    numberstyle = \tiny,
    numbersep = 1em,
    firstnumber = 1,
    stepnumber = 1,
}
\lstdefinestyle{llayout}{
    breaklines = true,
    tabsize = 2,
    columns = flexible,
}
\lstdefinestyle{lgeometry}{
    xleftmargin = 20pt,
    xrightmargin = 0pt,
    frame = tb,
    framesep = \fboxsep,
    framexleftmargin = 20pt,
}
\lstdefinestyle{lgeneral}{
    style = lfonts,
    style = lnumbers ,
    style = llayout,
    style = lgeometry,
}
\lstdefinestyle{python}{
    language = {Python},
    style = lgeneral,
}

\definecolor{deepred}{rgb}{0.6,0,0} % 深红色
\definecolor{darkblue}{rgb}{0,0,0.5} % 深蓝色

\lstdefinestyle{xmlfonts}{
    basicstyle = \footnotesize\ttfamily,
    stringstyle = \color{deepred}, % XML 中字符串使用深红色
    tagstyle = \color{darkblue}\bfseries, % XML 标签使用深蓝色并加粗
    commentstyle = \color{gray}\itshape, % 注释使用灰色，斜体
}

\lstdefinelanguage{XML}{
    morestring=[b]",
    morecomment=[s]{<?}{?>},
    morecomment=[s]{<!--}{-->},
    tag=[s],
    otherkeywords={/>},
}

\lstdefinestyle{xmlnumbers}{
    numbers = left,
    numberstyle = \tiny,
    numbersep = 1em,
    firstnumber = 1,
    stepnumber = 1,
}
\lstdefinestyle{xmllayout}{
    breaklines = true,
    tabsize = 2,
    columns = flexible,
}
\lstdefinestyle{xmlgeometry}{
    xleftmargin = 20pt,
    xrightmargin = 0pt,
    frame = tb,
    framesep = \fboxsep,
    framexleftmargin = 20pt,
}
\lstdefinestyle{xmlgeneral}{
    style = xmlfonts,
    style = xmlnumbers,
    style = xmllayout,
    style = xmlgeometry,
    language = XML,
}
\lstdefinestyle{xml}{
    style = xmlgeneral,
}

\newcommand{\cmark}{\textcolor{my_green}{\ding{51}}} % check 
\newcommand{\xmark}{\textcolor{my_red}{\ding{55}}} % X
\definecolor{my_green}{RGB}{51,102,0}
\definecolor{my_red}{RGB}{204, 0, 0}
\AtBeginDocument{%
  \providecommand\BibTeX{{%
    \normalfont B\kern-0.5em{\scshape i\kern-0.25em b}\kern-0.8em\TeX}}}

\copyrightyear{2024}
\acmYear{2024}
\setcopyright{acmlicensed}
\acmConference[KDD '24]{Proceedings of the 30th ACM SIGKDD Conference on Knowledge Discovery and Data Mining}{August 25--29, 2024}{Barcelona, Spain}
\acmBooktitle{Proceedings of the 30th ACM SIGKDD Conference on Knowledge Discovery and Data Mining (KDD '24), August 25--29, 2024, Barcelona, Spain}
\acmDOI{10.1145/3637528.3671801}
\acmISBN{979-8-4007-0490-1/24/08}

\begin{document}
% \title{
%     \begin{minipage}{0.1\textwidth}
%       \includegraphics[width=1.5cm]{assets/FinAgentLogo.jpg}
%     \end{minipage}
%     \begin{minipage}{0.85\textwidth}
%     \centering
%     A Multi-Modal Foundation Agent for Financial Trading: \\ Tool-Augmented, Diversified, and Generalist
%     \end{minipage}
% }

% \title[A Multimodal Foundation Agent for Financial Trading: Tool-Augmented, Diversified, and Generalist]{\parbox[c]{0.08\textwidth}{\includegraphics[width=\linewidth]{assets/FinAgentLogo.pdf}}%
% \parbox[c]{0.9\textwidth}{\centering{A Multimodal Foundation Agent for Financial Trading: Tool-Augmented, Diversified, and Generalist}}}

\title{A Multimodal Foundation Agent for Financial Trading: Tool-Augmented, Diversified, and Generalist}

\author{Wentao Zhang}
\affiliation{%
  \institution{Nanyang Technological University}
  \country{Singapore}
}
\email{wt.zhang@ntu.edu.sg}

\author{Lingxuan Zhao}
\authornote{Lingxuan Zhao and Haochong Xia contributed equally to this research.}
\author{Haochong Xia}
\authornotemark[1]
\affiliation{%
  \institution{Nanyang Technological University}
  \country{Singapore}
}
\email{{zhao0375,haochong001}@e.ntu.edu.sg}

% \author{Haochong Xia}
% \affiliation{%
%   \institution{Nanyang Technological University}
%   \country{Singapore}
% }
% \email{haochong001@e.ntu.edu.sg}

\author{Shuo Sun}
\affiliation{%
  \institution{Nanyang Technological University}
  \country{Singapore}
}
\email{shuo003@e.ntu.edu.sg}

\author{Jiaze Sun}
\affiliation{%
  \institution{National University of Singapore}
  \country{Singapore}
}
\email{e0564914@u.nus.edu}

\author{Molei Qin}
\author{Xinyi Li}
\author{Yuqing Zhao}
\affiliation{%
  \institution{Nanyang Technological University}
  \country{Singapore}
}
\email{{molei001,lixi0067,ZHAO0348}@e.ntu.edu.sg}

% \author{Xinyi Li}
% \affiliation{%
%   \institution{Nanyang Technological University}
%   \country{Singapore}
% }
% \email{lixi0067@e.ntu.edu.sg}

% \author{Yuqing Zhao}
% \affiliation{%
%   \institution{Nanyang Technological University}
%   \country{Singapore}
% }
% \email{ZHAO0348@e.ntu.edu.sg}

\author{Yilei Zhao}
\affiliation{%
  \institution{Zhejiang University}
  \country{China}
}
\email{yilei\_zhao@zju.edu.cn}

\author{Xinyu Cai}
\author{Longtao Zheng}
\affiliation{%
  \institution{Nanyang Technological University}
  \country{Singapore}
}
\email{{xinyu009,longtao001}@e.ntu.edu.sg}

% \author{Longtao Zheng}
% \affiliation{%
%   \institution{Nanyang Technological University}
%   \country{Singapore}
% }
% \email{longtao001@e.ntu.edu.sg}

\author{Xinrun Wang}
\authornote{Corresponding Authors.}
\affiliation{%
  \institution{Singapore Management University}
  \country{Singapore}
}
\email{xrwang@smu.edu.sg}

\author{Bo An}
\authornotemark[2]
\affiliation{%
    \institution{Nanyang Technological University \\ Skywork AI}
    \country{Singapore}
}
\email{boan@ntu.edu.sg}

\renewcommand{\shortauthors}{Wentao Zhang et al.}

\begin{abstract}
% Financial trading is fundamental for the financial market, where a good financial training agent need to handle the multimodal 
Financial trading is a crucial component of the markets, informed by a multimodal information landscape encompassing news, prices, and Kline charts, and encompasses diverse tasks such as quantitative trading and high-frequency trading with various assets. While advanced AI techniques like deep learning and reinforcement learning are extensively utilized in finance, their application in financial trading tasks often faces challenges due to inadequate handling of multimodal data and limited generalizability across various tasks. To address these challenges, we present \texttt{FinAgent}, a multimodal foundational agent with tool augmentation for financial trading. \texttt{FinAgent}'s market intelligence module processes a diverse range of data—numerical, textual, and visual—to accurately analyze the financial market. Its unique dual-level reflection module not only enables rapid adaptation to market dynamics but also incorporates a diversified memory retrieval system, enhancing the agent's ability to learn from historical data and improve decision-making processes. The agent's emphasis on reasoning for actions fosters trust in its financial decisions. Moreover, \texttt{FinAgent} integrates established trading strategies and expert insights, ensuring that its trading approaches are both data-driven and rooted in sound financial principles. With comprehensive experiments on 6 financial datasets, including stocks and Crypto, \texttt{FinAgent} significantly outperforms 12 state-of-the-art baselines in terms of 6 financial metrics  with over 36\% average improvement on profit. Specifically, a 92.27\% return (a 84.39\% relative improvement) is achieved on one dataset. Notably, \texttt{FinAgent} is the first advanced multimodal foundation agent designed for financial trading tasks.

\end{abstract}

% <concept_id>10002951.10003227.10003351</concept_id>
% <concept_desc>Information systems~Data mining</concept_desc>
% <concept_id>10010147.10010257</concept_id>
% <concept_desc>Computing methodologies~Machine learning</concept_desc>
% <concept_id>10010405.10003550</concept_id>
% <concept_desc>Applied computing~Electronic commerce</concept_desc>

% \ccsdesc[500]{Information systems~Data mining}
% \ccsdesc[500]{Computing methodologies~Machine learning}
% \ccsdesc[500]{Applied computing~Electronic commerce}

\begin{CCSXML}
<ccs2012>
<concept>
<concept_id>10002951.10003227.10003351</concept_id>
<concept_desc>Information systems~Data mining</concept_desc>
<concept_significance>500</concept_significance>
</concept>
<concept>
<concept_id>10010147.10010257</concept_id>
<concept_desc>Computing methodologies~Machine learning</concept_desc>
<concept_significance>500</concept_significance>
</concept>
<concept>
<concept_id>10010405.10003550</concept_id>
<concept_desc>Applied computing~Electronic commerce</concept_desc>
<concept_significance>500</concept_significance>
</concept>
</ccs2012>
\end{CCSXML}

\ccsdesc[500]{Information systems~Data mining}
\ccsdesc[500]{Computing methodologies~Machine learning}
\ccsdesc[500]{Applied computing~Electronic commerce}

\keywords{Large Language Models, Quantitative Trading, Financial AI Agents}



\maketitle

\section{Introduction}

Financial markets are essential for economic stability, facilitating capital allocation and risk management. Financial trading systems, developed from technical analysis strategies \cite{edwards2018technical}, enhance these markets by enabling efficient trading. Rule-based trading systems are rigid and struggle to adapt to market volatility, often resulting in underperformance in evolving markets. Reinforcement learning-based systems\cite{an2022deep} demonstrate enhanced adaptability but encounter substantial obstacles, such as the need for extensive training data and the inexplainability of decision-making processes. Additionally, they struggle with generalizing across diverse market conditions, are sensitive to market noise, and often fail to integrate multimodal market intelligence like news and reports into their analysis. The financial trading landscape demands more advanced machine-learning methods to address complex market dynamics, seeking to move beyond the limitations of rule-based and RL methods.

\begin{figure}[t]
\captionsetup{skip=2pt}
  \centering
    \includegraphics[width=0.45\textwidth]{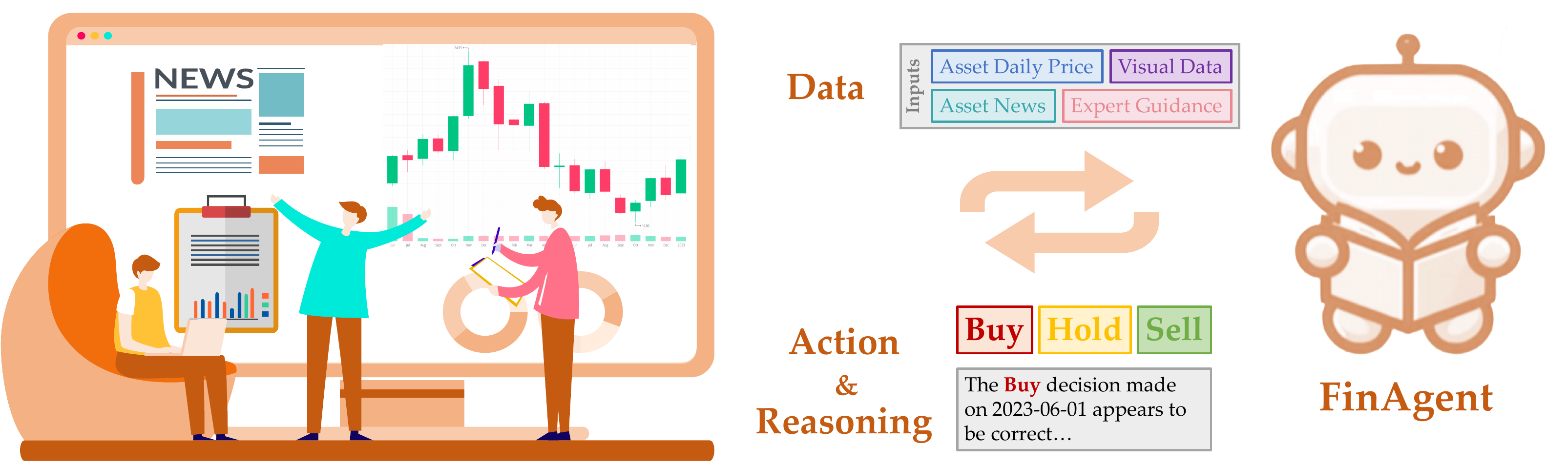}
    \caption{Overview of FinAgent.}
    \label{fig:network}
\vspace{-0.5cm}
\end{figure}

Recently, Large Language Models (LLMs) have showcased their potential in a range of decision-making tasks when applied in AI agents 
\cite{park2023generative,sumers2023cognitive,zhang2023appagent,wang2023voyager}, marking a significant expansion beyond natural language processing into more complex, task-specific functions. This advancement includes the integration of memory and planning modules, which enable these agents to adapt within dynamic environments, akin to human cognitive processes. This evolution has been further pushed by the advent of multimodal LLMs like GPT-4V \cite{openai2023gpt4v}, which enhances the capabilities of LLMs by processing both textual and visual data. Moreover, the integration of tool-augmented models like Toolformer \cite{Schick2023ToolformerLM} empowers LLMs to utilize external tools, thus elevating their decision-making abilities in complex scenarios. This combination of adaptability and enhanced processing capabilities offers new possibilities in fields such as fintech, where nuanced analysis and adaptation are important.

LLMs have demonstrated remarkable capabilities in analyzing and interpreting financial data, as evidenced by developments like BloombergGPT \cite{wu2023bloomberggpt}, and FinGPT \cite{yang2023fingpt}. However, there is a natural gap between QA tasks and sequential decision-making in trading. Although FinMEM \cite{yu2023finmem} is an LLM trading agent with a human-aligned memory mechanism and character design, the full capabilities of LLMs as comprehensive autonomous trading systems remain underexplored, particularly in their ability to interpret multimodal data and utilize diverse tools. The challenges in navigating the complexities of financial markets are identified as follows:

% \begin{itemize}[left=0cm]
%     \item \textbf{Ch1: Insufficient Multimodal Data Processing Ability}. Processing the extensive range of market intelligence, including numerical, textual, and visual information, presents significant difficulties. Advanced analytical methods are required to extract key insights from such diverse data to predict market trends. 
%     \item \textbf{Ch2: Lack of Targeted Information Retrieval}. Some agents mix up retrieval with their main tasks, relying only on brief target task summaries to retrieve historical data. This absence of precise searching introduces too much irrelevant data into the results, which detracts from overall performance.
%     \item \textbf{Ch3: Adaptability in Rapidly Evolving Markets}. Financial trading requires the ability to quickly adapt to fluctuating market conditions. Traditional methods often fall short, highlighting the necessity for models capable of responding to real-time data and adjusting strategies according to historical market trends.
%     \item \textbf{Ch4: Integration of Domain Knowledge.} Current models often struggle to integrate established methods such as expert guidance and advanced trading tools effectively, leading to a decline in both the effectiveness and depth of market analysis.
%     \item \textbf{Ch5. Reasoning for Actions}. The black-box nature of many sophisticated AI models, directly giving results of decisions without providing the reasoning process.
% \end{itemize}

\begin{itemize}[left=0cm]
    \item \textbf{Ch1: Insufficient Multimodal Data Processing Ability}. Processing numerical, textual, and visual market intelligence data significantly requires advanced analytical methods to extract key insights and predict market trends.
    \item \textbf{Ch2: Imprecise information retrieval}. Mixing retrieval with main tasks and relying on brief summaries causes imprecise searches, introducing irrelevant data and reducing performance.
    \item \textbf{Ch3: Adaptability in Rapidly Evolving Markets}. Financial trading requires the ability to quickly adapt to fluctuating market conditions. Traditional methods often fall short, highlighting the necessity for models capable of responding to real-time data and adjusting strategies according to historical market trends.
    \item \textbf{Ch4: Integration of Domain Knowledge.} Current models often struggle to integrate established methods such as expert guidance and advanced trading tools effectively, leading to a decline in both the effectiveness and depth of market analysis.
    \item \textbf{Ch5. Reasoning for Actions}. The black-box nature of many sophisticated AI models, directly giving results of decisions without providing the reasoning process.
\end{itemize}

To address the challenges of adapting the multimodal LLMs to the dynamic and information-rich financial trading tasks, we present \texttt{FinAgent}, a multimodal foundation agent that integrates both textual and visual information for a comprehensive analysis of market dynamics and historical trading patterns. 
Specifically, \texttt{FinAgent}'s market intelligence module processes multimodal data, such as numerical, textual, and visual, to provide precise analysis of financial market trends, offering insights for future trading tasks (\textbf{Ch1}). 
A uniquely designed dual-level reflection module is developed, capable of not only rapidly adapting to market dynamics but also enhancing the agent's ability to learn from historical data and improve its decision-making process (\textbf{Ch2}). 
\texttt{FinAgent} introduces a diversified memory retrieval system for the market intelligence and reflection modules, separating trading and retrieval tasks to enhance focus on their specific functions and minimize noise in the results (\textbf{Ch3}). Finally, the decision-making module incorporates expert knowledge, comprising both supplementary expert guidance and auxiliary expert strategies, to guide the agent's decisions. This emphasis on providing reasoned explanations for actions fosters trust in its financial decisions (\textbf{Ch4} \& \textbf{Ch5}). Specifically, our contributions are four-fold:
\begin{itemize}[left=0em]
\item We introduce the market intelligence module, which is able to extract key insights from multimodal datasets encompassing asset prices, visual representations, news, and expert analyses, offering a multifaceted view across various markets.
% \item We utilize a query input system that has two text fields and diverse retrieval types, separating summaries for trading decisions from specialized queries for memory retrieval. This approach facilitates a more targeted and efficient memory retrieval process.
\item We not only generate summaries for trading tasks but also provide query fields for retrieval tasks. These query texts include different retrieval types, tailored to enable focused retrieval of specific types of information.
\item Our duel-level reflection module combines a low-level reflection that analyzes market price movement for insights, while the high-level reflection assesses past trading decisions for improvement, emulating the learning process in decision-making. 
\item We employ a suite of tools in \texttt{FinAgent}, including expert guidance and technical indicator-based advanced trading strategies, to infuse domain knowledge in financial trading.
\end{itemize}
% We conduct a comprehensive experiments on 6 financial datasets, including stocks and Crypto, \texttt{FinAgent} significantly outperforms 9 state-of-the-art baselines in terms of 6 financial metrics. And the results demonstrate the effectiveness of our method.
With comprehensive experiments on 6 financial datasets, including stocks and Crypto, \texttt{FinAgent} significantly outperforms 12 state-of-the-art baselines in terms of 6 financial metrics  with over 36\% average improvement on profit. Specifically, a 92.27\% return (a 84.39\% relative improvement) is achieved on one dataset. Notably, \texttt{FinAgent} is the first advanced multimodal foundation agent designed for financial trading tasks.

\begin{table*}[htbp] 
\footnotesize
\centering
\caption{Comparison of \texttt{FinAgent} versus trading strategies and LLM agents. Brief introduction can be found in Section \ref{baselines}.}
\vspace{-0.2cm}
\renewcommand\tabcolsep{2pt} % column space
\renewcommand{\arraystretch}{0.5}
\resizebox{\textwidth}{!}{
\begin{tabular}{l|cccc|cc|ccccc} 
\toprule
\multirow{2}{*}{\textbf{Method}} & 
\multicolumn{4}{c|}{\textbf{Market Intelligence}} & 
\multicolumn{2}{c|}{\textbf{Tool Use}} & 
\multicolumn{5}{c}{\textbf{Inference \& Extension}} \\ 
\cmidrule(lr){2-5} \cmidrule(lr){6-7} \cmidrule(lr){8-12} 
& News & Reports & Price  & Visual Data  &  Info & Tools & Preference & Training Scheme & Planning & Explainability & Generalization \\
\midrule
Rule-based & \xmark & \xmark & \cmark & \xmark  & \xmark & \xmark & \xmark & Hyper-parameter Tuning & Myopic & - & Single trading task   \\
\midrule
RL method & \xmark & \xmark & \cmark & \xmark & \xmark & \xmark & \xmark & Model training & Sequential & \xmark & Single trading task   \\
\midrule
FinGPT & \cmark & \xmark & \cmark & \xmark & \xmark & \xmark & \xmark & LLM Fine-tuning & Myopic & \cmark & Limited trading tasks   \\
FinMem & \cmark & \cmark & \cmark & \xmark & \xmark & \xmark & \cmark & Reflection & Myopic & \cmark & Multiple trading tasks   \\
\midrule
\textbf{\texttt{FinAgent}} & \cmark & \cmark & \cmark & \cmark &\cmark & \cmark & \cmark & Reflection & Sequential & \cmark & Multiple trading tasks   \\
\bottomrule
\end{tabular}
}
\vspace{-0.2cm}
\label{tab:tool_comparison}
\end{table*}

\section{Related Work}
\subsection{LLM Agents for Decision Making}
% LLM
The field of artificial intelligence and natural language processing has reached a significant milestone with the emergence of LLMs like ChatGPT \cite{openai2021chatgpt} and GPT-4 \cite{openai2023gpt4}.  BloombergGPT \cite{wu2023bloomberggpt} introduced the first LLM in the finance domain, combining financial and text data, but without public access. FinGPT \cite{yang2023fingpt} proposed the first open-source finance LLMs, incorporating reinforcement learning with human feedback.

% LLM Agent for decision-making
While LLMs achieve impressive performance in NLP tasks \cite{touvron2023llama, chowdhery2023palm}, 
more works explored the capability of LLMs to function not just as language processors but as agents capable of performing complex tasks. Initiatives like AutoGPT \cite{yang2023autogpt} and MetaGPT \cite{hong2023metagpt}, Voyager \cite{wang2023voyager}, and AI agents \cite{sumers2023cognitive, park2023generative} expand LLMs' capabilities to complex tasks involving reasoning and collaboration, significantly advancing technology and impacting daily life. FinMEM \cite{yu2023finmem} presents an LLM agent with a human-aligned memory mechanism and character design for automated trading.

% LLM Agent with tool
Recently, there has been growing interest in enhancing LLM agents with external tools and modular methods as AI agents. Tool-augmented Language Models (TALM) \cite{parisi2022talm,thoppilan2022lamda,nakano2021webgpt,Schick2023ToolformerLM} have been evaluated through recent benchmarks, such as ScienceQA and TabMWP \cite{lu2022learn, lu2022dynamic, chen2023theoremqa, wang2023scibench, sun2023scieval, lu2023mathvista}, designed to assess their ability to tackle intricate reasoning challenges, particularly those requiring the use of external tools. These improvements enable LLMs to retrieve current information through web searches \cite{nakano2021webgpt} and to apply specialized knowledge from external sources \cite{yu2022generate}.

% Multimodality LMM
However, a major limitation of LLM agents is their dependence on text-based information, which limits their perception and interaction with the environment. Introducing models equipped with vision capabilities, such as the latest iteration of GPT-4V \cite{openai2023gpt4v}, marks a pivotal breakthrough. There has also been the emergence of multimodal agents \cite{zhang2023appagent, yang2023mmreact, lu2023chameleon} utilizing the visual capabilities of multimodal large language models to perform tasks previously unachievable by text-only agents. Most existing LLMs in finance focus on NLP tasks, and their potential in trading is not fully explored. \texttt{FinAgent} is a multi-modal, tool-augmented LLM foundation agent for financial trading to bridge the gap.

\subsection{AI for Financial Trading}
AI techniques have been widely used in various financial trading tasks. RNN-based such as GRU \cite{nelson2017stock} and LSTM \cite{wang19} models are popular for stock prediction since they are specifically designed to capture temporal patterns in sequential data. Another direction of work employs graph-based DL models to model pair-wise relations between stocks. For instance, \citet{feng2019temporal} enhance graph convolutional networks (GCNs) with temporal convolutions for mining inter-stock relations. \citet{sawhneydeep} focus on stock industry data and links between company CEOs. Tree-based models \cite{ke2017lightgbm} also achieve robust performance. \citet{xu2018stock} propose a variational autoencoder architecture to extract latent information from tweets. \citet{chen2019investment} enhance trading strategy design with the investment behaviors of professional fund managers. Other data sources such as economics news \cite{hu2018listening} and earning calls \cite{sawhney2020voltage} are also used to improve the prediction performance. \citet{sun2023mastering} introduce a novel three-stage ensemble learning method. Reinforcement learning \cite{sun2023reinforcement} has achieved success in finance with algorithms, platform \cite{sun2023trademaster}, and evaluation toolkits \cite{sun2023prudex}. However, most of these methods are hindered by their focus on price data and limited generalization, necessitating advanced techniques that can integrate multimodal intelligence and navigate complex market dynamics.

\section{Problem Formulation}
\label{problem_formulation}
We first introduce the Markov Decision Process (MDP) formulation of financial trading. Later on, we provide the formal formulation of \texttt{FinAgent}, which integrates LLMs into the RL pipeline to enable flexible reasoning and decision-making in financial trading. 

% TODO 这Section先不写，参考pangu agent的formulation（https://arxiv.org/abs/2312.14878）
\subsection{Financial Trading as MDP}
A financial trading task involves sequentially making investment decisions (e.g., buy/sell stocks) to maximize total profit under certain risk tolerance \cite{sun2023reinforcement}. We formulate it as an MDP under a classic RL scenario following \cite{sun2023trademaster, finrl2020}, where an agent (investor) interacts with an environment (the financial market) to make actions (investment decisions) at discrete time to earn rewards (profits). The MDP is constructed by a 5-tuple $(\mathit{S}, \mathcal{A}, \mathcal{T}, R, \gamma)$. Specifically, $\mathit{S}$ is a finite set of states. $\mathcal{A}$ is a finite set of actions. The state transition function $\mathcal{T}:\mathit{S} \times \mathcal{A} \times \mathit{S} \rightarrow [0, 1]$ encapsulates transition probabilities between states based on chosen actions. The reward function $R: \mathit{S} \times \mathcal{A} \rightarrow R$ quantifies the immediate reward of taking an action in a state. The discount factor is $\gamma \in [0, 1)$. A policy $\mathcal{\pi} : \mathit{S} \times \mathcal{A} \rightarrow [0, 1]$ assigns each state $s \in \mathit{S}$ a distribution over actions, where $a \in \mathcal{A}$ has probability $\pi(a|s)$. During training, the agent is in charge of making investment decisions at each time step through one whole trading period and tries to learn an optimal policy (investment strategy) that maximizes the expected sum of discounted reward (overall profit): $\pi_{\theta^{*}} = \arg\max_{\pi_{\theta}}\mathbb{E}_{\pi_{\theta}}\lbrack\sum^{T}_{i=0}\gamma^{i}r_{t+i}|s_{t}=s\rbrack$. 

Specifically, we focus on single asset (e.g., stock or Crypto) trading. A \textit{state} represents RL agents’ perception on the financial market based on price information, limited order book \cite{qin2023earnhft},  technical indicators, trend prediction \cite{ye2020reinforcement}, financial news \cite{sawhney2021quantitative}, experts’ investment behaviors \cite{ding2018investor} and overall market status \cite{wang2021deeptrader}. The \textit{action space} includes three choices to buy, sell or hold the asset \cite{deng2016deep, liu2020adaptive}. The \textit{reward function} leverages the change of market capitals (earned/lost money) \cite{liu2020adaptive} with consideration of commission fee \cite{sun2023trademaster,wang2021commission}.

\subsection{Problem Formulation}
We further integrate multimodal LLMs into the RL framework \cite{christianos2023panguagent}, enabling the flexible definition of the reasoning processes. In \texttt{FinAgent} formulation, we focus on the necessity of defining, learning, and applying these processes independently. We extend the classic RL optimization problem for \texttt{FinAgent} as follows:
% In this part of our discussion, we delve further into integrating LMMs into the RL framework. Our focus is on the necessity of defining, learning, and applying these processes independently. This approach enables users to customize and reconfigure nested structures in ways that are most beneficial for their specific tasks. Similar to Pangu-Agent \cite{christianos2023panguagent}, we can rewrite the LMMs agent optimization problem as follows:
\begin{equation}
\footnotesize
\pi_{\theta^{*}} = \arg \max_{\pi_{\theta}} \mathbb{E}_{\pi_{\theta}} \left[ \sum\nolimits^{T}_{i=0} \gamma^{i} r_{t+i} | s_{t}=s, \mu_{t} = \mu \right],
\end{equation}
where $r_{t}$ is the reward at the time step $t$ that depends on the environmental state $s_{t}$ and action $a_{t}$. $\mu(\cdot)$ are specialized modules that encapsulate beneficial internal reasoning processes. Note that a state contains multimodal information including textual, numerical, and visual data. Faced with a task $\lambda$ and equipped with a memory $Mem^{\lambda}_{t}$ and a tool $Tool^{\lambda}_{t}$, \texttt{FinAgent} acting as the multimodal LLM agent, determines its action $a_{t}$ through the following process:
% We have drawn inspiration from the approach of Tool-augmented Language Models to enhance the reasoning capabilities of \texttt{FinAgent} by utilizing expert knowledge and external tools. Specifically, the tools considered for \texttt{FinAgent} are auxiliary information (e.g., expert guidance from experienced investment professionals) and auxiliary agents (e.g., advanced strategies).
% Recent advances with LLMs agent and LMMs agent have moved us beyond strict task limits and intense engineering, introducing new frameworks like memory management and tool use to address complex problems. They support Chain of Thought (CoT) \cite{wei2022chain, tutunov2023can}  processes using LLMs for idea generation and refinement, and serve as dynamic memory systems for efficient information retrieval and use. Inspired by Tool-augmented Language Models, we've enhanced \texttt{FinAgent}'s reasoning with expert knowledge and tools, specifically incorporating auxiliary information (e.g., expert guidance) and auxiliary agents (e.g., advanced strategies). In principle, it's possible to integrate these specialized modules for LMMs agent into any phase of the RL pipeline. Faced with a task $\lambda$ and equipped with a memory $Mem^{\lambda}_{t}$ and a tool $Tool^{\lambda}_{t}$, \texttt{FinAgent} acting as the LMMs agent, determines its action $a_{t}$ through the following process:
\begin{equation}
\footnotesize
\label{equ:finagent}
    \begin{split}
    \pi_{\texttt{FinAgent}} \left(a_{t}|s_{t}, \mu_{t} \right) & \equiv \mathcal{D}^{\lambda} \left(LLM \left( \phi^{\lambda}_{D} \left( s_{t}, \mu_{t} \right) \right) \right) \\
        \mu_{t} & = \mu(s_{t}, Mem^{\lambda}_{t}, Tool^{\lambda}_{t})\\
    \end{split}
\end{equation}
where $\phi(\cdot)$ is a task-relevant prompt generator. The prompt is then passed to a multimodal LLM, from which a response is generated. Finally, the response is parsed through the task-specific action parsing function $\mathcal{D}^{\lambda}(\cdot)$ to perform compatible actions in the environment.

\texttt{FinAgent} is a multimodal LLMs agent in this framework specifically designed for financial trading, which contains five core modules, namely market intelligence module (\textit{M}), memory module (\textit{Mem}), low-level reflection module (\textit{L}), high-level reflection module (\textit{H}) and decision-making module (D). We can define the $\mu_{t}$ and other modules as follows:
\begin{equation}
\label{equ:mu_module}
\footnotesize
    \begin{split}
        \mu_{t} = \mu(s_{t}, Mem^{\lambda}_{t}, Tool^{\lambda}_{t}) & = \mu(M^{\lambda}_{t}, L^{\lambda}_{t}, H^{\lambda}_{t}, Tool^{\lambda}_{t}) \\
        M^{\lambda}_{t} = LLM(\phi^{\lambda}_{M} & (s_{t}, Mem^{M,\lambda}_{t})) \\
        L^{\lambda}_{t} = LLM(\phi^{\lambda}_{L} & (M^{\lambda}_{t}, KC_{t}, Mem^{L,\lambda}_{t})) \\
        H^{\lambda}_{t} = LLM(\phi^{\lambda}_{H} & (M^{\lambda}_{t}, TC_{t}, Mem^{H,\lambda}_{t})),
    \end{split}
\end{equation}
where $M$, $Mem$, $L$, $H$, $D$ correspond to each module respectively, $Mem^{*,\lambda}$ denotes the memory of $M$, $L$, and $H$. $KC$ and $TC$ represent the Kline chart and Trading chart. $\phi^{\lambda}_{*}$ denotes the prompt generator corresponding to each module associated with task $\lambda$.

Therefore, with the integration of memory mechanism, augmented tools, and several designed modules, the overall objective of \texttt{FinAgent} is to find policies as described in Eq. (\ref{equ:finagent}) to optimize total discounted returns:
\begin{equation}
\footnotesize
    \begin{split}
        \pi^{*}_{\texttt{FinAgent}} & = \arg \max_{\pi(\cdot), \bm{\mu}(\cdot)} \mathbb{E}_{\pi} \left[ \sum\nolimits^{T}_{i=0} \gamma^{i} r_{t+i} | s_{t}=s, \mu_{t} = \mu \right] \\
        \text{s.t.} \quad & \pi \left(a_{t}|s_{t}, \mu_{t} \right) = \mathcal{D}^{\lambda} \left(LLM \left( \phi^{\lambda}_{D} \left( s_{t}, \mu_{t} \right) \right) \right) \  with \  Eq. (\ref{equ:mu_module}) \quad \forall t. \\
    \end{split}
\end{equation}

% We have drawn inspiration from the approach of Tool-augmented Language Models to enhance the reasoning capabilities of \texttt{FinAgent} by utilizing expert knowledge and external tools. Specifically, the tools considered for \texttt{FinAgent} are: 
% \begin{itemize}[left=0em]
%     \item \textit{Auxiliary Information}: Expert guidance from experienced investment professionals as tools enhances trading strategies by providing informed decision-making,  adaptability to market changes, specialized knowledge, and a depth of analytical insight that cannot be accurately perceived by algorithmic methods.
%     \item \textit{Auxiliary Agents}: Incorporating auxiliary agents (e.g., conventional trading strategies) as tools brings the advantage of leveraging well-established, historically validated insights into market dynamics, thereby significantly improving the accuracy and reliability of decision-making in trading environments. This approach not only capitalizes on the depth of historical market analysis but also provides a solid foundation for strategic planning and risk management.
% \end{itemize}

% 介绍Tools包含两个部分，一个是aux information提供辅助信息，一个是aux agents提供辅助决策和决策解释
% https://arxiv.org/pdf/2304.09842.pdf Tool-Augmented language model \cite{lu2023chameleon}

\section{FinAgent Framework}
\label{sec:finagent}
As shown in Figure \ref{fig:network}, the \texttt{FinAgent} framework comprises five core modules. Specifically, the market intelligence module (\textbf{\S\ref{market_intelligence_module}}) is responsible for collecting, collating, summarizing, and analyzing market information, which includes daily updates on stock news, prices, and monthly and quarterly financial reports. The low-level reflection module (\textbf{\S\ref{low_level_reflection_module}}) establishes the inherent correlation between market intelligence and price changes. And the high-level reflection module (\textbf{\S\ref{high_level_reflection_module}}) involves reflecting on market conditions, price changes, and other factors in the context of outcomes from past trading decisions, which aims to derive insights from previous experiences and identify potential improvement in profitability by assessing the efficacy of historical decisions and offering recommendations for future decision-making processes. The primary role of the memory module (\textbf{\S\ref{memory_module}}) is to support the aforementioned three modules by offering storage capabilities and vector retrieve functions. The tool-augmented decision-making module (\textbf{\S\ref{decision_module}}) integrates the aforementioned information, along with augmented tools and trader preferences, to make final investment decisions with a comprehensive analysis.

%要解释market intelligence可能包含哪些内容，参考：https://www.marketingevolution.com/marketing-essentials/marketing-intelligence#:~:text=Marketing intelligence is everyday data,consumer trends%2C and market opportunities.
%这一部分主要想介绍Market Intelligence Module的作用，它主要用于收集和summarize市场情报。主要有三个部分latest market intelligence summary，diversified query和past market intelligence summary。对应于contribution 2和3
% 对于query设计：会生成两个字段，一个是query用于diversified query，一个是summary，要说明分成两个任务的必要性，因为query任务和trading任务是不一样的，所以需要生成对应的query和summary分别做两个任务
%diversified query的做法是生成的query字段的文本就体现出了不同的multi-perspective（例如short term，medium term和long term），可以使得query的新闻是不同感受尺度的

\begin{figure*}[htbp]
\vspace{-0.2cm}
\captionsetup{skip=2pt}
  \centering
    \includegraphics[width=1.0\textwidth]{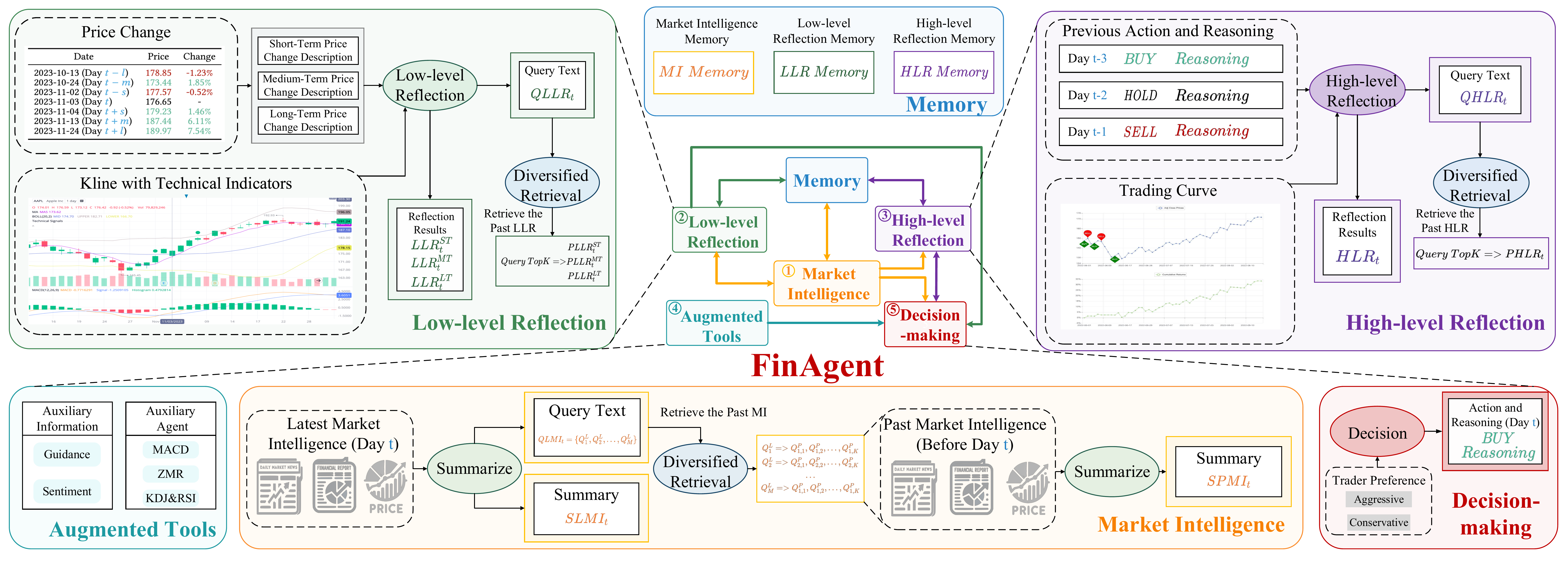}
    \caption{The overall architecture of FinAgent. The ordinal numbers in the figure represent the order of execution, where augmented tools are implemented with the decision-making module.}
    \label{fig:network}
\vspace{-0.3cm}
\end{figure*}

\subsection{Market Intelligence Module}
\label{market_intelligence_module}
To make profitable investment decisions, it is beneficial to collect, summarize, analyze, and extract key insights from various multimodal financial data sources.
We design the market intelligence module to achieve this goal. Market intelligence typically involves daily data about the macro environment, current market conditions or investors’ sentiments that inform investment and trading decisions. In \texttt{FinAgent}, we harness the power of both the latest and historical news, financial reports, and asset prices related to the targeted asset in order to inform and optimize trading decisions.
% As depicted in Figure \ref{fig:network}, the architecture of the three components: the \textit{Latest Market Intelligence} for day $t$, an archive of relevant \textit{Past Market Intelligence} up to day $t$, and the process of retrieving relevant past market intelligence utilizing key insights derived from the latest market intelligence through a \textit{Diversified Retrieval Operation}.

\noindent
\textbf{Latest Market Intelligence.}
This module mainly consists of asset news and daily asset prices. However, it is not confined to these elements alone. Any information impacting the market can be encompassed within our framework as part of the latest market intelligence. The objective of this component is to evaluate the sentiment\footnote{Market intelligence can be categorized as positive, negative, or neutral based on its impact on market perceptions and potential outcomes.} of each market intelligence item regarding its influence on future asset prices and to provide a detailed summary of whether the market has recently exhibited bearish or bullish tendencies, thereby assisting in informed decision-making.

Nevertheless, historical data can offer insights into patterns that might influence future pricing and potentially affect current and upcoming market dynamics. For instance, if a past product launch significantly boosted a company's stock, a recent launch might have a similar effect\footnote{Some news will detail the percentage increase or decrease in a company's stock price after some event occurs.}. We hope to incorporate these historical experiences and patterns into \texttt{FinAgent}'s considerations. This inspired us to add two additional functional layers: retrieving relevant information from past market intelligence and summarizing key insights and historical experiences from them.

\noindent
\textbf{Diversified Retrieval Operation.} A straightforward approach involves using the summary of the latest market intelligence as the query text and then employing an LLM to extract its semantically rich embeddings. This allows for retrieving past market intelligence with similar content through vector similarity. However, adopting this approach inevitably comes with two significant shortcomings: i) the summary of recent market intelligence is primarily aimed at supporting subsequent trading decision tasks, not for retrieval tasks. The significant gap between these two objectives can lead to unsatisfactory retrieval results; ii) some noise unrelated to the retrieval task may be contained in the summary, directly affecting the retrieval results. To address these challenges, diversified retrieval is implemented in \texttt{FinAgent}. Specifically, we have introduced an additional query text field to the output of the latest market intelligence component, which is dedicated to serving retrieval tasks in parallel with the summary that caters to trading tasks. It is worth emphasizing that we can define various retrieval types\footnote{The retrieval types include short-term, medium/long-term market impacts, asset price increase/decrease, market trends bearish/bullish, news/reports, etc.} to enable an agent to retrieve past market intelligence from multiple perspectives, in multiple senses, and with a purpose. As shown in Figure \ref{fig:network}, there are $M$ retrieval types, so retrieving top $K$ historical market intelligence separately can form a combination of $M \times K$ market intelligence in the past. This approach assigns specific retrieval types to each piece of historical information accompanying the summaries. This nuanced labeling facilitates a more targeted and efficient search and retrieval process.
% , enhancing the precision of historical data utilization.

\noindent
\textbf{Past Market Intelligence.} Once similar past market intelligence is searched, it undergoes the summarising step, delivering key insights tailored to augment trading decisions. This meticulous approach ensures that only the most relevant information is incorporated, mitigating the impact of noise and maximizing the utility of historical data in informing trading strategies.

% 四段内容
% 第一段是mi的详细内容包含哪些
% latest mi需要对他summary关键信息，市场情绪等。为后面trading服务。但是我们还需要过去一些相似的mi辅助我们决策，或者说检索一些相似的经历为后续决策提供参考。引出retrieve（query）这个概念
%详细介绍retrieve（三个部分：diversified retrieve的motivation；详细设计；他的优势和作用），做法是输出两个字段，一个“query”用于retrieve，一个summary用于trading decision
%past mi的summary，他就没有query字段了，只有summary用于未来决策提供辅助信息

% 这一部分主要介绍memory是vector storage，然后有三个组成部分，分别是market intelligence memory，low-level reflection memory和high-level reflection memory
% 基于query字段文本向量相似度做retrieve
\subsection{Memory Module}
\label{memory_module}

The memory mechanism \cite{park2023generative, zhang2023appagent,christianos2023panguagent} is crucial in LLM Agents for effectively handling extensive texts, grasping the context, ensuring the coherence of conversations, and improving the agent's comprehension and logical abilities. In the context of multimodal LLM agents for financial trading, memory mechanisms play a crucial role in three main aspects: i) \textit{Acuity}. This feature enables multimodal LLM agents to use market news, financial reports, and other information for better market forecasting. By analyzing historical data and current events, these agents can predict market trends and asset prices more accurately, aiding in effective trading decisions. ii) \textit{Adaptability}. As market conditions change rapidly, memory mechanisms allow multimodal LLM agents to quickly learn and adapt. By continuously analyzing market data and trading outcomes, these agents adjust their strategies to handle volatility and seize new opportunities. iii) \textit{Amendability}. It helps multimodal LLM agents learn from past mistakes and successful trades. By reflecting on these experiences, agents can avoid repeating errors and improve their trading strategies. This continuous learning enhances their performance and creates more robust, efficient trading strategies.

To realize the \textit{3A} superiority - \textit{Acuity}, \textit{Adaptability}, and \textit{Amendability} - in the memory mechanism, our development of the memory module employed a vector storage architecture. This module is composed of three main components: market intelligence memory (service for (\textbf{\S\ref{market_intelligence_module}})), low-level reflection memory (service for (\textbf{\S\ref{low_level_reflection_module}})), and high-level reflection memory (service for (\textbf{\S\ref{high_level_reflection_module}})). As shown in Figure \ref{fig:network}, the summarize operation creates a query text field for each module, enhancing memory storage and retrieval. The market intelligence module uniquely retrieves past data through query text, using vector representations for efficient matching based on the vector similarity. All analyses and summaries from the market intelligence, low-level reflection, and high-level reflection modules are stored in the memory module. This integration equips the agent with extensive market data and insights, improving its decision-making capabilities.

\begin{figure*}[ht]
\captionsetup{skip=3pt}
  \centering
    \includegraphics[width=1.0\textwidth]{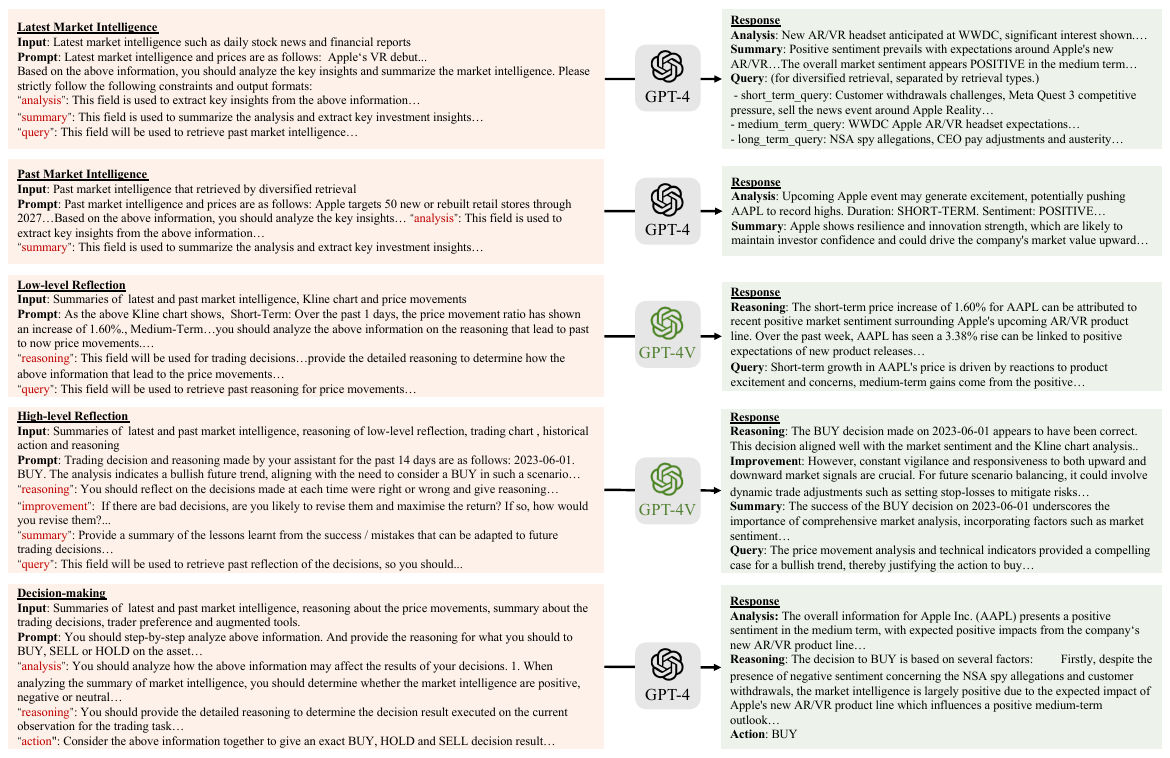}
    \caption{Case studies of FinAgent. We only display the partial prompt for brevity.}
    \label{fig:network}
\vspace{-0.2cm}
\end{figure*}

\subsection{Reflection Module}
% 介绍low-level reflection的步骤，详细说明它反思的是当前的observation和价格趋势变化的联系（价格趋势变化包含未来的价格趋势变化），并且这里用的plain query，不会再多尺度的query过去的reflection了
\label{sec:reflection}
A reflection module is incorporated into the agent's design to emulate the cognitive learning process inherent in human decision-making. The reflection framework is divided into low-level reflection and high-level reflection, each serving distinct purposes to enhance the agent's trading decisions. The low-level reflection module involves reflecting on the relationship between the agent's observations (e.g., news, financial reports, Kline chart and technical indicators) and the resultant price movements in the market, drawing connections between the provided information and the actual price changes. Whereas the high-level reflection step examines past decisions, tracking both the agent's actions and the subsequent price movements in order to learn from past successes or mistakes. 

\noindent
\textbf{Low-level Reflection Module}
\label{low_level_reflection_module}
The primary focus of the low-level reflection module is to analyze the connection between the given market intelligence together with the Kline chart and technical indicators and past and future price changes to enhance decision-making. After taking in the price change data, the module generates detailed analysis for varying temporal horizons, spanning short-term, medium-term to long-term perspectives. The emphasis is placed on identifying potential patterns in the price movements of the targeted stock and deriving insights from how the given market intelligence summaries and Kline chart analysis can lead to such price movements. In order to facilitate future access and reference, the module generates a query field containing a concise summary of learned lessons, ensuring efficient retrieval and application of insights in subsequent decision-making endeavors.

\begin{table}[htbp]
\vspace{-0.2cm}
\renewcommand{\arraystretch}{0.9}
\footnotesize
\centering
\setlength{\abovecaptionskip}{0.1cm}
\caption{Differences between reflection of low and high}
\begin{threeparttable}
\footnotesize
\renewcommand{\arraystretch}{0.9}
\setlength{\tabcolsep}{3pt}
\resizebox{\linewidth}{!}{
\begin{tabular}{p{2.5cm} c c c}
\toprule
\multicolumn{1}{p{2.5cm}}{Reflection} & \multicolumn{1}{c}{Low-level Reflection} & \multicolumn{1}{c}{High-level Reflection} \\
\midrule
\multicolumn{1}{p{2.5cm}}{Target} & \multicolumn{1}{c}{Price Movements} & \multicolumn{1}{c}{Trading Decisions} \\
\multicolumn{1}{p{2.5cm}}{Visual Data} & \multicolumn{1}{c}{Kline Chart} & \multicolumn{1}{c}{Trading Chart} \\
\multicolumn{1}{p{2.5cm}}{Market Understanding} & \multicolumn{1}{c}{Micro} & \multicolumn{1}{c}{Macro}  \\
\multicolumn{1}{p{2.5cm}}{Function} & \multicolumn{1}{c}{Adaptability} & \multicolumn{1}{c}{Amendability}  \\
\bottomrule
\end{tabular}
}
\end{threeparttable}
\label{tab:differences_of_low_and_high}
\vspace{-0.3cm}
\end{table}

% 介绍low-level reflection的步骤，详细说明它反思的是从历史决策中学习经验，看看是否做错和是否有提升的点。并且这里用的plain query，不会再多尺度的query过去的reflection了
\noindent
\textbf{High-level Reflection Module}
\label{high_level_reflection_module}
The high-level reflection module is designed to provide analysis and reflections on past trading decisions. Besides the past trading decisions and their underlying reasoning, this module incorporates a graphical representation of buy and sell points on a trading chart, coupled with a cumulative return plot, to offer an intuitive representation of the efficacy of historical decisions. The initial phase assesses each trading decision's correctness, identifying successes and mistakes. Subsequently, the module recommends improvements or corrective actions tailored to each identified mistake or success, fostering a continuous learning process. Beyond individual decision analysis, the module generates overarching lessons from both successes and mistakes, providing a summary that can be adapted to future trading decisions and a query text to facilitate the retrieval of relevant reflections. This iterative learning process equips the agent with a dynamic knowledge base that evolves with each decision and allows the trading agent to draw connections between similar scenarios, applying learned lessons for more informed decision-making.

\vspace{-0.3cm}
\subsection{Tool-Augmented Decision-making Module}
\label{decision_module}
The decision-making module integrates key inputs, including market intelligence summaries, low-level reflection about price movement analyses, and reflections on past decisions. Augmented tools with professional investment guidance and traditional trading strategies like MACD Crossover, KDJ with RSI Filter and Mean Reversion are also considered. The module analyzes sentiment in market intelligence, predicts bullish or bearish trends from price movements, reflects on lessons learned, and evaluates professional guidance and traditional indicators. Decisions are derived from combining insights from these analyses, also considering the current financial position, leading to a final decision—whether to buy, sell, or hold the asset. Leveraging the Chain-of-Thought (COT) approach and in-context learning principles, our trading decision-making module not only executes trades but also provides reasoning, ensuring that each decision is rooted in a comprehensive understanding of market dynamics and contextual knowledge.

\section{Experiment Setup}
Our research aims to conduct a thorough evaluation of \texttt{FinAgent}'s trading effectiveness, underscoring its unique capability to function efficiently with a significantly reduced historical data training window. This assessment also involves leveraging multimodal data inputs, incorporating both informational and agent-assistive augmented tools, along with a multi-perspective diversified retrieval. This approach is intended to enhance the understanding of market dynamics and sentiments, enabling more comprehensive and logical decision-making processes along with substantiated explanations. To validate its effectiveness, we have conducted a series of experiments to address the following research questions (\textbf{RQs}):
\begin{itemize}[leftmargin=*]
    \item \textbf{RQ1}: Is \texttt{FinAgent} outperforming current state-of-the-art trading agents and handling tasks that challenge other algorithms?
    \item \textbf{RQ2}: What is the effectiveness of each component of \texttt{FinAgent} in contributing to its overall performance?
    \item \textbf{RQ3}: Does the integration of augmented tools in \texttt{FinAgent} lead to a distinguishable improvement in its trading performance?
    \item \textbf{RQ4}: How effective is the diversified retrieval in \texttt{FinAgent}?
\end{itemize}

\subsection{Datasets}

\vspace{-0.1cm}
\begin{table}[htbp]
\vspace{-0.2cm}
\footnotesize
\renewcommand{\arraystretch}{1.0}
\centering
\setlength{\abovecaptionskip}{0.1cm}
\caption{Dataset statistics detailing the chronological period and the number of each data source for each asset.}
\begin{threeparttable}
\footnotesize
\renewcommand{\arraystretch}{0.9}
\setlength{\tabcolsep}{3pt}
\resizebox{1\linewidth}{!}{
\begin{tabular}{p{2.0cm}c c c c c c c}
\toprule
\multicolumn{1}{p{2.0cm}}{Asset} & \multicolumn{1}{c}{AAPL} & \multicolumn{1}{c}{AMZN} & \multicolumn{1}{c}{GOOGL} & \multicolumn{1}{c}{MSFT} & \multicolumn{1}{c}{TSLA} & \multicolumn{1}{c}{ETHUSD} \\
\midrule
\multicolumn{1}{p{2.0cm}}{Trading Date} & \multicolumn{6}{c}{From 2022-06-01 to 2024-01-01 (398 trading days)}\\
\multicolumn{1}{p{2.0cm}}{Asset Price} & \multicolumn{6}{c}{398 $\times$ (open, high, low, close, adj\_close)} \\
\multicolumn{1}{p{2.0cm}}{Visual Data} & \multicolumn{6}{c}{398 $\times$ (Kline\ Chart, Trading\ Chart)} \\
\midrule
\multicolumn{1}{p{2.0cm}}{Asset News} & 9748 & 10007 & 7923 & 8178 & 10076 & 2611 \\
\multicolumn{1}{p{2.0cm}}{Expert Guidance} & 593 & 509 & 488 & 393 & 600 & $-$ \\

\bottomrule
\end{tabular}
}
\end{threeparttable}
\label{tab:dataset}
\vspace{-0.2cm}
\end{table}

To conduct a thorough evaluation of \texttt{FinAgent}, we evaluate it across 6 real-world datasets. These included five datasets from the US stock markets, and one is the cryptocurrency. Each of them has multiple forms of data that come from various sources. Specifically, i) \textbf{Asset Price} at the day-level, including price data for open, high, low, close, and adj close. ii) \textbf{Visual Data} consists of historical Kline charts and trading charts, which are visual representations of asset market data and trading process on a daily basis. iii) \textbf{Asset News} coverage with daily updates from various esteemed sources such as Bloomberg Technology, Seeking Alpha and CNBC Television, ensuring a diverse and thorough perspective on the financial markets. iv) \textbf{Expert Guidance} provided by financial experts as the auxiliary information, aiming to furnish a thorough and well-rounded comprehension of market status. We summarize statistics of the 6 datasets in Table \ref{tab:dataset} and further elaborate on them in Appendix \ref{app:datasets}.

\definecolor{FBest}{RGB}{198,70,17} % first best results
% #E4A304
\definecolor{SBest}{RGB}{228,163,4} % second best results
% #548235
\definecolor{TBest}{RGB}{84,130,53} % third best results

\begin{table*}[thbp]
\renewcommand{\arraystretch}{0.9}
\footnotesize
\caption{Performance comparison of all methods on six profitable metrics. Results in red, yellow and green show the best, second best and third best results on each dataset. The improvement row is the FinAgent over the best-performing baselines.}
\vspace{-0.2cm}
\setlength{\abovecaptionskip}{0.1cm}
\centering
\begin{threeparttable}
\setlength{\tabcolsep}{2.5pt}
\resizebox{\linewidth}{!}{ 
\begin{tabular}{clccclccclccclccclccclccc}
\toprule
\multirow{3}{*}{Categories} & \multirow{3}{*}{Models} & \multicolumn{3}{c}{AAPL} & & \multicolumn{3}{c}{AMZN} & & \multicolumn{3}{c}{GOOGL} & & \multicolumn{3}{c}{MSFT} & & \multicolumn{3}{c}{TSLA} & & \multicolumn{3}{c}{ETHUSD}   \\ 
\cmidrule{3-5}\cmidrule{7-9}\cmidrule{11-13}\cmidrule{15-17}\cmidrule{19-21}\cmidrule{23-25}
&& ARR\%$\uparrow$  & SR$\uparrow$  & MDD\%$\downarrow$  & & ARR\%$\uparrow$  & SR$\uparrow$  & MDD\%$\downarrow$  & & ARR\%$\uparrow$  & SR$\uparrow$  & MDD\%$\downarrow$ & & ARR\%$\uparrow$  & SR$\uparrow$  & MDD\%$\downarrow$ & & ARR\%$\uparrow$  & SR$\uparrow$  & MDD\%$\downarrow$ & & ARR\%$\uparrow$  & SR$\uparrow$  & MDD\%$\downarrow$  \\ 
\midrule
Market 
& B\&H& 13.0& 0.6& 14.78&& \textcolor{TBest}{42.33}& 1.08& 17.38&& 22.47& 0.71& 12.97&& 22.49& 0.84& 12.92&& 37.4& 0.72& 32.65&& 29.26& 0.87& 23.21\\
\midrule
\multirow{3}{*}{\begin{tabular}[c]{@{}c@{}}Rule-based\end{tabular}}
& MACD& 11.86& 0.72& 10.38&& 14.27& 0.71& \textcolor{TBest}{7.84}&& -18.0& -0.89& 20.07&& 15.23& 0.77& 8.34&& -4.9& -0.02& 14.15&& 10.24& 0.47& 24.32\\
& KDJ\&RSI& 2.17& 0.17& 11.88&& 19.38& 0.65& 17.27&& 24.39& \textcolor{FBest}{2.13}& \textcolor{FBest}{2.03}&& 18.84& 1.06& 7.78&& 2.14& 0.17& 24.73&& 8.87& 0.51& 16.95\\
& ZMR& -3.91& -0.22& \textcolor{TBest}{8.88}&& 18.73& 0.84& 7.89&& 32.51& \textcolor{TBest}{1.45}& \textcolor{SBest}{5.38}&& 9.86& 0.71& \textcolor{TBest}{6.22}&& -7.28& -0.09& 19.9&& 29.35& \textcolor{TBest}{1.23}& 13.11\\
\midrule
\multirow{3}{*}{\begin{tabular}[c]{@{}c@{}}ML\&DL-based\end{tabular}}
& LGBM& 16.93& \textcolor{FBest}{1.47}& \textcolor{FBest}{2.52}&& 29.34& 0.72& 17.41&& 24.77& 0.7& 12.98&& 19.28& 0.67& 12.96&& 15.57& 0.84& \textcolor{FBest}{3.88}&& 24.91& 0.72& 22.96\\
& LSTM& 10.97& 0.54& 11.95&& 15.91& 0.46& 17.41&& 24.86& 0.7& 12.98&& 18.86& 0.68& 11.75&& 17.36& 0.78& \textcolor{SBest}{4.44}&& \textcolor{TBest}{36.09}& 1.03& 21.5\\
& Transformer& 17.11& 0.96& \textcolor{SBest}{7.53}&& 32.66& \textcolor{TBest}{1.11}& \textcolor{FBest}{4.96}&& 13.69& 0.46& 12.93&& 17.44& \textcolor{TBest}{1.46}& \textcolor{FBest}{2.59}&& 39.7& \textcolor{SBest}{1.04}& \textcolor{TBest}{8.17}&& 31.0& 1.02& 12.93\\
\midrule
\multirow{3}{*}{\begin{tabular}[c]{@{}c@{}} RL-based\end{tabular}}
& DQN& 7.92& 0.4& 14.88&& 27.43& \textcolor{SBest}{1.17}& \textcolor{SBest}{5.27}&& \textcolor{TBest}{34.4}& 1.39& \textcolor{TBest}{7.15}&& \textcolor{TBest}{30.44}& 1.18& 10.56&& 15.07& 0.44& 28.12&& 29.81& 1.18& \textcolor{FBest}{9.53}\\
& SAC& \textcolor{SBest}{24.84}& \textcolor{TBest}{1.12}& 11.98&& 38.33& 1.07& 13.84&& 23.8& 0.75& 13.07&& 22.02& 0.82& 12.92&& \textcolor{TBest}{42.22}& 0.87& 26.19&& 17.84& 0.76& \textcolor{SBest}{10.06}\\
& PPO& 13.26& 0.61& 14.78&& 21.17& 0.7& 13.84&& \textcolor{SBest}{38.29}& 1.3& 8.45&& 11.32& 0.48& 17.51&& 33.64& 0.78& 28.35&& 34.75& \textcolor{FBest}{1.31}& \textcolor{TBest}{11.12}\\
\midrule
\multirow{2}{*}{\begin{tabular}[c]{@{}c@{}} LLM-based \end{tabular}}
& FinGPT& -5.46& -0.17& 16.23&& \textcolor{SBest}{42.93}& 1.1& 18.94&& 12.28& 0.44& 13.0&& 25.1& 0.97& 9.84&& 38.43& 0.75& 31.47&& 21.57& 0.68& 25.56\\
& FinMem& \textcolor{TBest}{23.78}& 1.11& 10.39&& 40.07& 1.03& 18.53&& 31.27& 1.11& 8.97&& \textcolor{SBest}{40.58}& \textcolor{SBest}{1.5}& 7.48&& \textcolor{SBest}{50.04}& \textcolor{TBest}{0.92}& 25.77&& \textcolor{FBest}{44.72}& \textcolor{SBest}{1.27}& 13.59\\
\midrule
\multirow{1}{*}{\begin{tabular}[c]{@{}c@{}} Ours\end{tabular}}
& \textbf{\texttt{FinAgent}}& \textcolor{FBest}{31.9}& \textcolor{SBest}{1.43}& 10.4&& \textcolor{FBest}{65.1}& \textcolor{FBest}{1.61}& 13.2&& \textcolor{FBest}{56.15}& \textcolor{SBest}{1.78}& 8.45&& \textcolor{FBest}{44.74}& \textcolor{FBest}{1.79}& \textcolor{SBest}{5.57}&& \textcolor{FBest}{92.27}& \textcolor{FBest}{2.01}& 12.14&& \textcolor{SBest}{43.08}& 1.18& 12.72\\
\midrule

\multicolumn{2}{c}{Improvement(\%)} 
&   28.39   &  -     &   -    &       &    51.64   &   37.61    &    -   &       &   46.64    &   -    &     -  &       &   10.25    &   19.33    &   -    &       &   84.39    &   93.27    &   -    &       &    -   &   -    & - \\
\bottomrule
\end{tabular}
}
\end{threeparttable}
\label{tab:baselines}
\vspace{-0.2cm}
\end{table*}

Our diversified portfolio includes five major stocks: Apple Inc. (AAPL), Amazon.com Inc. (AMZN), Alphabet Inc. (GOOGL), Microsoft Corporation (MSFT), and Tesla Inc. (TSLA) and a prominent cryptocurrency named Ethereum (ETHUSD). This selection aims to showcase \texttt{FinAgent}'s versatility and consistency across various financial assets. Chosen for their extensive news coverage and representation of different market sectors, these data provide a robust basis for assessing \texttt{FinAgent}'s generalization capabilities across diverse financial environments. For dataset split, the data from the latter half of the year is allocated for testing (2023-06-01 $\sim$ 2024-01-01) purposes, while the data from the penultimate year is utilized for training (2022-06-01 $\sim$ 2023-06-01).

\subsection{Evaluation Metrics}
\label{evaluation_metrics}
We compare \texttt{FinAgent} and baselines in terms of 6 financial metrics following \cite{sun2023trademaster,qin2023earnhft}, which include 1 profit metric: annual return rate (ARR), 3 risk-adjusted profit metrics: Sharpe ratio (SR), Calmar ratio (CR), Sortino ratio (SOR), and 2 risk metrics: maximum drawdown (MDD), volatility (VOL). Definitions and formulas are as follows:
\begin{itemize}[left=0em]
    \item \textbf{Annual Rate of Return (ARR)} is the annualized average return rate, calculated as $ARR = \frac{V_T-V_0}{V_0} \times \frac{C}{T}$, where $T$ is the total number of trading days, and $C = 252$ is the number of trading days within a year. $V_T$ and $V_0$ represent the final and initial portfolio values.
    \item \textbf{Sharpe Ratio (SR)} measures risk-adjusted returns of portfolios. It is defined as $SR = \frac{\mathbb{E[\textbf{r}]}}{\sigma[\textbf{r}]}$, where $\mathbb{E}[\cdot]$ is the expectation, $\sigma{[\cdot]}$ is the standard deviation, $\textbf{r} = [{\frac{V_1-V_0}{V_0}}, {\frac{V_2-V_1}{V_1}}, ..., {\frac{V_T-V_{T-1}}{V_{T-1}}}]^{T}$ denotes the historical sequence of the return rate.
    \item \textbf{Volatility (VOL)} is the variation in an investment's return over time, measured as the standard deviation $\sigma[\textbf{r}]$.
    % \begin{gather*}
    %     MDD = \mathop{\max}_{i=0}^T \frac{P_i-R_i}{P_i} \\
    %     R_i = \prod_{i=1}^{T} {\frac{V_i}{V_{i-1}}} \\
    %     P_i = \mathop{\max}_{i=1}^T R_i
    % \end{gather*} 
    \item \textbf{Maximum Drawdown (MDD)} measures the largest loss from any peak to show the worst case. It is defined as: $MDD = \mathop{\max}_{i=0}^T \frac{P_i-R_i}{P_i}$, where $R_i = \prod_{i=1}^{T} {\frac{V_i}{V_{i-1}}} $ and $ P_i = \mathop{\max}_{i=1}^T R_i$.
    \item \textbf{Calmar Ratio (CR)} compares average annualized return to maximum drawdown, assessing risk-adjusted performance. It is defined as $CR = \frac{\mathbb{E}[\textbf{r}]}{MDD}$.
    % \item \textbf{Sortino Ratio (SoR)} is a risk-adjusted measure that focuses on the downside risk of an investment or portfolio. It considers the volatility of negative return to assess the excess return per unit of downside risk. It is defined as $SoR = \frac{\mathbb{E[\textbf{r}]}}{DD}$. $DD$ is the standard deviation of negative return of the whole trading period. It is defined as $\sigma[\{ r \mid r \in \textbf{r}, r < 0 \}]$.
    \item \textbf{Sortino Ratio (SoR)} is a risk-adjusted measure that focuses on the downside risk of a portfolio. It is defined as $SoR = \frac{\mathbb{E}[\textbf{r}]}{DD}$, where $DD$ is the standard deviation of negative return.
\end{itemize}

\subsection{Baselines}
\label{baselines}
We compare and evaluate the trading performance of \texttt{FinAgent} with four widely accepted conventional rule-based trading strategies (\textbf{B\&H}, \textbf{MACD}, \textbf{KDJ\&RSI} and \textbf{ZMR}) and eight advanced algorithms. Among these, 
price prediction models based on machine learning and deep learning (ML \& DL-based) include \textbf{LGBM}\cite{yang2020qlib}, \textbf{LSTM}\cite{yang2020qlib}, and \textbf{Transformer}\cite{yang2020qlib}. \textbf{SAC} \cite{haarnoja2018soft}, \textbf{PPO} \cite{schulman2017proximal} and \textbf{DQN} \cite{mnih2013playing} are three models employed deep reinforcement learning (RL-based) methods, \textbf{FinGPT} \cite{yang2023fingpt} is based on LLM, and another is \textbf{FinMem} \cite{yu2023finmem} that based on LLM Agents. The following will provide a brief introduction to each model:
\begin{itemize}[leftmargin=*]
    \item \textbf{Rule-based}
        \begin{itemize}[leftmargin=0cm]
            \item \textbf{Buy-and-Hold (B\&H)} involves holding assets for an extended period, regardless of short-term market fluctuations, assuming that long-term returns will be more favorable.
            \item \textbf{Moving Average Convergence Divergence (MACD)} is a technical analysis tool that uses MACD indicator and signal line crossovers to identify trading signals and market trends.
            \item \textbf{KDJ with RSI Filter (KDJ\&RSI)} integrates the KDJ indicator for detecting market extremes with the RSI indicator for momentum analysis to identify precise trading signals in financial markets.
            % \item \textbf{Stochastic Oscillator and Bollinger Bands (SO\&BB)} combines momentum based SO for overbought or oversold conditions with BB to assess market volatility and potential price reversals.
            \item  \textbf{Z-score Mean Reversion (ZMR)} assumes that the price will revert to its mean over time with the metric of Z-score.
        \end{itemize}
    \item \textbf{ML\&DL-based}
        \begin{itemize}[leftmargin=0cm]
            \item  \textbf{LGBM} \cite{yang2020qlib} uses a series of tree models to predict price fluctuations and provide buy and sell signals.
            \item \textbf{LSTM} \cite{yang2020qlib} utilizes long short-term memory to improve the accuracy of price predictions.
            \item \textbf{Transformer} \cite{yang2020qlib} models leverage self-attention mechanisms to enhance the precision of price forecasts.
        \end{itemize}
    \item \textbf{RL-based}
        \begin{itemize}[leftmargin=0cm]
            \item \textbf{SAC} \cite{haarnoja2018soft} is an off-policy actor-critic algorithm that optimizes trading strategies using entropy regularization and soft value functions in continuous action spaces.
            \item \textbf{PPO} \cite{schulman2017proximal} updates trading policies iteratively to balance exploration and exploitation, ensuring stability and sample efficiency.
            \item \textbf{DQN} \cite{mnih2013playing} uses deep neural networks to approximate the action-value function and make trading decisions from market data.
        \end{itemize}
    \item \textbf{LLM-based}
        \begin{itemize}[leftmargin=0cm]
            \item \textbf{FinGPT} \cite{yang2023fingpt} is an open-source LLM framework for converting financial news and prices into financial decisions.
            \item \textbf{FinMem} \cite{yu2023finmem} is an advanced LLM agent framework for automated trading, fine-tuned to boost investment returns.
        \end{itemize}
\end{itemize}

\subsection{Implementation Details}
Although \texttt{FinAgent}'s training and inference can be done without a GPU, we utilized a single NVIDIA RTX A6000 GPU for our benchmark methods. To ensure equitable comparison, all benchmarks are conducted within the same RL environment for both training and evaluation. The following experiments related to \texttt{FinAgent} all have diversified retrieval if not specifically noted. Details on the benchmark and experiments setup are provided in Appendix \ref{app:implementation}.

\begin{figure*}[htbp]
\captionsetup{skip=2pt}
\vspace{-0.1cm}
  \centering
    \includegraphics[width=1.0\textwidth, height=0.25\textheight]{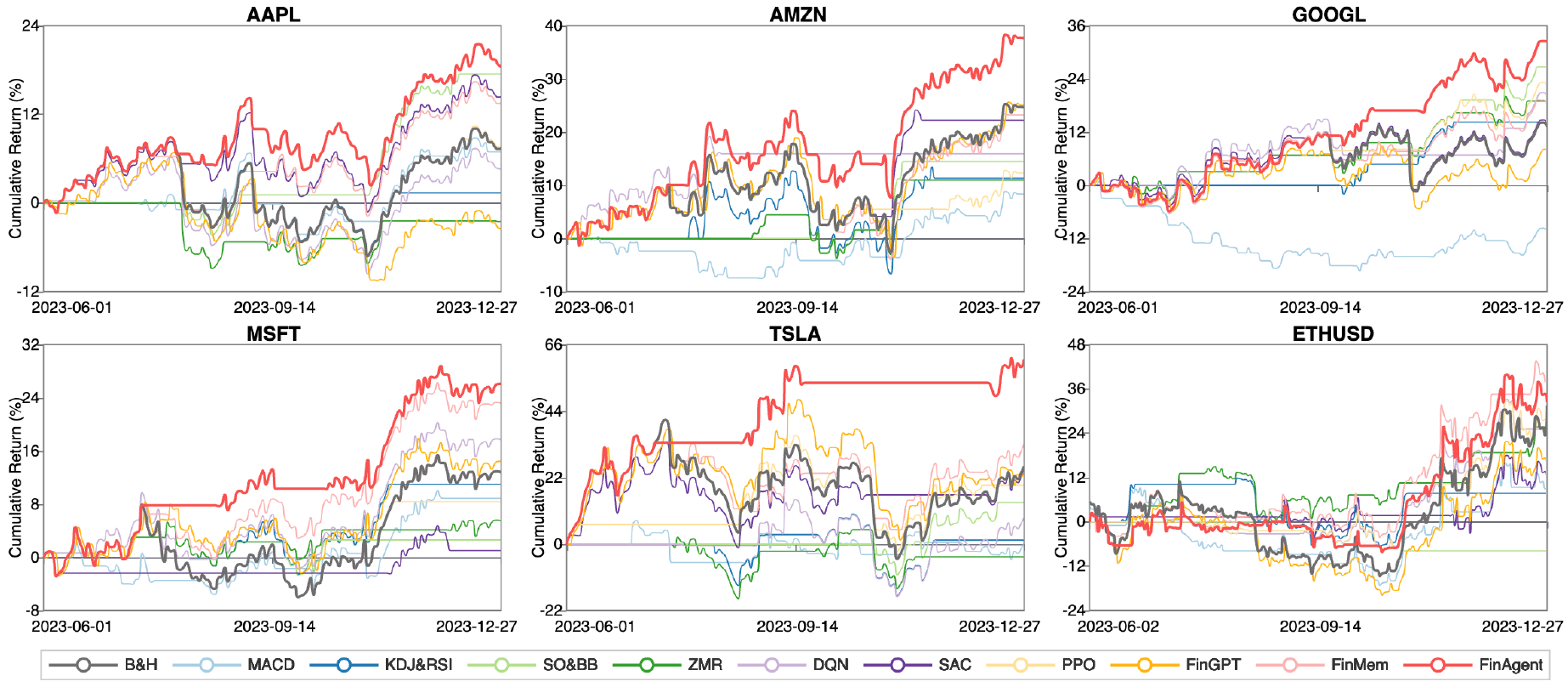}
    \caption{Performance comparison over time between \texttt{FinAgent} and other benchmarks across all assets.}
    \label{fig:baselines}
\vspace{-0.3cm}
\end{figure*}

\section{Experimental Results}

\textbf{Comparison with Baselines (RQ1).} We compared \texttt{FinAgent} with 9 baseline methods in terms of 6 financial metrics. Table \ref{tab:baselines} and Figure \ref{fig:baselines} demonstrate our method significantly outperforms existing baselines, especially remarkable improvements in profitability, and setting a new benchmark in the field. The full results and case studies of \texttt{FinAgent} are avaliable in Appendix \ref{app:comparison}. \texttt{FinAgent}'s performance on the five stocks, as measured by ARR\% and SR, with enhancements of at least 10\% and 19\%, compared to the best-performing baseline, respectively. Notably, its performance on the TSLA dataset stands out even more, achieving 84\% and 118\% improvement, significantly outperforming all other baselines. Across all datasets, FinAgent is the only method that consistently outperforms the broader market in terms of profitability.
In contrast, FinMem falls short on the AMZN dataset, where its ARR\% is 40\%, underperforming the market's Buy \& Hold (B\&H) strategy at 42\%. This underscores the superior stability and robustness of \texttt{FinAgent} compared to other baselines. We can also observe that rule-based methods are optimal in controlling risk, but not outstanding in capturing returns. This is because rule-based model methods are robust to outliers and noise in the data and thus can reduce decision risk. It is worth noting that high returns often come with high risks. Hence, \texttt{FinAgent} represents a slight compromise on risk control. 
This result relates to our chosen investor preference of an aggressive trader.
%This is also related to our chosen investor preference, which is that of an aggressive trader. 
Therefore, \texttt{FinAgent} can take on slightly higher risk to achieve substantially greater returns. It allows \texttt{FinAgent} to optimize performance by balancing risk and reward effectively.

Figure \ref{fig:baselines} illustrates that \texttt{FinAgent}'s performance surpasses other methods regarding cumulative returns, particularly on the TSLA dataset. Leveraging market intelligence and the reflection mechanism, \texttt{FinAgent} anticipates a significant stock price drop post-September 14, 2023. By taking a short position, it can effectively hedge against potential trading losses and generate high returns.
% By adopting short positions, it effectively dodges potential trading losses, thereby securing a high return.

It's important to note that our approach yields slightly lower returns than FinMem on the cryptocurrency ETH, primarily because our auxiliary agents are specialized strategies tailored for stocks, not for cryptocurrencies with higher trading frequency. Further insights from the ablation study section for \texttt{FinAgent} reveal that employing a generalized auxiliary agent for cryptocurrency could potentially increase returns to 54\%, compared to the current 44\%. This significant difference will be elaborated upon in the forthcoming ablation studies. 

\vspace{-0.1cm}
\section{Ablation Studies}
\subsection{Effectiveness of Each Component (RQ2)}
In Table \ref{tab:ablation}, we study the effectiveness of market intelligence (\textit{M}), low-level reflection (\textit{L}), high-level reflection (\textit{H}) and augmented tools (\textit{T}). When compared to using solely \textit{M} and \textit{ML}, the integration of the low-level reflection module leads to an impressive increase in ARR\% by 45\% to 101\% for TSLA, and ETHUSD, and cutting risk by 14\% to 44\%. When comparing the \textit{ML} and \textit{MLH}, the addition of the high-level reflection module significantly enhances the ARR\% and SR, while notably reducing risk. This improvement comes with a minor trade-off: a slight 7\% rise in MDD\% for TSLA. Compared to \textit{MLH} and \textit{MLHT}, there's a minor improvement in stock profitability. However, the performance of ETH cryptocurrency dropped by over 20\% due to the introduction of rule-based methods as auxiliary agents, which are specialized only for stocks.

\definecolor{Decrease}{RGB}{84,130,53}
\definecolor{Increase}{RGB}{198,70,17}

\begin{table}[thb]
\vspace{-0.3cm}
\caption{Ablation studies over different components. \(\surd\) indicates adding the component to FinAgent. Red and green indicate performance improvement and reduction.}
\setlength{\abovecaptionskip}{0.2cm}
\setlength{\tabcolsep}{1pt}
\centering
\resizebox{\linewidth}{!}{ 
\begin{tabular}{cccl|ccclccc}
\toprule
\multirow{3}{*}{M} & \multirow{3}{*}{L} & \multirow{3}{*}{H} & \multirow{3}{*}{T} &  \multicolumn{3}{c}{TSLA} & & \multicolumn{3}{c}{ETHUSD} \\ 
\cmidrule{5-7}\cmidrule{9-11}
& & & & ARR\%$\uparrow$  & SR$\uparrow$  & MDD\%$\downarrow$  & & ARR\%$\uparrow$  & SR$\uparrow$  & MDD\%$\downarrow$  \\ 

\midrule
& & & \(\surd\) & 39.01 & 0.90  & 22.54 &       & 16.21 & 0.63  & 15.93 \\

\midrule

% \multicolumn{4}{c}{No-finetuned} & 38.02	& 1.30 & 12.06 & & 23.40	& 0.79 & 23.23 \\

\(\surd\) & & & & 39.27 & 0.77  & 30.15 &       & 25.97 & 0.77  & 24.43 \\

\(\surd\) & \(\surd\) & & & 57.16(\textcolor{Increase}{+45.56\%}) & 1.02 (\textcolor{Increase}{+33.14\%})  & 25.77(\textcolor{Increase}{-14.52\%}) &       & 52.33(\textcolor{Increase}{+101.48\%}) & 1.34(\textcolor{Increase}{+72.99\%})  & 13.59(\textcolor{Increase}{-44.39\%}) \\

\(\surd\) & \(\surd\) &  \(\surd\) & & 89.25(\textcolor{Increase}{+56.14\%}) & 1.46(\textcolor{Increase}{+42.86\%})  & 27.62(\textcolor{Decrease}{+7.18\%}) &       & 54.80(\textcolor{Increase}{+4.73\%}) & 1.40(\textcolor{Increase}{+5.09\%})  & 11.74(\textcolor{Increase}{-13.57\%}) \\

\(\surd\) & \(\surd\) &  \(\surd\) &  \(\surd\) & 92.27(\textcolor{Increase}{+3.38\%}) & 2.01(\textcolor{Increase}{+37.84\%})  & 12.14(\textcolor{Increase}{-56.04\%}) &       & 43.08(\textcolor{Decrease}{-21.39\%}) & 1.18(\textcolor{Decrease}{-16.09\%})  & 12.72(\textcolor{Decrease}{+8.30\%}) \\

\bottomrule
\end{tabular}
}
\label{tab:ablation}
\vspace{-0.3cm}
\end{table}

\subsection{Effectiveness of Augmented Tools (RQ3)}
As previously discussed, while the addition of auxiliary agents to stock investments results in profit improvements, it causes a considerable performance decline in cryptocurrencies. Thus, we conduct the experiment that decisions are made solely by augmented tools, such as rule-based methods serving as auxiliary agents. We conducted the experiment in which various auxiliary agents provided both decisions and their explanations. These inputs are directly integrated into \texttt{FinAgent}'s decision-making module without other modules' involvement in the final decision process. As shown in Table \ref{tab:baselines} and Table \ref{tab:ablation}, the 16\% ARR\% for solely \textit{T} method starkly contrasts with the 29\% ARR\% of B\&H in ETHUSD, highlighting the inefficacy of the stock-specific rule-based methods for cryptocurrencies and demonstrating that introducing to \texttt{FinAgent} significantly affects performance. This suggests that investors should not indiscriminately add auxiliary agents for investment support. Instead, they must meticulously select agents that match the characteristics of the market to avoid detrimental impact on performance.

\subsection{Effectiveness of Diversified Retrieval (RQ4)}

\begin{figure}[htbp]
\captionsetup{skip=2pt}
\vspace{-0.3cm}
  \centering
    \includegraphics[width=0.45\textwidth]{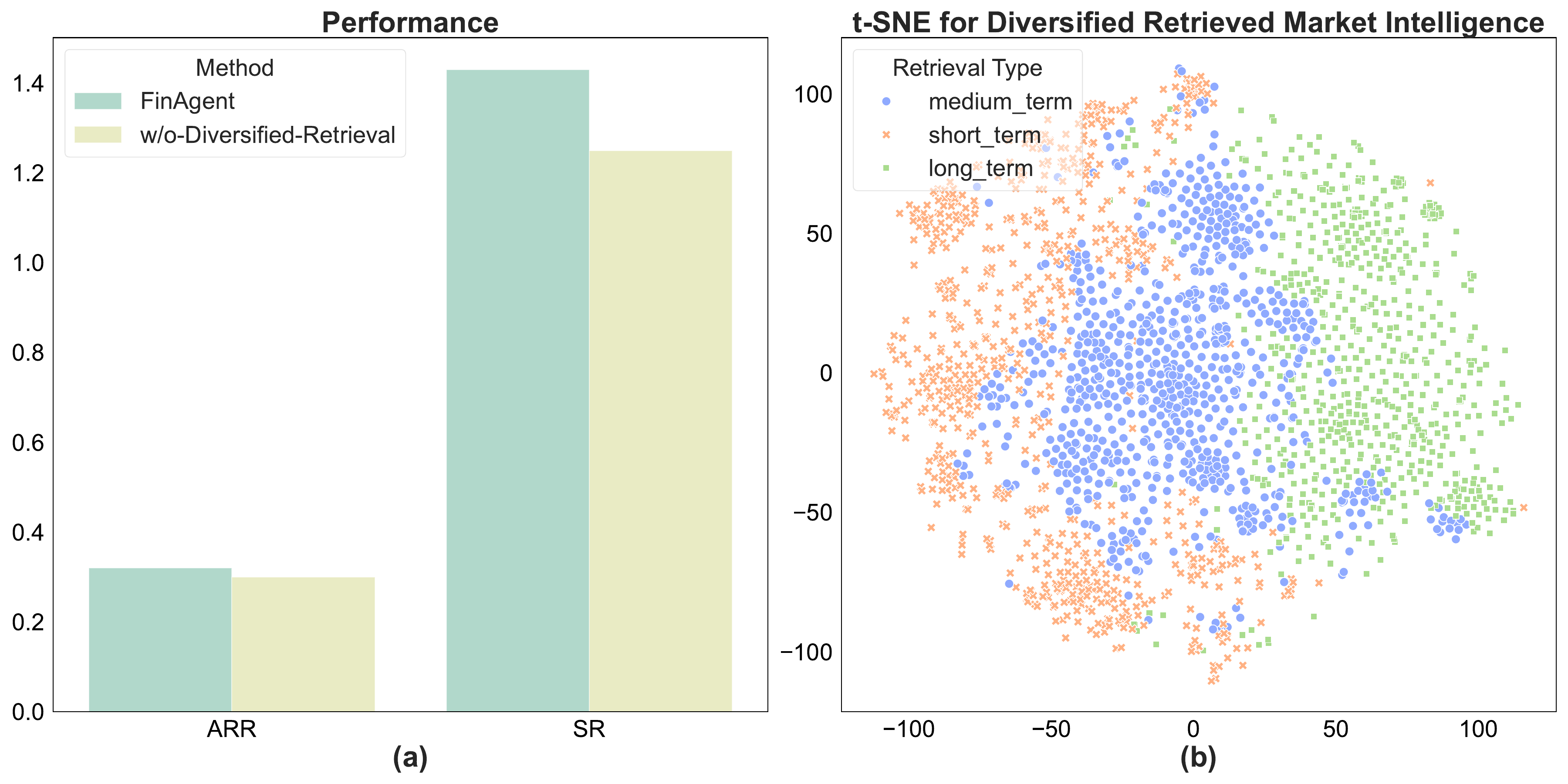}
    \caption{(a) Performance of FinAgent with/without diversified retrieval on AAPL. (b) Visualization of diversified retrieved market intelligence embedding by t-SNE on AAPL.}
    \label{fig:dr}
\vspace{-0.3cm}
\end{figure}

As shown in Figure \ref{fig:dr}(a), we compare the performance of FinAgent with or without diversified retrieval on AAPL, and find that the use of diversified retrieval can contribute an obvious improvement in ARR and SR. As shown in Figure \ref{fig:dr}(b), we extract different types of market intelligence that AAPL diversified retrieve to daily on the validation set and filter out individuals with the same content under the same type. We perform t-SNE visualization of its LLM extracted embedding, and we can find that the LLM extracted embedding has a clear distinction between different retrieval types, which proves the effectiveness of our method.

\vspace{-0.1cm}
\section{Conclusion and Future Work}
This paper introduces \texttt{FinAgent}, a financial trading agent powered by LLM that exhibits high reasoning ability and generalizability. \texttt{FinAgent} is a multimodal agent that integrates both textual and visual data, enabling a comprehensive understanding of market dynamics and historical trading behaviors. It is designed to independently leverage auxiliary tools for detailed market data analysis over different time scales. With its multi-perspective and diverse retrieval approach, \texttt{FinAgent} effectively identifies correlations between current market conditions and past market patterns and trends and integrates market information to make final and effective decisions. For future research directions, we will apply \texttt{FinAgent} to other financial tasks, such as portfolio management, where LLM is used to rank each stock according to the observed market intelligence and make the stock selection.

% significantly enhancing the effectiveness and reasoning process of portfolio.


\section{Acknowledgments}
This project is supported by the National Research Foundation, Singapore under its Industry Alignment Fund – Pre-positioning (IAF-PP) Funding Initiative. Any opinions, findings and conclusions or recommendations expressed in this material are those of the author(s) and do not reflect the views of National Research Foundation, Singapore. 

\clearpage
\balance
\bibliographystyle{ACM-Reference-Format}
\bibliography{citation}


\begin{thebibliography}{56}

%%% the Web address.

\ifx \showCODEN    \undefined \def \showCODEN     #1{\unskip}     \fi
\ifx \showDOI      \undefined \def \showDOI       #1{#1}\fi
\ifx \showISBNx    \undefined \def \showISBNx     #1{\unskip}     \fi
\ifx \showISBNxiii \undefined \def \showISBNxiii  #1{\unskip}     \fi
\ifx \showISSN     \undefined \def \showISSN      #1{\unskip}     \fi
\ifx \showLCCN     \undefined \def \showLCCN      #1{\unskip}     \fi
\ifx \shownote     \undefined \def \shownote      #1{#1}          \fi
\ifx \showarticletitle \undefined \def \showarticletitle #1{#1}   \fi
\ifx \showURL      \undefined \def \showURL       {\relax}        \fi
% invisible to TeX
\providecommand\bibfield[2]{#2}
\providecommand\bibinfo[2]{#2}
\providecommand\natexlab[1]{#1}
\providecommand\showeprint[2][]{arXiv:#2}

\bibitem[An et~al\mbox{.}(2022)]%
        {an2022deep}
\bibfield{author}{\bibinfo{person}{Bo An}, \bibinfo{person}{Shuo Sun}, {and} \bibinfo{person}{Rundong Wang}.} \bibinfo{year}{2022}\natexlab{}.
\newblock \showarticletitle{Deep reinforcement learning for quantitative trading: Challenges and opportunities}.
\newblock \bibinfo{journal}{\emph{IEEE Intelligent Systems}} \bibinfo{volume}{37}, \bibinfo{number}{2} (\bibinfo{year}{2022}), \bibinfo{pages}{23--26}.
\newblock


\bibitem[Chen et~al\mbox{.}(2019)]%
        {chen2019investment}
\bibfield{author}{\bibinfo{person}{Chi Chen}, \bibinfo{person}{Li Zhao}, \bibinfo{person}{Jiang Bian}, \bibinfo{person}{Chunxiao Xing}, {and} \bibinfo{person}{Tie-Yan Liu}.} \bibinfo{year}{2019}\natexlab{}.
\newblock \showarticletitle{Investment behaviors can tell what inside: Exploring stock intrinsic properties for stock trend prediction}. In \bibinfo{booktitle}{\emph{Proceedings of the 25th ACM SIGKDD International Conference on Knowledge Discovery \& Data Mining}}. \bibinfo{pages}{2376--2384}.
\newblock


\bibitem[Chen et~al\mbox{.}(2023)]%
        {chen2023theoremqa}
\bibfield{author}{\bibinfo{person}{Wenhu Chen}, \bibinfo{person}{Ming Yin}, \bibinfo{person}{Max Ku}, \bibinfo{person}{Pan Lu}, \bibinfo{person}{Yixin Wan}, \bibinfo{person}{Xueguang Ma}, \bibinfo{person}{Jianyu Xu}, \bibinfo{person}{Xinyi Wang}, {and} \bibinfo{person}{Tony Xia}.} \bibinfo{year}{2023}\natexlab{}.
\newblock \showarticletitle{Theoremqa: A theorem-driven question answering dataset}.
\newblock \bibinfo{journal}{\emph{arXiv preprint arXiv:2305.12524}} (\bibinfo{year}{2023}).
\newblock


\bibitem[Chowdhery et~al\mbox{.}(2023)]%
        {chowdhery2023palm}
\bibfield{author}{\bibinfo{person}{Aakanksha Chowdhery}, \bibinfo{person}{Sharan Narang}, \bibinfo{person}{Jacob Devlin}, \bibinfo{person}{Maarten Bosma}, \bibinfo{person}{Gaurav Mishra}, \bibinfo{person}{Adam Roberts}, \bibinfo{person}{Paul Barham}, \bibinfo{person}{Hyung~Won Chung}, \bibinfo{person}{Charles Sutton}, \bibinfo{person}{Sebastian Gehrmann}, {et~al\mbox{.}}} \bibinfo{year}{2023}\natexlab{}.
\newblock \showarticletitle{Palm: Scaling language modeling with pathways}.
\newblock \bibinfo{journal}{\emph{Journal of Machine Learning Research}} \bibinfo{volume}{24}, \bibinfo{number}{240} (\bibinfo{year}{2023}), \bibinfo{pages}{1--113}.
\newblock


\bibitem[Christianos et~al\mbox{.}(2023)]%
        {christianos2023panguagent}
\bibfield{author}{\bibinfo{person}{Filippos Christianos}, \bibinfo{person}{Georgios Papoudakis}, \bibinfo{person}{Matthieu Zimmer}, \bibinfo{person}{Thomas Coste}, \bibinfo{person}{Zhihao Wu}, \bibinfo{person}{Jingxuan Chen}, \bibinfo{person}{Khyati Khandelwal}, \bibinfo{person}{James Doran}, \bibinfo{person}{Xidong Feng}, \bibinfo{person}{Jiacheng Liu}, \bibinfo{person}{Zheng Xiong}, \bibinfo{person}{Yicheng Luo}, \bibinfo{person}{Jianye Hao}, \bibinfo{person}{Kun Shao}, \bibinfo{person}{Haitham Bou-Ammar}, {and} \bibinfo{person}{Jun Wang}.} \bibinfo{year}{2023}\natexlab{}.
\newblock \bibinfo{title}{Pangu-Agent: A fine-tunable generalist agent with structured reasoning}.
\newblock
\newblock
\showeprint[arxiv]{2312.14878}~[cs.AI]


\bibitem[Deng et~al\mbox{.}(2016)]%
        {deng2016deep}
\bibfield{author}{\bibinfo{person}{Yue Deng}, \bibinfo{person}{Feng Bao}, \bibinfo{person}{Youyong Kong}, \bibinfo{person}{Zhiquan Ren}, {and} \bibinfo{person}{Qionghai Dai}.} \bibinfo{year}{2016}\natexlab{}.
\newblock \showarticletitle{Deep direct reinforcement learning for financial signal representation and trading}.
\newblock \bibinfo{journal}{\emph{IEEE Transactions on Neural Networks and Learning Systems}} \bibinfo{volume}{28}, \bibinfo{number}{3} (\bibinfo{year}{2016}), \bibinfo{pages}{653--664}.
\newblock


\bibitem[Ding et~al\mbox{.}(2018)]%
        {ding2018investor}
\bibfield{author}{\bibinfo{person}{Yi Ding}, \bibinfo{person}{Weiqing Liu}, \bibinfo{person}{Jiang Bian}, \bibinfo{person}{Daoqiang Zhang}, {and} \bibinfo{person}{Tie-Yan Liu}.} \bibinfo{year}{2018}\natexlab{}.
\newblock \showarticletitle{Investor-imitator: A framework for trading knowledge extraction}. In \bibinfo{booktitle}{\emph{Proceedings of the 24th ACM SIGKDD International Conference on Knowledge Discovery \& Data Mining}}. \bibinfo{pages}{1310--1319}.
\newblock


\bibitem[Edwards et~al\mbox{.}(2018)]%
        {edwards2018technical}
\bibfield{author}{\bibinfo{person}{Robert~D Edwards}, \bibinfo{person}{John Magee}, {and} \bibinfo{person}{WH~Charles Bassetti}.} \bibinfo{year}{2018}\natexlab{}.
\newblock \bibinfo{booktitle}{\emph{Technical analysis of stock trends}}.
\newblock \bibinfo{publisher}{CRC press}.
\newblock


\bibitem[Feng et~al\mbox{.}(2019)]%
        {feng2019temporal}
\bibfield{author}{\bibinfo{person}{Fuli Feng}, \bibinfo{person}{Xiangnan He}, \bibinfo{person}{Xiang Wang}, \bibinfo{person}{Cheng Luo}, \bibinfo{person}{Yiqun Liu}, {and} \bibinfo{person}{Tat-Seng Chua}.} \bibinfo{year}{2019}\natexlab{}.
\newblock \showarticletitle{Temporal relational ranking for stock prediction}.
\newblock \bibinfo{journal}{\emph{ACM Transactions on Information Systems (TOIS)}} \bibinfo{volume}{37}, \bibinfo{number}{2} (\bibinfo{year}{2019}), \bibinfo{pages}{1--30}.
\newblock


\bibitem[Haarnoja et~al\mbox{.}(2018)]%
        {haarnoja2018soft}
\bibfield{author}{\bibinfo{person}{Tuomas Haarnoja}, \bibinfo{person}{Aurick Zhou}, \bibinfo{person}{Kristian Hartikainen}, \bibinfo{person}{George Tucker}, \bibinfo{person}{Sehoon Ha}, \bibinfo{person}{Jie Tan}, \bibinfo{person}{Vikash Kumar}, \bibinfo{person}{Henry Zhu}, \bibinfo{person}{Abhishek Gupta}, \bibinfo{person}{Pieter Abbeel}, {et~al\mbox{.}}} \bibinfo{year}{2018}\natexlab{}.
\newblock \showarticletitle{Soft actor-critic algorithms and applications}.
\newblock \bibinfo{journal}{\emph{arXiv preprint arXiv:1812.05905}} (\bibinfo{year}{2018}).
\newblock


\bibitem[Hong et~al\mbox{.}(2023)]%
        {hong2023metagpt}
\bibfield{author}{\bibinfo{person}{Sirui Hong}, \bibinfo{person}{Mingchen Zhuge}, \bibinfo{person}{Jonathan Chen}, \bibinfo{person}{Xiawu Zheng}, \bibinfo{person}{Yuheng Cheng}, \bibinfo{person}{Ceyao Zhang}, \bibinfo{person}{Jinlin Wang}, \bibinfo{person}{Zili Wang}, \bibinfo{person}{Steven Ka~Shing Yau}, \bibinfo{person}{Zijuan Lin}, \bibinfo{person}{Liyang Zhou}, \bibinfo{person}{Chenyu Ran}, \bibinfo{person}{Lingfeng Xiao}, \bibinfo{person}{Chenglin Wu}, {and} \bibinfo{person}{Jürgen Schmidhuber}.} \bibinfo{year}{2023}\natexlab{}.
\newblock \bibinfo{title}{MetaGPT: Meta programming for a multi-agent collaborative framework}.
\newblock
\newblock
\showeprint[arxiv]{2308.00352}~[cs.AI]


\bibitem[Hu et~al\mbox{.}(2018)]%
        {hu2018listening}
\bibfield{author}{\bibinfo{person}{Ziniu Hu}, \bibinfo{person}{Weiqing Liu}, \bibinfo{person}{Jiang Bian}, \bibinfo{person}{Xuanzhe Liu}, {and} \bibinfo{person}{Tie-Yan Liu}.} \bibinfo{year}{2018}\natexlab{}.
\newblock \showarticletitle{Listening to chaotic whispers: A deep learning framework for news-oriented stock trend prediction}. In \bibinfo{booktitle}{\emph{Proceedings of the 11th ACM International Conference on Web Search and Data Mining (WSDM)}}. \bibinfo{pages}{261--269}.
\newblock


\bibitem[Ke et~al\mbox{.}(2017)]%
        {ke2017lightgbm}
\bibfield{author}{\bibinfo{person}{Guolin Ke}, \bibinfo{person}{Qi Meng}, \bibinfo{person}{Thomas Finley}, \bibinfo{person}{Taifeng Wang}, \bibinfo{person}{Wei Chen}, \bibinfo{person}{Weidong Ma}, \bibinfo{person}{Qiwei Ye}, {and} \bibinfo{person}{Tie-Yan Liu}.} \bibinfo{year}{2017}\natexlab{}.
\newblock \showarticletitle{Lightgbm: A highly efficient gradient boosting decision tree}.
\newblock \bibinfo{journal}{\emph{Advances in neural information processing systems}}  \bibinfo{volume}{30} (\bibinfo{year}{2017}).
\newblock


\bibitem[Liu et~al\mbox{.}(2020b)]%
        {finrl2020}
\bibfield{author}{\bibinfo{person}{Xiao-Yang Liu}, \bibinfo{person}{Hongyang Yang}, \bibinfo{person}{Qian Chen}, \bibinfo{person}{Runjia Zhang}, \bibinfo{person}{Liuqing Yang}, \bibinfo{person}{Bowen Xiao}, {and} \bibinfo{person}{Christina~Dan Wang}.} \bibinfo{year}{2020}\natexlab{b}.
\newblock \showarticletitle{{FinRL}: A deep reinforcement learning library for automated stock trading in quantitative finance}.
\newblock \bibinfo{journal}{\emph{Deep RL Workshop, NeurIPS 2020}} (\bibinfo{year}{2020}).
\newblock


\bibitem[Liu et~al\mbox{.}(2020a)]%
        {liu2020adaptive}
\bibfield{author}{\bibinfo{person}{Yang Liu}, \bibinfo{person}{Qi Liu}, \bibinfo{person}{Hongke Zhao}, \bibinfo{person}{Zhen Pan}, {and} \bibinfo{person}{Chuanren Liu}.} \bibinfo{year}{2020}\natexlab{a}.
\newblock \showarticletitle{Adaptive quantitative trading: An imitative deep reinforcement learning approach}. In \bibinfo{booktitle}{\emph{Proceedings of the AAAI Conference on Artificial Intelligence}}, Vol.~\bibinfo{volume}{34}. \bibinfo{pages}{2128--2135}.
\newblock


\bibitem[Lu et~al\mbox{.}(2023a)]%
        {lu2023mathvista}
\bibfield{author}{\bibinfo{person}{Pan Lu}, \bibinfo{person}{Hritik Bansal}, \bibinfo{person}{Tony Xia}, \bibinfo{person}{Jiacheng Liu}, \bibinfo{person}{Chunyuan Li}, \bibinfo{person}{Hannaneh Hajishirzi}, \bibinfo{person}{Hao Cheng}, \bibinfo{person}{Kai-Wei Chang}, \bibinfo{person}{Michel Galley}, {and} \bibinfo{person}{Jianfeng Gao}.} \bibinfo{year}{2023}\natexlab{a}.
\newblock \showarticletitle{Mathvista: Evaluating mathematical reasoning of foundation models in visual contexts}.
\newblock \bibinfo{journal}{\emph{arXiv preprint arXiv:2310.02255}} (\bibinfo{year}{2023}).
\newblock


\bibitem[Lu et~al\mbox{.}(2022a)]%
        {lu2022learn}
\bibfield{author}{\bibinfo{person}{Pan Lu}, \bibinfo{person}{Swaroop Mishra}, \bibinfo{person}{Tanglin Xia}, \bibinfo{person}{Liang Qiu}, \bibinfo{person}{Kai-Wei Chang}, \bibinfo{person}{Song-Chun Zhu}, \bibinfo{person}{Oyvind Tafjord}, \bibinfo{person}{Peter Clark}, {and} \bibinfo{person}{Ashwin Kalyan}.} \bibinfo{year}{2022}\natexlab{a}.
\newblock \showarticletitle{Learn to explain: Multimodal reasoning via thought chains for science question answering}.
\newblock \bibinfo{journal}{\emph{Advances in Neural Information Processing Systems}}  \bibinfo{volume}{35} (\bibinfo{year}{2022}), \bibinfo{pages}{2507--2521}.
\newblock


\bibitem[Lu et~al\mbox{.}(2023b)]%
        {lu2023chameleon}
\bibfield{author}{\bibinfo{person}{Pan Lu}, \bibinfo{person}{Baolin Peng}, \bibinfo{person}{Hao Cheng}, \bibinfo{person}{Michel Galley}, \bibinfo{person}{Kai-Wei Chang}, \bibinfo{person}{Ying~Nian Wu}, \bibinfo{person}{Song-Chun Zhu}, {and} \bibinfo{person}{Jianfeng Gao}.} \bibinfo{year}{2023}\natexlab{b}.
\newblock \bibinfo{title}{Chameleon: Plug-and-Play compositional reasoning with large language models}.
\newblock
\newblock
\showeprint[arxiv]{2304.09842}~[cs.CL]


\bibitem[Lu et~al\mbox{.}(2022b)]%
        {lu2022dynamic}
\bibfield{author}{\bibinfo{person}{Pan Lu}, \bibinfo{person}{Liang Qiu}, \bibinfo{person}{Kai-Wei Chang}, \bibinfo{person}{Ying~Nian Wu}, \bibinfo{person}{Song-Chun Zhu}, \bibinfo{person}{Tanmay Rajpurohit}, \bibinfo{person}{Peter Clark}, {and} \bibinfo{person}{Ashwin Kalyan}.} \bibinfo{year}{2022}\natexlab{b}.
\newblock \showarticletitle{Dynamic prompt learning via policy gradient for semi-structured mathematical reasoning}.
\newblock \bibinfo{journal}{\emph{arXiv preprint arXiv:2209.14610}} (\bibinfo{year}{2022}).
\newblock


\bibitem[Mnih et~al\mbox{.}(2013)]%
        {mnih2013playing}
\bibfield{author}{\bibinfo{person}{Volodymyr Mnih}, \bibinfo{person}{Koray Kavukcuoglu}, \bibinfo{person}{David Silver}, \bibinfo{person}{Alex Graves}, \bibinfo{person}{Ioannis Antonoglou}, \bibinfo{person}{Daan Wierstra}, {and} \bibinfo{person}{Martin Riedmiller}.} \bibinfo{year}{2013}\natexlab{}.
\newblock \showarticletitle{Playing atari with deep reinforcement learning}.
\newblock \bibinfo{journal}{\emph{arXiv preprint arXiv:1312.5602}} (\bibinfo{year}{2013}).
\newblock


\bibitem[Nakano et~al\mbox{.}(2021)]%
        {nakano2021webgpt}
\bibfield{author}{\bibinfo{person}{Reiichiro Nakano}, \bibinfo{person}{Jacob Hilton}, \bibinfo{person}{Suchir Balaji}, \bibinfo{person}{Jeff Wu}, \bibinfo{person}{Long Ouyang}, \bibinfo{person}{Christina Kim}, \bibinfo{person}{Christopher Hesse}, \bibinfo{person}{Shantanu Jain}, \bibinfo{person}{Vineet Kosaraju}, \bibinfo{person}{William Saunders}, {et~al\mbox{.}}} \bibinfo{year}{2021}\natexlab{}.
\newblock \showarticletitle{Webgpt: Browser-assisted question-answering with human feedback}.
\newblock \bibinfo{journal}{\emph{arXiv preprint arXiv:2112.09332}} (\bibinfo{year}{2021}).
\newblock


\bibitem[Nelson et~al\mbox{.}(2017)]%
        {nelson2017stock}
\bibfield{author}{\bibinfo{person}{David~MQ Nelson}, \bibinfo{person}{Adriano~CM Pereira}, {and} \bibinfo{person}{Renato~A de Oliveira}.} \bibinfo{year}{2017}\natexlab{}.
\newblock \showarticletitle{Stock market's price movement prediction with LSTM neural networks}. In \bibinfo{booktitle}{\emph{2017 International Joint Conference on Neural Networks (IJCNN)}}. \bibinfo{pages}{1419--1426}.
\newblock


\bibitem[OpenAI(2021)]%
        {openai2021chatgpt}
\bibfield{author}{\bibinfo{person}{OpenAI}.} \bibinfo{year}{2021}\natexlab{}.
\newblock \bibinfo{title}{Chatgpt}.
\newblock
\newblock
\urldef\tempurl%
\url{https://openai.com/ research/chatgpt}
\showURL{%
\tempurl}


\bibitem[OpenAI(2023a)]%
        {openai2023gpt4}
\bibfield{author}{\bibinfo{person}{OpenAI}.} \bibinfo{year}{2023}\natexlab{a}.
\newblock \bibinfo{title}{GPT-4 Technical Report}.
\newblock
\newblock
\showeprint[arxiv]{2303.08774}~[cs.AI]


\bibitem[OpenAI(2023b)]%
        {openai2023gpt4v}
\bibfield{author}{\bibinfo{person}{OpenAI}.} \bibinfo{year}{2023}\natexlab{b}.
\newblock \bibinfo{title}{GPT-4V(ision) system card}.
\newblock
\newblock
\urldef\tempurl%
\url{https://openai.com/research/gpt-4v-system-card}
\showURL{%
\tempurl}


\bibitem[Parisi et~al\mbox{.}(2022)]%
        {parisi2022talm}
\bibfield{author}{\bibinfo{person}{Aaron Parisi}, \bibinfo{person}{Yao Zhao}, {and} \bibinfo{person}{Noah Fiedel}.} \bibinfo{year}{2022}\natexlab{}.
\newblock \showarticletitle{Talm: Tool augmented language models}.
\newblock \bibinfo{journal}{\emph{arXiv preprint arXiv:2205.12255}} (\bibinfo{year}{2022}).
\newblock


\bibitem[Park et~al\mbox{.}(2023)]%
        {park2023generative}
\bibfield{author}{\bibinfo{person}{Joon~Sung Park}, \bibinfo{person}{Joseph~C. O'Brien}, \bibinfo{person}{Carrie~J. Cai}, \bibinfo{person}{Meredith~Ringel Morris}, \bibinfo{person}{Percy Liang}, {and} \bibinfo{person}{Michael~S. Bernstein}.} \bibinfo{year}{2023}\natexlab{}.
\newblock \bibinfo{title}{Generative Agents: Interactive simulacra of human behavior}.
\newblock
\newblock
\showeprint[arxiv]{2304.03442}~[cs.HC]


\bibitem[Qin et~al\mbox{.}(2023)]%
        {qin2023earnhft}
\bibfield{author}{\bibinfo{person}{Molei Qin}, \bibinfo{person}{Shuo Sun}, \bibinfo{person}{Wentao Zhang}, \bibinfo{person}{Haochong Xia}, \bibinfo{person}{Xinrun Wang}, {and} \bibinfo{person}{Bo An}.} \bibinfo{year}{2023}\natexlab{}.
\newblock \showarticletitle{Earnhft: Efficient hierarchical reinforcement learning for high frequency trading}.
\newblock \bibinfo{journal}{\emph{arXiv preprint arXiv:2309.12891}} (\bibinfo{year}{2023}).
\newblock


\bibitem[Sawhney et~al\mbox{.}(2020a)]%
        {sawhneydeep}
\bibfield{author}{\bibinfo{person}{Ramit Sawhney}, \bibinfo{person}{Shivam Agarwal}, \bibinfo{person}{Arnav Wadhwa}, {and} \bibinfo{person}{Rajiv Shah}.} \bibinfo{year}{2020}\natexlab{a}.
\newblock \showarticletitle{Deep attentive learning for stock movement prediction from social media text and company correlations}. In \bibinfo{booktitle}{\emph{Proceedings of the 2020 Conference on Empirical Methods in Natural Language Processing (EMNLP)}}. \bibinfo{pages}{8415--8426}.
\newblock


\bibitem[Sawhney et~al\mbox{.}(2020b)]%
        {sawhney2020voltage}
\bibfield{author}{\bibinfo{person}{Ramit Sawhney}, \bibinfo{person}{Piyush Khanna}, \bibinfo{person}{Arshiya Aggarwal}, \bibinfo{person}{Taru Jain}, \bibinfo{person}{Puneet Mathur}, {and} \bibinfo{person}{Rajiv Shah}.} \bibinfo{year}{2020}\natexlab{b}.
\newblock \showarticletitle{VolTAGE: Volatility forecasting via text-audio fusion with graph convolution networks for earnings calls}. In \bibinfo{booktitle}{\emph{Proceedings of the 2020 Conference on Empirical Methods in Natural Language Processing (EMNLP)}}. \bibinfo{pages}{8001--8013}.
\newblock


\bibitem[Sawhney et~al\mbox{.}(2021)]%
        {sawhney2021quantitative}
\bibfield{author}{\bibinfo{person}{Ramit Sawhney}, \bibinfo{person}{Arnav Wadhwa}, \bibinfo{person}{Shivam Agarwal}, {and} \bibinfo{person}{Rajiv Shah}.} \bibinfo{year}{2021}\natexlab{}.
\newblock \showarticletitle{Quantitative day trading from natural language using reinforcement learning}. In \bibinfo{booktitle}{\emph{Proceedings of the 2021 Conference of the North American Chapter of the Association for Computational Linguistics: Human Language Technologies}}. \bibinfo{pages}{4018--4030}.
\newblock


\bibitem[Schick et~al\mbox{.}(2023)]%
        {Schick2023ToolformerLM}
\bibfield{author}{\bibinfo{person}{Timo Schick}, \bibinfo{person}{Jane Dwivedi-Yu}, \bibinfo{person}{Roberto Dess{\`\i}}, \bibinfo{person}{Roberta Raileanu}, \bibinfo{person}{Maria Lomeli}, \bibinfo{person}{Luke Zettlemoyer}, \bibinfo{person}{Nicola Cancedda}, {and} \bibinfo{person}{Thomas Scialom}.} \bibinfo{year}{2023}\natexlab{}.
\newblock \showarticletitle{Toolformer: Language models can teach themselves to use tools}.
\newblock \bibinfo{journal}{\emph{arXiv preprint arXiv:2302.04761}} (\bibinfo{year}{2023}).
\newblock


\bibitem[Schulman et~al\mbox{.}(2017)]%
        {schulman2017proximal}
\bibfield{author}{\bibinfo{person}{John Schulman}, \bibinfo{person}{Filip Wolski}, \bibinfo{person}{Prafulla Dhariwal}, \bibinfo{person}{Alec Radford}, {and} \bibinfo{person}{Oleg Klimov}.} \bibinfo{year}{2017}\natexlab{}.
\newblock \showarticletitle{Proximal policy optimization algorithms}.
\newblock \bibinfo{journal}{\emph{arXiv preprint arXiv:1707.06347}} (\bibinfo{year}{2017}).
\newblock


\bibitem[Sumers et~al\mbox{.}(2023)]%
        {sumers2023cognitive}
\bibfield{author}{\bibinfo{person}{Theodore~R Sumers}, \bibinfo{person}{Shunyu Yao}, \bibinfo{person}{Karthik Narasimhan}, {and} \bibinfo{person}{Thomas~L Griffiths}.} \bibinfo{year}{2023}\natexlab{}.
\newblock \showarticletitle{Cognitive architectures for language agents}.
\newblock \bibinfo{journal}{\emph{arXiv preprint arXiv:2309.02427}} (\bibinfo{year}{2023}).
\newblock


\bibitem[Sun et~al\mbox{.}(2023a)]%
        {sun2023scieval}
\bibfield{author}{\bibinfo{person}{Liangtai Sun}, \bibinfo{person}{Yang Han}, \bibinfo{person}{Zihan Zhao}, \bibinfo{person}{Da Ma}, \bibinfo{person}{Zhennan Shen}, \bibinfo{person}{Baocai Chen}, \bibinfo{person}{Lu Chen}, {and} \bibinfo{person}{Kai Yu}.} \bibinfo{year}{2023}\natexlab{a}.
\newblock \showarticletitle{Scieval: A multi-level large language model evaluation benchmark for scientific research}.
\newblock \bibinfo{journal}{\emph{arXiv preprint arXiv:2308.13149}} (\bibinfo{year}{2023}).
\newblock


\bibitem[Sun et~al\mbox{.}(2023b)]%
        {sun2023prudex}
\bibfield{author}{\bibinfo{person}{Shuo Sun}, \bibinfo{person}{Molei Qin}, \bibinfo{person}{Xinrun Wang}, {and} \bibinfo{person}{Bo An}.} \bibinfo{year}{2023}\natexlab{b}.
\newblock \showarticletitle{{PRUDEX}-Compass: Towards systematic evaluation of reinforcement learning in financial markets}.
\newblock \bibinfo{journal}{\emph{Transactions on Machine Learning Research}} (\bibinfo{year}{2023}).
\newblock
\showISSN{2835-8856}


\bibitem[Sun et~al\mbox{.}(2023c)]%
        {sun2023trademaster}
\bibfield{author}{\bibinfo{person}{Shuo Sun}, \bibinfo{person}{Molei Qin}, \bibinfo{person}{wentao zhang}, \bibinfo{person}{Haochong Xia}, \bibinfo{person}{Chuqiao Zong}, \bibinfo{person}{Jie Ying}, \bibinfo{person}{Yonggang Xie}, \bibinfo{person}{Lingxuan Zhao}, \bibinfo{person}{Xinrun Wang}, {and} \bibinfo{person}{Bo An}.} \bibinfo{year}{2023}\natexlab{c}.
\newblock \showarticletitle{TradeMaster: A holistic quantitative trading platform empowered by reinforcement learning}. In \bibinfo{booktitle}{\emph{Thirty-seventh Conference on Neural Information Processing Systems Datasets and Benchmarks Track}}.
\newblock


\bibitem[Sun et~al\mbox{.}(2023d)]%
        {sun2023reinforcement}
\bibfield{author}{\bibinfo{person}{Shuo Sun}, \bibinfo{person}{Rundong Wang}, {and} \bibinfo{person}{Bo An}.} \bibinfo{year}{2023}\natexlab{d}.
\newblock \showarticletitle{Reinforcement learning for quantitative trading}.
\newblock \bibinfo{journal}{\emph{ACM Transactions on Intelligent Systems and Technology}} \bibinfo{volume}{14}, \bibinfo{number}{3} (\bibinfo{year}{2023}), \bibinfo{pages}{1--29}.
\newblock


\bibitem[Sun et~al\mbox{.}(2023e)]%
        {sun2023mastering}
\bibfield{author}{\bibinfo{person}{Shuo Sun}, \bibinfo{person}{Xinrun Wang}, \bibinfo{person}{Wanqi Xue}, \bibinfo{person}{Xiaoxuan Lou}, {and} \bibinfo{person}{Bo An}.} \bibinfo{year}{2023}\natexlab{e}.
\newblock \showarticletitle{Mastering stock markets with efficient mixture of diversified trading experts}. In \bibinfo{booktitle}{\emph{Proceedings of the 29th ACM SIGKDD Conference on Knowledge Discovery and Data Mining}} \emph{(\bibinfo{series}{KDD '23})}. \bibinfo{pages}{2109–2119}.
\newblock
\showISBNx{9798400701030}


\bibitem[Thoppilan et~al\mbox{.}(2022)]%
        {thoppilan2022lamda}
\bibfield{author}{\bibinfo{person}{Romal Thoppilan}, \bibinfo{person}{Daniel De~Freitas}, \bibinfo{person}{Jamie Hall}, \bibinfo{person}{Noam Shazeer}, \bibinfo{person}{Apoorv Kulshreshtha}, \bibinfo{person}{Heng-Tze Cheng}, \bibinfo{person}{Alicia Jin}, \bibinfo{person}{Taylor Bos}, \bibinfo{person}{Leslie Baker}, \bibinfo{person}{Yu Du}, {et~al\mbox{.}}} \bibinfo{year}{2022}\natexlab{}.
\newblock \showarticletitle{Lamda: Language models for dialog applications}.
\newblock \bibinfo{journal}{\emph{arXiv preprint arXiv:2201.08239}} (\bibinfo{year}{2022}).
\newblock


\bibitem[Touvron et~al\mbox{.}(2023)]%
        {touvron2023llama}
\bibfield{author}{\bibinfo{person}{Hugo Touvron}, \bibinfo{person}{Thibaut Lavril}, \bibinfo{person}{Gautier Izacard}, \bibinfo{person}{Xavier Martinet}, \bibinfo{person}{Marie-Anne Lachaux}, \bibinfo{person}{Timoth{\'e}e Lacroix}, \bibinfo{person}{Baptiste Rozi{\`e}re}, \bibinfo{person}{Naman Goyal}, \bibinfo{person}{Eric Hambro}, \bibinfo{person}{Faisal Azhar}, {et~al\mbox{.}}} \bibinfo{year}{2023}\natexlab{}.
\newblock \showarticletitle{Llama: Open and efficient foundation language models}.
\newblock \bibinfo{journal}{\emph{arXiv preprint arXiv:2302.13971}} (\bibinfo{year}{2023}).
\newblock


\bibitem[Wang et~al\mbox{.}(2023b)]%
        {wang2023voyager}
\bibfield{author}{\bibinfo{person}{Guanzhi Wang}, \bibinfo{person}{Yuqi Xie}, \bibinfo{person}{Yunfan Jiang}, \bibinfo{person}{Ajay Mandlekar}, \bibinfo{person}{Chaowei Xiao}, \bibinfo{person}{Yuke Zhu}, \bibinfo{person}{Linxi Fan}, {and} \bibinfo{person}{Anima Anandkumar}.} \bibinfo{year}{2023}\natexlab{b}.
\newblock \showarticletitle{Voyager: An open-ended embodied agent with large language models}.
\newblock \bibinfo{journal}{\emph{arXiv preprint arXiv: Arxiv-2305.16291}} (\bibinfo{year}{2023}).
\newblock


\bibitem[Wang et~al\mbox{.}(2019)]%
        {wang19}
\bibfield{author}{\bibinfo{person}{Jia Wang}, \bibinfo{person}{Tong Sun}, \bibinfo{person}{Benyuan Liu}, \bibinfo{person}{Yu Cao}, {and} \bibinfo{person}{Hongwei Zhu}.} \bibinfo{year}{2019}\natexlab{}.
\newblock \showarticletitle{CLVSA: A convolutional LSTM based variational sequence-to-sequence model with attention for predicting trends of financial markets}. In \bibinfo{booktitle}{\emph{Proceedings of the 28th International Joint Conference on Artificial Intelligence (IJCAI)}}. \bibinfo{pages}{3705–3711}.
\newblock


\bibitem[Wang et~al\mbox{.}(2021b)]%
        {wang2021commission}
\bibfield{author}{\bibinfo{person}{Rundong Wang}, \bibinfo{person}{Hongxin Wei}, \bibinfo{person}{Bo An}, \bibinfo{person}{Zhouyan Feng}, {and} \bibinfo{person}{Jun Yao}.} \bibinfo{year}{2021}\natexlab{b}.
\newblock \showarticletitle{Commission fee is not enough: A hierarchical reinforced framework for portfolio management}. In \bibinfo{booktitle}{\emph{Proceedings of the AAAI Conference on Artificial Intelligence}}, Vol.~\bibinfo{volume}{35}. \bibinfo{pages}{626--633}.
\newblock


\bibitem[Wang et~al\mbox{.}(2023a)]%
        {wang2023scibench}
\bibfield{author}{\bibinfo{person}{Xiaoxuan Wang}, \bibinfo{person}{Ziniu Hu}, \bibinfo{person}{Pan Lu}, \bibinfo{person}{Yanqiao Zhu}, \bibinfo{person}{Jieyu Zhang}, \bibinfo{person}{Satyen Subramaniam}, \bibinfo{person}{Arjun~R Loomba}, \bibinfo{person}{Shichang Zhang}, \bibinfo{person}{Yizhou Sun}, {and} \bibinfo{person}{Wei Wang}.} \bibinfo{year}{2023}\natexlab{a}.
\newblock \showarticletitle{Scibench: Evaluating college-level scientific problem-solving abilities of large language models}.
\newblock \bibinfo{journal}{\emph{arXiv preprint arXiv:2307.10635}} (\bibinfo{year}{2023}).
\newblock


\bibitem[Wang et~al\mbox{.}(2021a)]%
        {wang2021deeptrader}
\bibfield{author}{\bibinfo{person}{Zhicheng Wang}, \bibinfo{person}{Biwei Huang}, \bibinfo{person}{Shikui Tu}, \bibinfo{person}{Kun Zhang}, {and} \bibinfo{person}{Lei Xu}.} \bibinfo{year}{2021}\natexlab{a}.
\newblock \showarticletitle{DeepTrader: a deep reinforcement learning approach for risk-return balanced portfolio management with market conditions Embedding}. In \bibinfo{booktitle}{\emph{Proceedings of the AAAI Conference on Artificial Intelligence}}, Vol.~\bibinfo{volume}{35}. \bibinfo{pages}{643--650}.
\newblock


\bibitem[Wu et~al\mbox{.}(2023)]%
        {wu2023bloomberggpt}
\bibfield{author}{\bibinfo{person}{Shijie Wu}, \bibinfo{person}{Ozan Irsoy}, \bibinfo{person}{Steven Lu}, \bibinfo{person}{Vadim Dabravolski}, \bibinfo{person}{Mark Dredze}, \bibinfo{person}{Sebastian Gehrmann}, \bibinfo{person}{Prabhanjan Kambadur}, \bibinfo{person}{David Rosenberg}, {and} \bibinfo{person}{Gideon Mann}.} \bibinfo{year}{2023}\natexlab{}.
\newblock \showarticletitle{Bloomberggpt: A large language model for finance}.
\newblock \bibinfo{journal}{\emph{arXiv preprint arXiv:2303.17564}} (\bibinfo{year}{2023}).
\newblock


\bibitem[Xu and Cohen(2018)]%
        {xu2018stock}
\bibfield{author}{\bibinfo{person}{Yumo Xu} {and} \bibinfo{person}{Shay~B Cohen}.} \bibinfo{year}{2018}\natexlab{}.
\newblock \showarticletitle{Stock movement prediction from tweets and historical prices}. In \bibinfo{booktitle}{\emph{Proceedings of the 56th Annual Meeting of the Association for Computational Linguistics (ACL)}}. \bibinfo{pages}{1970--1979}.
\newblock


\bibitem[Yang et~al\mbox{.}(2023b)]%
        {yang2023fingpt}
\bibfield{author}{\bibinfo{person}{Hongyang Yang}, \bibinfo{person}{Xiao-Yang Liu}, {and} \bibinfo{person}{Christina~Dan Wang}.} \bibinfo{year}{2023}\natexlab{b}.
\newblock \showarticletitle{FinGPT: Open-Source Financial Large Language Models}.
\newblock \bibinfo{journal}{\emph{arXiv preprint arXiv:2306.06031}} (\bibinfo{year}{2023}).
\newblock


\bibitem[Yang et~al\mbox{.}(2023c)]%
        {yang2023autogpt}
\bibfield{author}{\bibinfo{person}{Hui Yang}, \bibinfo{person}{Sifu Yue}, {and} \bibinfo{person}{Yunzhong He}.} \bibinfo{year}{2023}\natexlab{c}.
\newblock \bibinfo{title}{Auto-GPT for Online Decision Making: Benchmarks and Additional Opinions}.
\newblock
\newblock
\showeprint[arxiv]{2306.02224}~[cs.AI]


\bibitem[Yang et~al\mbox{.}(2020)]%
        {yang2020qlib}
\bibfield{author}{\bibinfo{person}{Xiao Yang}, \bibinfo{person}{Weiqing Liu}, \bibinfo{person}{Dong Zhou}, \bibinfo{person}{Jiang Bian}, {and} \bibinfo{person}{Tie-Yan Liu}.} \bibinfo{year}{2020}\natexlab{}.
\newblock \showarticletitle{Qlib: An AI-oriented quantitative investment platform}.
\newblock \bibinfo{journal}{\emph{arXiv preprint arXiv:2009.11189}} (\bibinfo{year}{2020}).
\newblock


\bibitem[Yang et~al\mbox{.}(2023a)]%
        {yang2023mmreact}
\bibfield{author}{\bibinfo{person}{Zhengyuan Yang}, \bibinfo{person}{Linjie Li}, \bibinfo{person}{Jianfeng Wang}, \bibinfo{person}{Kevin Lin}, \bibinfo{person}{Ehsan Azarnasab}, \bibinfo{person}{Faisal Ahmed}, \bibinfo{person}{Zicheng Liu}, \bibinfo{person}{Ce Liu}, \bibinfo{person}{Michael Zeng}, {and} \bibinfo{person}{Lijuan Wang}.} \bibinfo{year}{2023}\natexlab{a}.
\newblock \bibinfo{title}{MM-REACT: Prompting ChatGPT for multimodal reasoning and action}.
\newblock
\newblock
\showeprint[arxiv]{2303.11381}~[cs.CV]


\bibitem[Ye et~al\mbox{.}(2020)]%
        {ye2020reinforcement}
\bibfield{author}{\bibinfo{person}{Yunan Ye}, \bibinfo{person}{Hengzhi Pei}, \bibinfo{person}{Boxin Wang}, \bibinfo{person}{Pin-Yu Chen}, \bibinfo{person}{Yada Zhu}, \bibinfo{person}{Ju Xiao}, {and} \bibinfo{person}{Bo Li}.} \bibinfo{year}{2020}\natexlab{}.
\newblock \showarticletitle{Reinforcement-learning based portfolio management with augmented asset movement prediction states}. In \bibinfo{booktitle}{\emph{Proceedings of the AAAI Conference on Artificial Intelligence}}, Vol.~\bibinfo{volume}{34}. \bibinfo{pages}{1112--1119}.
\newblock


\bibitem[Yu et~al\mbox{.}(2022)]%
        {yu2022generate}
\bibfield{author}{\bibinfo{person}{Wenhao Yu}, \bibinfo{person}{Dan Iter}, \bibinfo{person}{Shuohang Wang}, \bibinfo{person}{Yichong Xu}, \bibinfo{person}{Mingxuan Ju}, \bibinfo{person}{Soumya Sanyal}, \bibinfo{person}{Chenguang Zhu}, \bibinfo{person}{Michael Zeng}, {and} \bibinfo{person}{Meng Jiang}.} \bibinfo{year}{2022}\natexlab{}.
\newblock \showarticletitle{Generate rather than retrieve: Large language models are strong context generators}.
\newblock \bibinfo{journal}{\emph{arXiv preprint arXiv:2209.10063}} (\bibinfo{year}{2022}).
\newblock


\bibitem[Yu et~al\mbox{.}(2023)]%
        {yu2023finmem}
\bibfield{author}{\bibinfo{person}{Yangyang Yu}, \bibinfo{person}{Haohang Li}, \bibinfo{person}{Zhi Chen}, \bibinfo{person}{Yuechen Jiang}, \bibinfo{person}{Yang Li}, \bibinfo{person}{Denghui Zhang}, \bibinfo{person}{Rong Liu}, \bibinfo{person}{Jordan~W. Suchow}, {and} \bibinfo{person}{Khaldoun Khashanah}.} \bibinfo{year}{2023}\natexlab{}.
\newblock \bibinfo{title}{FinMem: A performance-enhanced LLM trading agent with layered memory and character design}.
\newblock
\newblock
\showeprint[arxiv]{2311.13743}~[q-fin.CP]


\bibitem[Zhang et~al\mbox{.}(2023)]%
        {zhang2023appagent}
\bibfield{author}{\bibinfo{person}{Chi Zhang}, \bibinfo{person}{Zhao Yang}, \bibinfo{person}{Jiaxuan Liu}, \bibinfo{person}{Yucheng Han}, \bibinfo{person}{Xin Chen}, \bibinfo{person}{Zebiao Huang}, \bibinfo{person}{Bin Fu}, {and} \bibinfo{person}{Gang Yu}.} \bibinfo{year}{2023}\natexlab{}.
\newblock \bibinfo{title}{AppAgent: Multimodal agents as smartphone users}.
\newblock
\newblock
\showeprint[arxiv]{2312.13771}~[cs.CV]


\end{thebibliography}

\onecolumn
\appendix
\section{Details of Notations}
\label{app:notations}

We provide the main notations in Table \ref{app:table_notation}. 

\begin{table}[htbp]
\centering
\caption{Notations in the paper.}
\footnotesize
\label{app:table_notation}
\centering
\resizebox{0.6\textwidth}{!}{
\begin{tabular}{p{3cm}p{5cm}}
\toprule   
Notation & Description \\ 
\midrule
$t$                 &   Current day  \\
$T$                 &   Total trading days \\
$t-s, t+s$          &   Short-term price analysis from $t-s$ to $t$ and $t$ to $s$ \\
$t-m, t+m$          &   Medium-term price analysis from $t-m$ to $t$ and $t$ to $m$ \\
$t-l, t+l$          &   Long-term price analysis from $t-l$ to $t$ and $t$ to $l$ \\
$\mathit{S}$        &   A finite set of states \\
$s_{t}$             &   State of day $t$ \\
$\mathcal{A}$       &   A finite set of actions \\
$a_{t}$             &   Action of day $t$ \\
$\mathcal{T}$       &   Transition function \\
$R$                 &   Reward function \\
$r_{t}$             &   Reward of day $t$ with $s_{t}$ and $a_{t}$ \\
$\gamma$            &   Discount factor \\
$\pi$               &   Policy \\
$\mu(\cdot)$        &   Specialized modules for reasoning \\
$\mu_{t}$           &   Specialized modules of day $t$ \\
$\lambda$           &   Financial trading task \\
$Mem^{\lambda}_{t}$ &   Memory of day $t$ in the task $\lambda$ \\
$Tool^{\lambda}_{t}$&   Tool of day $t$ in the task $\lambda$ \\
$\phi(\cdot)$       &   Task-relevant prompt generator \\
$\mathcal{D}^{\lambda}_{t}$ &  Action parsing function \\
$M^{\lambda}_{t}, L^{\lambda}_{t}, H^{\lambda}_{t}$ & M, L, H modules \\
$\phi^{\lambda}_{M}, \phi^{\lambda}_{L}, \phi^{\lambda}_{H}$ & Prompt generator for M, L, H \\
$Mem^{M, \lambda}_{t}$, $Mem^{L, \lambda}_{t}$, $Mem^{H, \lambda}_{t}$ & Memory of M, L, H modules of day $t$ in the task $\lambda$ \\
$KC_{t}$            & Kline chart of day $t$ \\
$TC_{t}$            & Trading chart of day $t$ \\
$SLMI_{t}$          & Summary of latest market intelligence of day $t$ \\
$QLMI_{t} = \{Q^{L}_{1}, ..., Q^{L}_{M}  \}$          & $M$ query texts for retrieving past market intelligence of day $t$ \\
$K$                 & Retrieved topk items \\
$Q^{P}_{i,j}$       & Retrieval type $i$ and top $j$ retrieved past market latest intelligence \\
$SPMI_{t}$          & Summary of past market intelligence of day $t$ \\
$LLR^{ST}_{t}, LLR^{MT}_{t}, LLR^{LT}_{t}$ & Low-level reflection results at short term, medium term and long term impact \\
$QLLR_{t}$          & Query text for low-level reflection of day $t$ \\
$PLLR^{ST}_{t}, PLLR^{MT}_{t}, PLLR^{LT}_{t}$ & retrieved topk low-level reflection in short term, medium term and long term \\
$HLR_{t}$           & High-level reflection results of day $t$ \\
$QHLR_{t}$          & Query text for high-level reflection of day $t$ \\
$PHLR_{t}$          & Retrieved topk high-level reflection of day $t$ \\
\bottomrule  
\end{tabular} 
}
\end{table}

\section{Details of Datasets and Processing}
\label{app:datasets}
To conduct a thorough evaluation of \texttt{FinAgent}, we evaluate it across 6 real-world datasets. These included five datasets from the US stock markets and one is the cryptocurrency. Each of them have multiple forms of data that come from various sources. Specifically, i) \textbf{Asset Price} at the day-level, including price data for open, high, low, close, and adj close; ii) \textbf{Visual Data} consists of historical Kline charts and trading charts, which are visual representations of asset market data and trading process on a daily basis; iii) \textbf{Asset News} coverage with daily updates from various esteemed sources, including Bloomberg Technology, Seeking Alpha, CNBC Television, and more, ensuring a diverse and thorough perspective on the financial markets; iv) \textbf{Expert Guidance} provided by financial experts as the auxiliary information, aiming to furnish a thorough and well-rounded comprehension of market status. We summarize statistics of the 6 datasets in Table \ref{tab:dataset} and further elaborate on them as follows:

\textbf{Asset}. We selected a varied portfolio comprising five stocks Apple Inc. (AAPL), Amazon.com Inc. (AMZN), Alphabet Inc. (GOOGL), Microsoft Corporation (MSFT), and Tesla Inc. (TSLA), a foreign exchange pair, and a prominent cryptocurrency, Ethereum (ETH). This selection aims to showcase \texttt{FinAgent}'s versatility and consistency across various financial assets. Chosen for their extensive news coverage and representation of different market sectors, these data provide a robust basis for assessing \texttt{FinAgent}'s generalization capabilities across diverse financial environments.

\textbf{Price and News}. We acquired price and news data for all assets from Financial Modeling Prep\footnote{FMP API provides data about stock historical price and news, company financial statements, and cryptocurrencies. Entry is https://site.financialmodelingprep.com.} (FMP), wherein the price data encompasses including open, high, low, clos, and adj close. The news data is sourced from renowned market analysis and stock research platforms, notably including Seeking Alpha and so on. This selection ensures a comprehensive dataset, integrating both quantitative financial metrics and qualitative market insights.

\textbf{Visual Data}. Within the textual data framework, we furnish \texttt{FinAgent} with visual information, specifically including historical Kline charts and trading line charts, to enhance its analytical capabilities. The tool employed for this plotting task is the pyecharts \footnote{https://github.com/pyecharts/pyecharts}, a specialized library for financial data visualization.

\textbf{Expert Guidance}. Expert Guidance is provided as a distinct component of the auxiliary information by augmented tools. This selection ensures a comprehensive dataset, integrating professional analysts and individual investors insights, fostering a diverse range of perspectives in the investment community. We obtained the expert professional analysis from Seeking Alpha\footnote{https://seekingalpha.com/}. Seeking Alpha is a popular platform among investors and financial analysts, is renowned for its diverse professional analysis, providing valuable insights from seasoned analysts across the financial market spectrum. 

\textbf{Trading Date}. For dataset split, the data from the latter half of the year is allocated for testing (2023-06-01\footnote{Dates follow the YYYY-MM-DD format, e.g., "2023-06-01" for June 1st, 2023.} $\sim$ 2024-01-01) purposes, while the data from the penultimate year is utilized for training (2022-06-01 $\sim$ 2023-06-01). 

\section{Details of Comparison with Baselines}
\label{app:comparison}
We compared FinAgent with 9 baseline methods in terms of 6 financial metrics. Table \ref{tab:app_baselines} and Figure \ref{fig:baselines} demonstrate our method significantly outperforms existing baselines, especially remarkable improvements in profitability, and setting a new benchmark in the field.

\begin{table*}[!t]
\caption{Performance comparison of all methods on six profitable metrics. Results in red, yellow and green show the best, second best and third best results on each dataset. Improvement is the FinAgent over the best-performing baselines.}
\centering
\footnotesize
\renewcommand{\arraystretch}{0.8}
\resizebox{\textwidth}{!}{ 
\begin{tabular}{clccclccclccclccclccclccc}
\toprule
\multirow{3}{*}{Categories} & \multirow{3}{*}{Models} & \multicolumn{3}{c}{AAPL} & & \multicolumn{3}{c}{AMZN} & & \multicolumn{3}{c}{GOOGL} & & \multicolumn{3}{c}{MSFT} & & \multicolumn{3}{c}{TSLA} & & \multicolumn{3}{c}{ETHUSD}   \\ 
\cmidrule{3-5}\cmidrule{7-9}\cmidrule{11-13}\cmidrule{15-17}\cmidrule{19-21}\cmidrule{23-25}
&& ARR\%$\uparrow$  & SR$\uparrow$  & MDD\%$\downarrow$  & & ARR\%$\uparrow$  & SR$\uparrow$  & MDD\%$\downarrow$  & & ARR\%$\uparrow$  & SR$\uparrow$  & MDD\%$\downarrow$ & & ARR\%$\uparrow$  & SR$\uparrow$  & MDD\%$\downarrow$ & & ARR\%$\uparrow$  & SR$\uparrow$  & MDD\%$\downarrow$ & & ARR\%$\uparrow$  & SR$\uparrow$  & MDD\%$\downarrow$  \\ 
\midrule
Market 
& B\&H& 13.0024& 0.5998& 14.7809&& \textcolor{TBest}{42.3337}& 1.0834& 17.3848&& 22.4726& 0.7108& 12.9705&& 22.4942& 0.8373& 12.9214&& 37.4009& 0.7239& 32.6523&& 29.2588& 0.8655& 23.2077\\
\midrule
\multirow{4}{*}{\begin{tabular}[c]{@{}c@{}}Rule-based\end{tabular}}
& MACD& 11.8642& 0.7221& 10.3799&& 14.2748& 0.7056& 7.841&& -18.0034& -0.8867& 20.0718&& 15.2322& 0.7704& 8.3445&& -4.8974& -0.0203& 14.1546&& 10.236& 0.4689& 24.3238\\
& KDJ\&RSI& 2.1737& 0.1746& 11.8789&& 19.3757& 0.6495& 17.2746&& 24.391& \textcolor{FBest}{2.1282}& \textcolor{FBest}{2.03}&& 18.8415& 1.0587& 7.7806&& 2.137& 0.1695& 24.727&& 8.8745& 0.5098& 16.9536\\
& ZMR& -3.9084& -0.2186& 8.8819&& 18.7289& 0.8412& 7.8938&& 32.5112& \textcolor{TBest}{1.4533}& \textcolor{TBest}{5.3845}&& 9.8637& 0.7106& \textcolor{TBest}{6.221}&& -7.2806& -0.0863& 19.9048&& 29.3519& 1.2294& 13.1098\\
\midrule
\multirow{4}{*}{\begin{tabular}[c]{@{}c@{}}ML\&DL-based\end{tabular}}
& LGBM& 16.9268& \textcolor{SBest}{1.4708}& \textcolor{FBest}{2.5204}&& 29.3395& 0.7187& 17.414&& 24.7746& 0.6958& 12.9814&& 19.2771& 0.6668& 12.9616&& 15.575& 0.843& \textcolor{FBest}{3.8844}&& 24.9111& 0.7154& 22.9568\\
& LSTM& 10.9742& 0.5363& 11.9535&& 15.9051& 0.4588& 17.414&& 24.8583& 0.6989& 12.9814&& 18.8603& 0.6779& 11.7544&& 17.3617& 0.7796& \textcolor{SBest}{4.4384}&& \textcolor{TBest}{36.0865}& 1.0254& 21.5043\\
& Transformer& 17.115& 0.957& \textcolor{SBest}{7.5295}&& 32.6621& \textcolor{TBest}{1.1134}& \textcolor{FBest}{4.9593}&& 13.692& 0.4571& 12.9253&& 17.4417& \textcolor{TBest}{1.4553}& \textcolor{FBest}{2.5895}&& 39.701& \textcolor{SBest}{1.0445}& \textcolor{TBest}{8.1721}&& 31.0038& 1.0205& 12.9309\\
\midrule
\multirow{3}{*}{\begin{tabular}[c]{@{}c@{}} RL-based \end{tabular}}
& DQN& 7.9236& 0.401& 14.8785&& 27.4305& \textcolor{SBest}{1.1701}& \textcolor{TBest}{5.2736}&& \textcolor{TBest}{34.4026}& 1.3859& 7.1473&& \textcolor{TBest}{30.4406}& 1.1782& 10.5612&& 15.0693& 0.443& 28.1204&& 29.8052& 1.1826& \textcolor{FBest}{9.5297}\\
& SAC& \textcolor{SBest}{24.8449}& \textcolor{TBest}{1.1234}& 11.9776&& 38.3318& 1.0733& 13.8432&& 23.8034& 0.7506& 13.0667&& 22.0218& 0.8177& 12.9214&& \textcolor{TBest}{42.2209}& 0.8727& 26.1947&& 17.8439& 0.7635& \textcolor{SBest}{10.0587}\\
& PPO& 13.2619& 0.6096& 14.7809&& 21.1745& 0.6965& 13.8432&& \textcolor{SBest}{38.2907}& 1.2982& 8.4536&& 11.3219& 0.4831& 17.5054&& 33.6444& 0.7767& 28.3527&& 34.7469& \textcolor{SBest}{1.3096}& \textcolor{TBest}{11.1171}\\
\midrule
\multirow{2}{*}{\begin{tabular}[c]{@{}c@{}} LLM-based \end{tabular}}
& FinGPT& -5.4632& -0.1731& 16.2268&& \textcolor{SBest}{42.9331}& 1.1026& 18.9359&& 12.277& 0.4444& 13.0013&& 25.1012& 0.9667& 9.8426&& 38.4338& 0.7504& 31.474&& 21.5746& 0.6801& 25.562\\
& FinMem& \textcolor{TBest}{23.7809}& 1.1073& 10.3872&& 40.07& 1.034& 18.5279&& 31.2716& 1.1073& 8.9706&& \textcolor{SBest}{40.5757}& \textcolor{SBest}{1.4989}& 7.4838&& \textcolor{SBest}{50.0353}& \textcolor{TBest}{0.9233}& 25.7714&& \textcolor{SBest}{44.717}& \textcolor{TBest}{1.2738}& 13.587\\
\midrule
\multirow{5}{*}{\begin{tabular}[c]{@{}c@{}}\textbf{\texttt{FinAgent}}\\(Our)\end{tabular}}
& No-finetuned& -2.0047& -0.0216& 16.7567&& 41.6548& 1.3959& \textcolor{SBest}{5.2678}&& 21.108& 0.682& 13.0719&& 22.6483& 0.8373& 12.9214&& 38.0164& 1.2951& 12.0609&& 23.4038& 0.7924& 23.231\\
& w/o-MLH& 18.5186& 0.9882& 11.8842&& 62.3106& 1.6379& 11.3999&& 37.3308& 1.2657& \textcolor{SBest}{4.0}&& 18.297& 0.8945& 9.6792&& 39.0087& 0.8971& 22.5376&& 16.2106& 0.6329& 15.9325\\
& w/o-LHT& 12.6872& 0.5849& 14.7806&& 43.2195& 1.1057& 14.6786&& 17.414& 0.5763& 13.006&& 18.4571& 0.7108& 14.2551&& 39.2704& 0.7662& 30.1507&& 25.9708& 0.7718& 24.4314\\
& w/o-HT& 21.3044& 0.9777& 13.6424&& 47.3916& 1.2032& 15.5038&& 29.3235& 1.0177& 8.9706&& 39.9073& 1.4777& 7.4838&& 57.1638& 1.0201& 25.7714&& 52.3265& 1.3351& 13.587\\
& w/o-T& \textcolor{FBest}{33.7509}& \textcolor{FBest}{1.5205}& \textcolor{TBest}{8.1783}&& 63.8116& \textcolor{FBest}{1.7009}& 9.7325&& 52.1066& \textcolor{SBest}{1.8228}& 8.3106&& 42.5213& 1.4909& 7.6998&& 89.2532& 1.4573& 27.6213&& \textcolor{FBest}{54.804}& \textcolor{FBest}{1.403}& 11.7427\\
& \texttt{FinAgent}& 31.8972& 1.4326& 10.4032&& \textcolor{FBest}{65.0998}& 1.6096& 13.198&& \textcolor{FBest}{56.1542}& 1.7786& 8.4532&& \textcolor{FBest}{44.7359}& \textcolor{FBest}{1.7884}& \textcolor{SBest}{5.5732}&& \textcolor{FBest}{92.2677}& \textcolor{FBest}{2.0088}& 12.143&& 43.0822& 1.1773& 12.7171\\
\midrule

\multicolumn{2}{c}{Improvement(\%)}
&   35.8464    &  3.3791     &    -   &       &    51.6308   &   45.3636    &    -   &       &   46.6523    &    -   &   -    &       &    10.2529   &   19.3142    &   -    &       &     84.4052  &   92.3217    &   -    &       &    22.5574   &    7.1319   & - \\
\bottomrule
\toprule
\multirow{3}{*}{Categories} & \multirow{3}{*}{Models} & \multicolumn{3}{c}{AAPL} & & \multicolumn{3}{c}{AMZN} & & \multicolumn{3}{c}{GOOGL} & & \multicolumn{3}{c}{MSFT} & & \multicolumn{3}{c}{TSLA} & & \multicolumn{3}{c}{ETHUSD}   \\ 
\cmidrule{3-5}\cmidrule{7-9}\cmidrule{11-13}\cmidrule{15-17}\cmidrule{19-21}\cmidrule{23-25}
&& SOR$\uparrow$  & CR$\uparrow$ & VOL$\downarrow$ && SOR$\uparrow$  & CR$\uparrow$ & VOL$\downarrow$ && SOR$\uparrow$  & CR$\uparrow$ & VOL$\downarrow$ && SOR$\uparrow$  & CR$\uparrow$ & VOL$\downarrow$ && SOR$\uparrow$  & CR$\uparrow$ & VOL$\downarrow$ && SOR$\uparrow$  & CR$\uparrow$ & VOL$\downarrow$ \\ 
\midrule
Market 
& B\&H& 16.5846& 0.9589& 0.0114&& \textcolor{SBest}{35.1804}& 2.4319& 0.0188&& 18.5186& 1.9025& 0.0167&& 26.5133& 1.8135& 0.0135&& 23.3319& 1.3856& 0.0301&& 23.2235& 1.3831& 0.0222\\
\midrule
\multirow{4}{*}{\begin{tabular}[c]{@{}c@{}}Rule-based\end{tabular}}
& MACD& 13.7755& 1.1877& 0.0082&& 19.2452& 1.9176& \textcolor{FBest}{0.0103}&& -20.917& -0.8883& \textcolor{SBest}{0.0097}&& 18.7485& 1.8966& 0.0099&& -0.4067& -0.0544& 0.0182&& 10.8951& 0.5433& 0.0166\\
& KDJ\&RSI& 3.3994& 0.2578& 0.0084&& 16.6915& 1.2471& 0.016&& 36.6655& \textcolor{FBest}{11.4076}& \textcolor{FBest}{0.0052}&& 19.5346& 2.4169& \textcolor{TBest}{0.0085}&& 3.3617& 0.2737& 0.0192&& 7.7786& 0.6106& \textcolor{FBest}{0.012}\\
& ZMR& -2.9977& -0.37& \textcolor{SBest}{0.0072}&& 9.5853& 2.4473& \textcolor{SBest}{0.011}&& 35.1125& \textcolor{TBest}{5.7782}& \textcolor{TBest}{0.0103}&& 12.9275& 1.6385& \textcolor{SBest}{0.0069}&& -1.8214& -0.1642& 0.0182&& 21.4433& 2.1715& \textcolor{SBest}{0.0137}\\
\midrule
\multirow{4}{*}{\begin{tabular}[c]{@{}c@{}}ML\&DL-based\end{tabular}}
& LGBM& \textcolor{SBest}{45.2444}& \textcolor{FBest}{6.5825}& \textcolor{FBest}{0.0049}&& 26.2943& 1.8469& 0.0193&& 19.7497& 2.082& 0.0167&& 22.6593& 1.5993& 0.0134&& 12.1981& \textcolor{TBest}{4.0837}& \textcolor{FBest}{0.0081}&& 21.8037& 1.2503& 0.0217\\
& LSTM& 14.5498& 0.9915& 0.0095&& 14.7442& 1.1241& 0.0184&& 19.5489& 2.0865& 0.0167&& 21.5244& 1.7144& 0.0128&& 16.1653& 4.0391& \textcolor{SBest}{0.0099}&& 27.1716& 1.7031& 0.0193\\
& Transformer& 28.3604& \textcolor{TBest}{2.288}& \textcolor{TBest}{0.0078}&& 27.8371& \textcolor{SBest}{6.5258}& 0.0125&& 11.5053& 1.2619& 0.0154&& 19.7493& \textcolor{SBest}{6.6012}& \textcolor{FBest}{0.0051}&& \textcolor{SBest}{34.1884}& \textcolor{SBest}{4.8624}& 0.0164&& \textcolor{TBest}{27.864}& 2.4209& 0.0166\\
\midrule
\multirow{3}{*}{\begin{tabular}[c]{@{}c@{}} RL-based \end{tabular}}
& DQN& 10.3705& 0.6266& 0.0111&& 29.7698& \textcolor{TBest}{5.1156}& \textcolor{TBest}{0.0111}&& \textcolor{TBest}{37.7389}& 4.619& 0.0114&& \textcolor{TBest}{32.8642}& 2.8493& 0.0122&& 11.0394& 0.7399& 0.0225&& 24.6279& \textcolor{TBest}{3.0545}& \textcolor{TBest}{0.0143}\\
& SAC& \textcolor{TBest}{33.5676}& 2.0552& 0.0105&& 32.432& 2.766& 0.0112&& 19.3937& 1.9708& 0.0165&& 26.0662& 1.7829& 0.0135&& \textcolor{TBest}{26.2869}& 1.441& 0.0251&& 13.3918& 1.9199& 0.0146\\
& PPO& 16.766& 0.9747& 0.0113&& 20.3167& 1.6735& 0.0159&& \textcolor{SBest}{42.7201}& 4.3691& 0.0136&& 14.7808& 0.7484& 0.013&& 19.1547& 1.3203& 0.0231&& 27.2416& 2.9836& 0.0147\\
\midrule
\multirow{2}{*}{\begin{tabular}[c]{@{}c@{}} LLM-based\end{tabular}}
& FinGPT& -4.6731& -0.246& 0.0111&& \textcolor{TBest}{34.8082}& 2.2545& 0.0186&& 11.6596& 1.1842& 0.0167&& 30.1935& 2.5867& 0.0127&& 23.1813& 1.4402& 0.0291&& 18.5817& 1.0131& 0.0221\\
& FinMem& 29.8819& 2.2731& 0.0102&& 33.0779& 2.1843& 0.0188&& 34.7826& 3.4572& 0.0134&& \textcolor{SBest}{47.1061}& \textcolor{TBest}{5.1266}& 0.0123&& 25.8819& 2.0887& 0.028&& \textcolor{SBest}{34.1492}& \textcolor{SBest}{3.1349}& 0.0194\\
\midrule
\multirow{5}{*}{\begin{tabular}[c]{@{}c@{}}\textbf{\texttt{FinAgent}}\\(Our)\end{tabular}} 
& No-finetuned& -0.5635& -0.0295& 0.011&& 46.9773& \textcolor{FBest}{7.5705}& 0.0127&& 17.5637& 1.7864& 0.0164&& 26.6949& 1.8259& 0.0135&& 17.2111& 3.0426& \textcolor{TBest}{0.0136}&& 19.6504& 1.1081& 0.0188\\
& w/o-MLH& 20.8183& 1.5678& 0.009&& \textcolor{FBest}{53.4328}& 4.9724& 0.0166&& 24.9489& 2.9016& 0.0137&& 22.0401& 1.9273& 0.01&& 20.3649& 1.8355& 0.0221&& 14.0758& 1.1917& 0.0174\\
& w/o-LHT& 16.15& 0.9396& 0.0114&& 35.1674& 2.9256& 0.0186&& 15.1339& 1.552& 0.0168&& 22.4806& 1.3883& 0.0133&& 22.2082& 1.5162& 0.0286&& 21.0778& 1.2131& 0.0222\\
& w/o-HT& 29.5194& 1.576& 0.0105&& 40.4896& 2.9758& 0.0184&& 33.6191& 3.2906& 0.0139&& 46.4396& 5.054& 0.0123&& 28.3015& 2.2956& 0.0278&& 37.172& 3.6027& 0.0212\\
& w/o-T& \textcolor{FBest}{46.6145}& \textcolor{SBest}{3.9301}& 0.0101&& 50.1096& 5.9255& 0.0162&& 62.2508& 5.7201& 0.0125&& 40.3937& 5.2137& 0.0129&& 41.5642& 2.9306& 0.0266&& \textcolor{FBest}{37.4619}& \textcolor{FBest}{4.2958}& 0.0208\\
& FinAgent& 44.2812& 2.9424& 0.0102&& 52.5602& 4.4841& 0.0176&& \textcolor{FBest}{62.2992}& \textcolor{SBest}{6.0365}& 0.0138&& \textcolor{FBest}{49.6249}& \textcolor{FBest}{7.4209}& 0.0111&& \textcolor{FBest}{45.4139}& \textcolor{FBest}{6.4543}& 0.0187&& 31.0159& 3.3078& 0.0207\\
\midrule
\multicolumn{2}{c}{Improvement(\%)}
&   3.0282   &   -    &    -   &       &   51.8823    &   16.0088    &    -   &       &   45.8311    &    -   &   -    &       &   5.3471    &   12.4174    &   -    &       &   32.8342    &   32.7390    &   -    &       &   9.7007    &   37.0315    & - \\
\bottomrule
\end{tabular}
}
\label{tab:app_baselines}
\end{table*}

\section{Details of Benchmark Methods}
\label{app:benchmark_methods}
We compare and evaluate the trading performance of \texttt{FinAgent} with four widely accepted conventional benchmark trading strategies (\textbf{B\&H}, \textbf{MACD}, \textbf{KDJ\&RSI} and \textbf{ZMR}) and five advanced algorithms. Among these, \textbf{SAC} \cite{haarnoja2018soft}, \textbf{PPO} \cite{schulman2017proximal} and \textbf{DQN} \cite{mnih2013playing} are three models employed deep reinforcement learning (RL) methods, \textbf{FinGPT} \cite{yang2023fingpt} is based on LLM, and another is \textbf{FinMem} \cite{yu2023finmem} that based on LLM Agents. The following will provide a brief introduction to each model:

\begin{itemize}[leftmargin=*]
    \item \textbf{Rule-based}
        \begin{itemize}[leftmargin=0cm]
            \item \textbf{Buy-and-Hold (B\&H)} involves holding assets for an extended period, regardless of short-term market fluctuations, assuming that long-term returns will be more favorable.
            \item \textbf{Moving Average Convergence Divergence (MACD)} is a technical analysis tool that uses MACD indicator and signal line crossovers to identify trading signals and market trends.
            \item \textbf{KDJ with RSI Filter (KDJ\&RSI)} integrates the KDJ indicator for detecting market extremes with the RSI indicator for momentum analysis to identify precise trading signals in financial markets.
            % \item \textbf{Stochastic Oscillator and Bollinger Bands (SO\&BB)} combines momentum based SO for overbought or oversold conditions with BB to assess market volatility and potential price reversals.
            \item  \textbf{Z-score Mean Reversion (ZMR)} assumes that the price will revert to its mean over time with the metric of Z-score.
        \end{itemize}
    \item \textbf{RL-based}
        \begin{itemize}[leftmargin=0cm]
            \item \textbf{SAC} \cite{haarnoja2018soft} is an off-policy actor-critic algorithm that optimizes trading strategies using entropy regularization and soft value functions in continuous action spaces.
            \item \textbf{PPO} \cite{schulman2017proximal} updates trading policies iteratively to balance exploration and exploitation, ensuring stability and sample efficiency.
            \item \textbf{DQN} \cite{mnih2013playing} uses deep neural networks (DNNs) to make trading decisions by approximating the action-value function based on market data.
        \end{itemize}
    \item \textbf{LLM-based}
        \begin{itemize}[leftmargin=0cm]
            \item \textbf{FinGPT} \cite{yang2023fingpt} is an open-source LLM framework designed to transform textual and numerical inputs into insightful financial decisions, asserting its advantage over conventional B\&H strategies.
            \item \textbf{FinMem} \cite{yu2023finmem} is an advanced LLM agent framework for automated trading, optimized through fine-tuning the agent's perceptual span and character settings, significantly enhancing trading performance and boosting cumulative investment returns.
        \end{itemize}
\end{itemize}

\section{Details of Workflow of FinAgent}
\label{app:workflow}
In this section we focus on FinAgent's workflow and code implementation.

\subsection{Main Entry}
We follow the RL process, as shown in the code below, where we initialize the dataset and construct environment for subsequent training and inference.

\begin{lstlisting}[style = python]
# load config
cfg = Config.fromfile(config_path)
# build dataset
dataset = DATASET.build(cfg.dataset)
# build environment
env = ENVIRONMENT.build(cfg.environment)
# init environment
state, info = env.reset()
# execute steps
while True:
    action = run_step(cfg, state, info, ...)
    state, reward, done, truncated, info = env.step(action)
    if done:
        break
# done
\end{lstlisting}

% \begin{algorithm}
% \caption{Main Entry}
% \begin{algorithmic}
% \State Steps
% \end{algorithmic}
% \lstset{language=Python}
% \begin{lstlisting}
% # load config
% cfg = Config.fromfile(config_path)
% # build dataset
% dataset = DATASET.build(cfg.dataset)
% # build environment
% env = ENVIRONMENT.build(cfg.environment)
% # init environment
% state, info = env.reset()
% # execute steps
% while True:
%     action = run_step(cfg, state, info, ...)
%     state, reward, done, truncated, info = env.step(action)
%     if done:
%         break
% # done
% \end{lstlisting}
% \end{algorithm}

\subsection{Run Step}
The whole process is mainly through the global \texttt{params} storage and transimit parameters, we do not list the parameter transmition process in detail here. To execute each step, the following primary procedures are adhered to:

\begin{lstlisting}[style = python]
# global params
params = dict()
# plot Kline chart
kline_path = plots.plot_kline(state, info, ...)
params.update({"kline_path": kline_path})
# prepare tools params
tools_params = prepared_tools_params(state, info, ...)
params.update(tools_params)
# 01 - latest market intelligence
template = read_resource_file(...) # load latest market intelligence prompt template
lmi = PROMPT.build(...) # build instance
lmi_res = lmi.run(state, info, params, template,...) # run
# 02 - retrieve the past market intelligence
retrieved_params = retrieve_pmi(state, info, params, memory, diverse_query, ...) 
params.update(retrieved_params)
# 03 - add latest market intelligence to memory
lmi.add_to_memory(lmi_res, memory, ...)
# 04 - past market intelligence
template = read_resource_file(...) # load past market intelligence prompt template
pmi = PROMPT.build(...) # build instance
pmi_res = pmi.run(state, info, params, template,...) # run
# 05 - low-level reflection
template = read_resource_file(...) # load low level reflection prompt template
llr = PROMPT.build(...) # build instance
llr_res = llr.run(state, info, params, template,...) # run
# 06 - retrieve the past low-level reflection
retrieved_params = retrieve_pllr(state, info, params, memory, diverse_query, ...) 
params.update(retrieved_params)
# 07 - add low-level reflection to memory
llr.add_to_memory(llr_res, memory, ...)
# plot trading chart
trading_path = plots.plot_trading(state, info, ...)
params.update({"trading_path": trading_path})
# 08 - high-level reflection
template = read_resource_file(...) # load high level reflection prompt template
hlr = PROMPT.build(...) # build instance
hlr_res = hlr.run(state, info, params, template,...) # run
# 09 - retrieve the past high-level reflection
retrieved_params = retrieve_phlr(state, info, params, memory, diverse_query, ...) 
params.update(retrieved_params)
# 10 - add high-level reflection to memory
hlr.add_to_memory(hlr_res, memory, ...)
# 11 - decision-making
template = read_resource_file(...) # load decision-making prompt template
decision = PROMPT.build(...) # build instance
decision_res = decision.run(state, info, params, template,...) # run
action, reasoning = decision_res["action"], decision_res["reasoing"]
\end{lstlisting}

\section{Details of Prompt Design}
\label{app:prompt_design}
Our prompt templates are designed modularly, featuring separate templates for the latest and past market intelligence, low-level reflection, high-level reflection, and decision-making modules. These templates include variables marked as "\$\$key\$\$" within the "params" dictionary. Our template utilizes HTML for its ability to combine formatting with user-friendly visualizations. In our testing, we find that JSON's strict formatting requirements frequently lead to errors. As a result, we opt for XML, a format with more flexible standards. XML is easy to parse, and extracting fields is simpler, making it an ideal output format for GPT-4. Next, we use the "params" to populate the template with the relevant fields. Following this, we employ an HTML parsing tool to create the JSON message format GPT-4 API demands. We then make API requests to obtain the response output. An example of low level reflection template running is shown as the following Figure \ref{fig:app_template_format}.

\begin{figure*}[htbp]
\captionsetup{skip=2pt}
  \centering
    \includegraphics[width=1.0\textwidth]{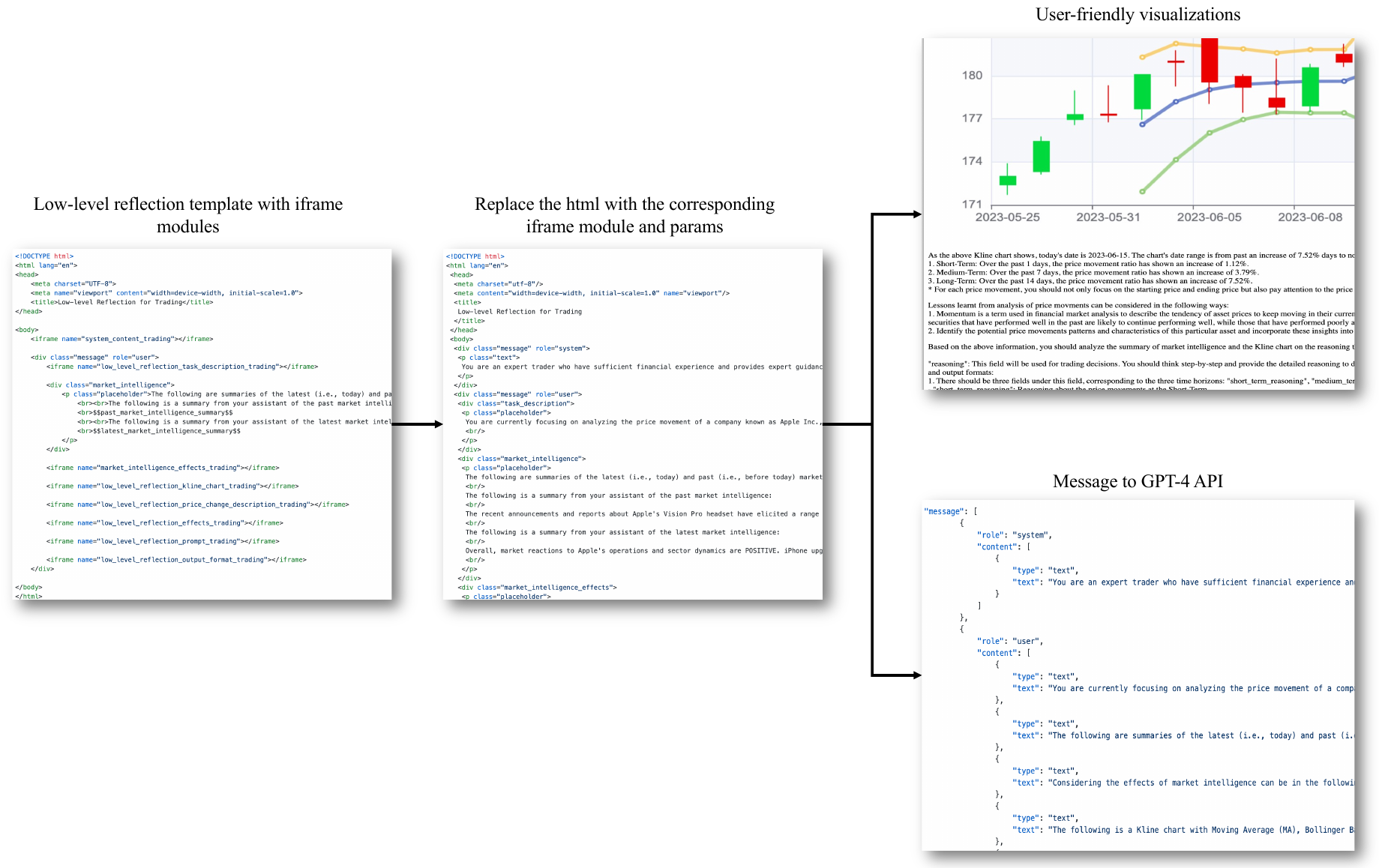}
    \caption{An example of low level reflection template running.}
    \label{fig:app_template_format}
\end{figure*}

\begin{figure*}[htbp]
\captionsetup{skip=2pt}
  \centering
    \includegraphics[width=1.0\textwidth]{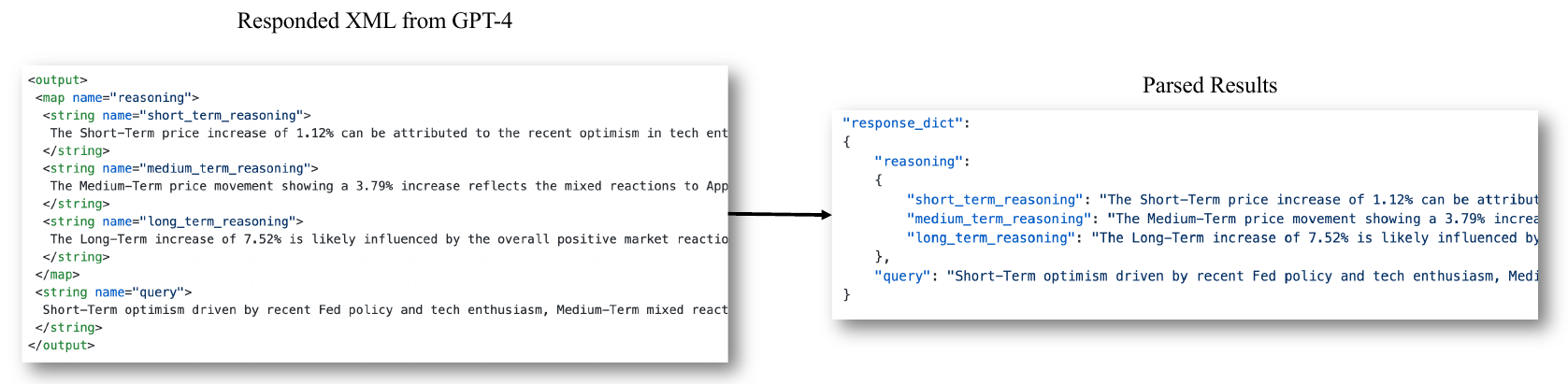}
    \caption{Example of parsing the results of a low-level reflection template running that call to GPT-4 get responded XML.}
    \label{fig:app_result_parse}
\end{figure*}

Lastly, we extract the necessary field information using an XML parsing tool as depicted in Figure \ref{fig:app_result_parse}. Each of these iframe modules is carefully designed by us, and we will show each of them in the following subsections.

\clearpage
\subsection{Market Intelligence}
Market intelligence consists of 2 templates and 7 iframe modules, the XML templates of which are detailed as follows:

\noindent
\begin{figure*}[htbp]
\paragraph{Latest Market Intelligence Template}\mbox{}\\
\begin{lstlisting}[style = xml]
<!DOCTYPE html>
<html lang="en">
<head>
  <meta charset="UTF-8">
  <meta name="viewport" content="width=device-width, initial-scale=1.0">
  <title>Summary of Latest Market Intelligence</title>
</head>
<body>

    <iframe name="system_content_trading"></iframe>

    <div class="message" role="user">
        <iframe name="market_intelligence_task_description_trading"></iframe>

        <div class="market_intelligence">
            <p class="placeholder">The following market intelligence (e.g., news, financial reports) contains latest (i.e., today) information related to $$asset_symbol$$, including the corresponding dates, headlines, and contents, with each item distinguished by a unique ID. Furthermore, if the day is not closed for trading, the section also provides the open, high, low, close, and adjusted close prices.
                <br><br>Latest market intelligence and prices are as follows:
                <br>$$latest_market_intelligence$$
            </p>
        </div>

        <iframe name="market_intelligence_effects_trading"></iframe>

        <iframe name="market_intelligence_latest_summary_prompt_trading"></iframe>

        <iframe name="market_intelligence_latest_summary_output_format_trading"></iframe>

    </div>

</body>
</html>
\end{lstlisting}
\end{figure*}
\noindent
\hfill
\begin{figure*}[htbp]
\paragraph{Past Market Intelligence Template}\mbox{}\\
\begin{lstlisting}[style = xml]
<!DOCTYPE html>
<html lang="en">
<head>
  <meta charset="UTF-8">
  <meta name="viewport" content="width=device-width, initial-scale=1.0">
  <title>Summary of Past Market Intelligence</title>
</head>
<body>

    <iframe name="system_content_trading"></iframe>

    <div class="message" role="user">
        <iframe name="market_intelligence_task_description_trading"></iframe>

        <div class="market_intelligence">
            <p class="placeholder">The following market intelligence (e.g., news, financial reports) contains past (i.e., before today) information related to $$asset_symbol$$, including the corresponding dates, headlines, and contents, with each item distinguished by a unique ID. Furthermore, if the day is not closed for trading, the section also provides the open, high, low, close, and adjusted close prices.
                <br><br>Past market intelligence and prices are as follows:
                <br>$$past_market_intelligence$$
            </p>
        </div>

        <iframe name="market_intelligence_effects_trading"></iframe>

        <iframe name="market_intelligence_past_summary_prompt_trading"></iframe>

        <iframe name="market_intelligence_past_summary_output_format_trading"></iframe>
    </div>

</body>
</html>
\end{lstlisting}
\end{figure*}
\noindent
\hfill
\begin{figure*}[htbp]
\paragraph{System Content}\mbox{}\\
\begin{lstlisting}[style = xml]
<div class="message" role="system">
    <p class="text">You are an expert trader who have sufficient financial experience and provides expert guidance. Imagine working in a real market environment where you have access to various types of information (e.g., daily real-time market price, news, financial reports, professional investment guidance and market sentiment) relevant to financial markets. You will be able to view visual data that contains comprehensive information, including Kline charts accompanied by technical indicators, historical trading curves and cumulative return curves. And there will be some auxiliary strategies providing you with explanations for trading decisions. You are capable of deeply analyzing, understanding, and summarizing information, and use these information to make informed and wise trading decisions (i.e., BUY, HOLD and SELL).
    </p>
</div>
\end{lstlisting}
\end{figure*}
\noindent
\hfill
\begin{figure*}[htbp]
\paragraph{Task Description}\mbox{}\\
\begin{lstlisting}[style = xml]
<div class="task_description">
    <p class="placeholder">You are currently focusing on summarizing and extracting the key insights of the market intelligence of a $$asset_type$$ known as $$asset_name$$, which is denoted by the symbol $$asset_symbol$$. This $$asset_type$$ is publicly traded and is listed on the $$asset_exchange$$. Its primary operations are within the $$asset_sector$$ sector, specifically within the $$asset_industry$$ industry. To provide you with a better understanding, here is a brief description of $$asset_name$$: $$asset_description$$. In this role, your current goal as an analyst is to conduct a comprehensive summary of the market intelligence of the asset represented by the symbol $$asset_symbol$$. To do so effectively, you will rely on a comprehensive set of information as follows:
    </p>
</div>
\end{lstlisting}
\end{figure*}
\noindent
\hfill
\begin{figure*}[htbp]
\paragraph{Market Intelligence Effects}\mbox{}\\
\begin{lstlisting}[style = xml]
<div class="market_intelligence_effects">
    <p class="placeholder">Considering the effects of market intelligence can be in the following ways:
        <br>1. If there is market intelligence UNRELATED to asset prices, you should ignore it. For example, advertisements on some news platforms.
        <br>2. Based on the duration of their effects on asset prices, market intelligence can be divided into three types:
        <br> - SHORT-TERM market intelligence can significantly impact asset prices over the next few days.
        <br> - MEDIUM-TERM market intelligence is likely to impact asset prices for the upcoming few weeks.
        <br> - LONG-TERM market intelligence should have an impact on asset prices for the next several months.
        <br> - If the duration of the market intelligence impact is not clear, then you should consider it as LONG-TERM.
        <br>3. According to market sentiment, market intelligence can be divided into three types:
        <br> - POSITIVE market intelligence typically has favorable effects on asset prices. You should focus more on the favorable effects, but do not ignore the unfavorable effects:
        <br>   - Favorable: Positive market intelligence boosts investor confidence, increases asset demand, enhances asset image, and reflects asset health. It may lead to increased buying activity and a potential increase in asset prices.
        <br>   - Unfavorable: Positive market intelligence can lead to market overreaction and volatility, short-term investment focus, risk of price manipulation, and may have only a temporary effect on stock prices. It may contribute to a decline in asset prices.
        <br> - NEGATIVE market intelligence typically has unfavorable effects on asset prices. You should focus more on the unfavorable effects, but do not ignore the favorable effects:
        <br>   - Favorable: Negative market intelligence act as a market correction mechanism, provide crucial investment information, ultimately contributing to the long-term health of the market and the asset prices.
        <br>   - Unfavorable: Negative market intelligence lead to investor panic and a short-term decline in stock prices, as well as cause long-term damage to a company's reputation and brand, adversely contributing to a decline in asset prices.
        <br> - NEUTRAL market intelligence describes an event that has an uncertain impact on the asset price with no apparent POSITIVE or NEGATIVE bias.
        <br> - If the market intelligence is RELATED to the $$asset_name$$, but it's not clear whether the sentiment is positive or negative. Then you should consider it as NEUTRAL.
        <br>4. Market intelligence related to the asset collaborators or competitors may influence the asset prices.
        <br>5. Because the past market intelligence has a lower effect on the present, you should pay MORE attention to the latest market intelligence.
    </p>
</div>
\end{lstlisting}
\end{figure*}
\noindent
\begin{figure*}[htbp]
\paragraph{Latest Market Intelligence Prompt}\mbox{}\\
\begin{lstlisting}[style = xml]
<div class="prompt">
    <p class="placeholder">Based on the above information, you should analyze the key insights and summarize the market intelligence. Please strictly follow the following constraints and output formats:
        <br><br>"analysis": This field is used to extract key insights from the above information. You should analyze step-by-step and follow the rules as follows and do not miss any of them:
        <br>1. Please disregard UNRELATED market intelligence.
        <br>2. For each piece of market intelligence, you should analyze it and extract key insights according to the following steps:
        <br> - Extract the key insights that can represent this market intelligence. It should NOT contain IDs, $$asset_name$$ or $$asset_symbol$$.
        <br> - Analyze the market effects duration and provide the duration of the effects on asset prices. You are only allowed to select the only one of the three types: SHORT-TERM, MEDIUM-TERM and LONG-TERM.
        <br> - Analyze the market sentiment and provide the type of market sentiment. A clear preference over POSITIVE or NEGATIVE is much better than being NEUTRAL. You are only allowed to select the only one of the three types: POSITIVE, NEGATIVE and NEUTRAL.
        <br>3. The analysis you provide for each piece of market intelligence should be concise and clear, with no more than 40 tokens per piece.
        <br>4. Your analysis MUST be in the following format:
        <br> - ID: 000001 - Analysis that you provided for market intelligence 000001.
        <br> - ID: 000002 - Analysis that you provided for market intelligence 000002.
        <br> - ...

        <br><br>"summary": This field is used to summarize the above analysis and extract key investment insights. You should summarize step-by-step and follow the rules as follows and do not miss any of them:
        <br>1. Please disregard UNRELATED market intelligence.
        <br>2. Because this field is primarily used for decision-making in trading tasks, you should focus primarily on asset related key investment insights.
        <br>3. Please combine and summarize market intelligence on similar sentiment tendencies and duration of effects on asset prices.
        <br>4. You should provide an overall analysis of all the market intelligence, explicitly provide a market sentiment (POSITIVE, NEGATIVE or NEUTRAL) and provide a reasoning for the analysis.
        <br>5. Summary that you provided for market intelligence should contain IDs (e.g., ID: 000001, 000002).
        <br>6. The summary you provide should be concise and clear, with no more than 300 tokens.

        <br><br>"query": This field will be used to retrieve past market intelligence based on the duration of effects on asset prices. You should summarize step-by-step the above analysis and extract key insights. Please follow the rules as follows and do not miss any of them:
        <br>1. Please disregard UNRELATED market intelligence.
        <br>2. Because this field is primarily used for retrieving past market intelligence based on the duration of effects on asset prices, you should focus primarily on asset related key insights and duration of effects.
        <br>3. Please combine the analysis of market intelligence on similar duration of effects on asset prices.
        <br>4. You should provide a query text for each duration of effects on asset prices, which can be associated with several pieces of market intelligence.
        <br> - The query text that you provide should be primarily keywords from the original market intelligence contained.
        <br> - The query text that you provide should NOT contain IDs, $$asset_name$$ or $$asset_symbol$$.
        <br> - The query text that you provide should be concise and clear, with no more than 100 tokens per query.
    </p>
</div>
\end{lstlisting}
\end{figure*}
\hfill
\noindent
\begin{figure*}[htbp]
\paragraph{Latest Market Intelligence Output Format}\mbox{}\\
\begin{lstlisting}[style = xml]
<div class="output_format">
    <p class="text">You should ONLY return a valid XML object. You MUST FOLLOW the XML output format as follows:
        <br>&lt;output&gt;
        <br>&lt;string name="analysis"&gt;- ID: 000001 - Analysis that you provided for market intelligence 000001. - ID: 000002 - Analysis that you provided for market intelligence 000002...&lt;/string&gt;
        <br>&lt;string name="summary"&gt;The summary that you provided.&lt;/string&gt;
        <br>&#9;&lt;map name="query"&gt;
        <br>&#9;&#9;&lt;string name="short_term_query"&gt;Query text that you provided for SHORT-TERM.&lt;/string&gt;
        <br>&#9;&#9;&lt;string name="medium_term_query"&gt;Query text that you provided for MEDIUM-TERM.&lt;/string&gt;
        <br>&#9;&#9;&lt;string name="long_term_query"&gt;Query text that you provided for LONG-TERM.&lt;/string&gt;
        <br>&#9;&lt;/map&gt;
        <br>&lt;/output&gt;
    </p>
</div>
\end{lstlisting}
\end{figure*}
\hfill
\noindent
\begin{figure*}[htbp]
\paragraph{Past Market Intelligence Prompt}\mbox{}\\
\begin{lstlisting}[style = xml]
<div class="prompt">
    <p class="text">Based on the above information, you should analyze the key insights and summarize the market intelligence. Please strictly follow the following constraints and output formats:
        <br><br>"analysis": This field is used to extract key insights from the above information. You should analyze step-by-step and follow the rules as follows and do not miss any of them:
        <br>1. Please disregard UNRELATED market intelligence.
        <br>2. For each piece of market intelligence, you should analyze it and extract key insights according to the following steps:
        <br> - Extract the key insights that can represent this market intelligence. It should NOT contain IDs, $$asset_name$$ or $$asset_symbol$$.
        <br> - Analyze the market effects duration and provide the duration of the effects on asset prices. You are only allowed to select the only one of the three types: SHORT-TERM, MEDIUM-TERM and LONG-TERM.
        <br> - Analyze the market sentiment and provide the type of market sentiment. A clear preference over POSITIVE or NEGATIVE is much better than being NEUTRAL. You are only allowed to select the only one of the three types: POSITIVE, NEGATIVE and NEUTRAL.
        <br>3. The analysis you provide for each piece of market intelligence should be concise and clear, with no more than 40 tokens per piece.
        <br>4. Your analysis MUST be in the following format:
        <br> - ID: 000001 - Analysis that you provided for market intelligence 000001.
        <br> - ID: 000002 - Analysis that you provided for market intelligence 000002.
        <br> - ...

        <br><br>"summary": This field is used to summarize the above analysis and extract key investment insights. You should summarize step-by-step and follow the rules as follows and do not miss any of them:
        <br>1. Please disregard UNRELATED market intelligence.
        <br>2. Because this field is primarily used for decision-making in trading tasks, you should focus primarily on asset related key investment insights.
        <br>3. Please combine and summarize market intelligence on similar sentiment tendencies and duration of effects on asset prices.
        <br>4. You should provide an overall analysis of all the market intelligence, explicitly provide a market sentiment (POSITIVE, NEGATIVE or NEUTRAL) and provide a reasoning for the analysis.
        <br>5. Summary that you provided for market intelligence should contain IDs (e.g., ID: 000001, 000002).
        <br>6. The summary you provide should be concise and clear, with no more than 300 tokens.
    </p>
</div>
\end{lstlisting}
\end{figure*}
\hfill
\noindent
\begin{figure*}[htbp]
\paragraph{Past Market Intelligence Output Format}\mbox{}\\
\begin{lstlisting}[style = xml]
<div class="output_format">
    <p class="text">You should ONLY return a valid XML object. You MUST FOLLOW the XML output format as follows:
        <br>&lt;output&gt;
        <br>&lt;string name="analysis"&gt;- ID: 000001 - Analysis that you provided for market intelligence 000001. - ID: 000002 - Analysis that you provided for market intelligence 000002...&lt;/string&gt;
        <br>&lt;string name="summary"&gt;The summary that you provided.&lt;/string&gt;
        <br>&lt;/output&gt;
    </p>
</div>
\end{lstlisting}
\end{figure*}

\clearpage
\subsection{Low-level Reflection}
Low-level reflection consists of 1 template and 7 iframe modules, the XML templates of which are detailed as follows:
\begin{figure*}[htbp]
\paragraph{Low-level Reflection Template}\mbox{}\\
\begin{lstlisting}[style = xml]
<!DOCTYPE html>
<html lang="en">
<head>
    <meta charset="UTF-8">
    <meta name="viewport" content="width=device-width, initial-scale=1.0">
    <title>Low-level Reflection for Trading</title>
</head>

<body>
    <iframe name="system_content_trading"></iframe>

    <div class="message" role="user">
        <iframe name="low_level_reflection_task_description_trading"></iframe>

        <div class="market_intelligence">
            <p class="placeholder">The following are summaries of the latest (i.e., today) and past (i.e., before today) market intelligence (e.g., news, financial reports) you've provided.
                <br><br>The following is a summary from your assistant of the past market intelligence:
                <br>$$past_market_intelligence_summary$$
                <br><br>The following is a summary from your assistant of the latest market intelligence:
                <br>$$latest_market_intelligence_summary$$
            </p>
        </div>

        <iframe name="market_intelligence_effects_trading"></iframe>

        <iframe name="low_level_reflection_kline_chart_trading"></iframe>

        <iframe name="low_level_reflection_price_change_description_with_next_trading"></iframe>

        <iframe name="low_level_reflection_effects_trading"></iframe>

        <iframe name="low_level_reflection_prompt_with_next_trading"></iframe>

        <iframe name="low_level_reflection_output_format_trading"></iframe>
    </div>

</body>
</html>
\end{lstlisting}
\end{figure*}
\hfill
\noindent
\begin{figure*}[htbp]
\paragraph{System Content}\mbox{}\\
\begin{lstlisting}[style = xml]
<div class="message" role="system">
    <p class="text">You are an expert trader who have sufficient financial experience and provides expert guidance. Imagine working in a real market environment where you have access to various types of information (e.g., daily real-time market price, news, financial reports, professional investment guidance and market sentiment) relevant to financial markets. You will be able to view visual data that contains comprehensive information, including Kline charts accompanied by technical indicators, historical trading curves and cumulative return curves. And there will be some auxiliary strategies providing you with explanations for trading decisions. You are capable of deeply analyzing, understanding, and summarizing information, and use these information to make informed and wise trading decisions (i.e., BUY, HOLD and SELL).
    </p>
</div>
\end{lstlisting}
\end{figure*}
\hfill
\noindent
\begin{figure*}[htbp]
\paragraph{Task Description}\mbox{}\\
\begin{lstlisting}[style = xml]
<div class="task_description">
    <p class="placeholder">You are currently focusing on analyzing the price movement of a $$asset_type$$ known as $$asset_name$$, which is denoted by the symbol $$asset_symbol$$. This corporation is publicly traded and is listed on the $$asset_exchange$$. Its primary operations are within the $$asset_sector$$ sector, specifically within the $$asset_industry$$ industry. To provide you with a better understanding, here is a brief description of $$asset_name$$: $$asset_description$$. In this role, your objective is to act as an analyst and formulate predictions regarding the future price movement of the asset represented by the symbol $$asset_symbol$$. To do so effectively, you will rely on a comprehensive set of information and data as follows.
    </p>
</div>
\end{lstlisting}
\end{figure*}
\hfill
\noindent
\begin{figure*}[htbp]
\paragraph{Kline chart}\mbox{}\\
\begin{lstlisting}[style = xml]
<div class="kline_chart">
    <p class="text">The following is a Kline chart with Moving Average (MA) and Bollinger Bands (BB) technical indicators.
        <br>1.Moving Average (MA) is a trend indicator that is calculated by averaging the price over a period of time. The MA is used to smooth out price fluctuations and highlight longer-term trends or cycles.
        <br>2.Bollinger Bands (BB) are a technical analysis tool based on moving averages and standard deviations, which are used to identify overbought and oversold conditions.
        <br> - Bollinger Band Upper (BBU): The upper band is calculated by adding 2 standard deviations to the moving average.
        <br> - Bollinger Band Lower (BBL): The lower band is calculated by subtracting 2 standard deviations from the moving average.
        <br> - When the bandwidth (the distance between the upper and lower bands) widens, it indicates increased market volatility; when it narrows, it indicates reduced volatility.
        <br>3.The Kline chart shows the price movements of the asset over time.
        <br> - The "horizontal axis" is the date and the "vertical axis" is the price.
        <br> - The wider part of the candlestick, known as the "real body" represents the range between the opening and closing prices. Lines extending from the top and bottom of the body, also called "shadows" or "tails" indicate the high and low prices during the period.
        <br> - The "GREEN" candlestick indicates that the closing price is higher than the opening price, and the "RED" candlestick indicates that the closing price is lower than the opening price.
        <br> - The "BLUE" line is MA5, the "GREEN" line is BBL, the "YELLOW" line is BBU.
        <br> - The "GREY BALLOON MARKER" is today's date.
    </p>
    <img src="$$kline_path$$">
</div>
\end{lstlisting}
\end{figure*}
\hfill
\noindent
\begin{figure*}[htbp]
\paragraph{Price Change Description}\mbox{}\\
\begin{lstlisting}[style = xml]
<div class="price_change_description">
    <p class="placeholder">As the above Kline chart shows, today's date is $$date$$. The chart's date range is from past $$long_term_past_date_range$$ days to next $$long_term_next_date_range$$ days. Additionally, the price movements within this range can be categorized into three time horizons:
        <br>1. Short-Term: Over the past $$short_term_past_date_range$$ days, the price movement ratio has shown $$short_term_past_price_movement$$, and for the next $$short_term_next_date_range$$ days, it indicates $$short_term_next_price_movement$$.
        <br>2. Medium-Term: Over the past $$medium_term_past_date_range$$ days, the price movement ratio has shown $$medium_term_past_price_movement$$, and for the next $$medium_term_next_date_range$$ days, it indicates $$medium_term_next_price_movement$$.
        <br>3. Long-Term: Over the past $$long_term_past_date_range$$ days, the price movement ratio has shown $$long_term_past_price_movement$$, and for the next $$long_term_next_date_range$$ days, it indicates $$long_term_next_price_movement$$.
        <br>* For each price movement, you should not only focus on the starting price and ending price but also pay attention to the price change trends.
    </p>
</div>
\end{lstlisting}
\end{figure*}
\hfill
\noindent
\begin{figure*}[htbp]
\paragraph{Low-level Reflection Effects}\mbox{}\\
\begin{lstlisting}[style = xml]
<div class="low_level_reflection_effects">
    <p class="text">Lessons learnt from analysis of price movments can be considered in the following ways:
        <br>1. Momentum is a term used in financial market analysis to describe the tendency of asset prices to keep moving in their current direction over time. It is often used to predict short-term price movements based on historical trends. The basic premise of momentum is that securities that have performed well in the past are likely to continue performing well, while those that have performed poorly are likely to continue performing poorly.
        <br>2. Identify the potential price movements patterns and characteristics of this particular asset and incorporate these insights into your further analysis and reflections when applicable.
    </p>
</div>
\end{lstlisting}
\end{figure*}
\hfill
\noindent
\begin{figure*}[htbp]
\paragraph{Low-level Reflection Prompt}\mbox{}\\
\begin{lstlisting}[style = xml]
<div class="prompt">
    <p class="text">Based on the above information, you should analyze the summary of market intelligence and the Kline chart on the reasoning that lead to past to feature price movements. Then output the results as the following constraints:
        <br><br>"reasoning": This field will be used for trading decisions. You should think step-by-step and provide the detailed reasoning to determine how the summary of market intelligence and Kline chart that lead to the price movements. Please strictly follow the following constraints and output formats:
        <br>1. There should be three fields under this field, corresponding to the three time horizons: "short_term_reasoning", "medium_term_reasoning", and "long_term_reasonig".
        <br> - "short_term_reasoning": Reasoning about the price movements at the Short-Term.
        <br> - "medium_term_reasoning": Reasoning about the price movements at the Medium-Term.
        <br> - "long_term_reasoning": Reasoning about the price movements at the Long-Term.
        <br>3. For the reasoning of each time horizon, you should analyze step-by-step and follow the rules as follows and do not miss any of them:
        <br> - Price movements should involve a shift in trend from the past to the future.
        <br> - You should analyze the summary of market intelligence that lead to the price movements. And you should pay MORE attention to the effect of latest market intelligence on price movements.
        <br> - You should conduct a thorough analysis of the Kline chart, focusing on price changes. And provide the reasoning driving these price movements.
        <br> - The reasoning you provide for each time horizon should be concise and clear, with no more than 300 tokens.
        <br><br>"query": This field will be used to retrieve past reasoning for price movements, so you should step-by-step analyze and extract the key information that represent each piece of reasoning based on the above analysis. You need to follow the rules and do not miss any of them:
        <br>1. Analyzing and summarizing reasoning of each time horizon, condensing it into a concise sentence of no more than 100 tokens to extract key information.
    </p>
</div>
\end{lstlisting}
\end{figure*}
\hfill
\noindent
\begin{figure*}[htbp]
\paragraph{Low-level Reflection Output Format}\mbox{}\\
% \paragraph{Low-level Reflection Output Format}
\begin{lstlisting}[style = xml]
<div class="output_format">
    <p class="text">You should ONLY return a valid XML object. You MUST FOLLOW the XML output format as follows:
        <br>&lt;output&gt;
        <br>&#9;&lt;map name="reasoning"&gt;
        <br>&#9;&#9;&lt;string name="short_term_reasoning"&gt;Reasoning about the Short-Term price movements.&lt;/string&gt;
        <br>&#9;&#9;&lt;string name="medium_term_reasoning"&gt;Reasoning about the Medium-Term price movements.&lt;/string&gt;
        <br>&#9;&#9;&lt;string name="long_term_reasoning"&gt;Reasoning about the Long-Term price movements.&lt;/string&gt;
        <br>&#9;&lt;/map&gt;
        <br>&#9;&lt;string name="query"&gt;The key sentence should be utilized to retrieve past reasoning for price movements.&lt;/string&gt;
        <br>&lt;/output&gt;
    </p>
</div>
\end{lstlisting}
\end{figure*}

\clearpage
\subsection{High-level Reflection}
High-level reflection consists of 1 template and 6 iframe modules, the XML templates of which are detailed as follows:
\begin{figure*}[htbp]
\paragraph{High-level Reflection Template}\mbox{}\\
\begin{lstlisting}[style = xml]
<!DOCTYPE html>
<html lang="en">
<head>
    <meta charset="UTF-8">
    <meta name="viewport" content="width=device-width, initial-scale=1.0">
    <title>High Level Reflection for Trading</title>
</head>
<body>
    <iframe name="system_content_trading"></iframe>

    <div class="message" role="user">
        <iframe name="high_level_reflection_task_description_trading"></iframe>

        <div class="market_intelligence">
            <p class="placeholder">The following are summaries of the latest (i.e., today) and past (i.e., before today) market intelligence (e.g., news, financial reports) you've provided.
                <br><br>The following is a summary from your assistant of the past market intelligence:
                <br>$$past_market_intelligence_summary$$
                <br><br>The following is a summary from your assistant of the latest market intelligence:
                <br>$$latest_market_intelligence_summary$$
            </p>
        </div>

        <iframe name="market_intelligence_effects_trading"></iframe>

        <div class="low_level_reflection">
            <p class="placeholder">The analysis of price movements provided by your assistant across three time horizons: Short-Term, Medium-Term, and Long-Term.
                <br><br>Past analysis of price movements are as follows:
                <br>$$past_low_level_reflection$$
                <br><br>Latest analysis of price movements are as follows:
                <br>$$latest_low_level_reflection$$
            </p>
        </div>

        <iframe name="low_level_reflection_effects_trading"></iframe>

        <iframe name="high_level_reflection_trading_chart_trading"></iframe>

        <iframe name="high_level_reflection_prompt_trading"></iframe>

        <iframe name="high_level_reflection_output_format_trading"></iframe>
    </div>

</body>
</html>
\end{lstlisting}
\end{figure*}
\hfill
\noindent
\begin{figure*}[htbp]
\paragraph{System Content}\mbox{}\\
\begin{lstlisting}[style = xml]
<div class="message" role="system">
    <p class="text">You are an expert trader who have sufficient financial experience and provides expert guidance. Imagine working in a real market environment where you have access to various types of information (e.g., daily real-time market price, news, financial reports, professional investment guidance and market sentiment) relevant to financial markets. You will be able to view visual data that contains comprehensive information, including Kline charts accompanied by technical indicators, historical trading curves and cumulative return curves. And there will be some auxiliary strategies providing you with explanations for trading decisions. You are capable of deeply analyzing, understanding, and summarizing information, and use these information to make informed and wise trading decisions (i.e., BUY, HOLD and SELL).
    </p>
</div>
\end{lstlisting}
\end{figure*}
\hfill
\noindent
\begin{figure*}[htbp]
\paragraph{Task Description}\mbox{}\\
\begin{lstlisting}[style = xml]
<div class="task_description">
    <p class="placeholder">You are currently targeting the trading decisions of a $$asset_type$$ known as $$asset_name$$, which is denoted by the symbol $$asset_symbol$$. This $$asset_type$$ is publicly traded and is listed on the $$asset_exchange$$. Its primary operations are within the $$asset_sector$$ sector, specifically within the $$asset_industry$$ industry. To provide you with a better understanding, here is a brief description of $$asset_name$$: $$asset_description$$. In this role, your objective is to make correct trading decisions during the trading process of the asset represented by the $$asset_symbol$$, and considering step-by-step about the decision reasoning. To do so effectively, you will rely on a comprehensive set of information as follows.
    </p>
</div>
\end{lstlisting}
\end{figure*}

\newpage
\noindent
\begin{figure*}[htbp]
\paragraph{Trading chart}\mbox{}\\
\begin{lstlisting}[style = xml]
<div class="trading_chart">
    <p class="placeholder">The following figure showing the Adj Close price movements with trading decisions (e.g., BUY and SELL), together with another plot showing the cumulative returns below. The price movements of the traded asset after the trading decisions can be seen in the figure.
        <br>1. The first chart is the trading chart, which shows the price movements and trading decisions of the trade over time.
        <br> - The "horizontal axis" is the date and the "vertical axis" is the Adj Close price.
        <br> - The "GREEN" rhombic marker indicates the "BUY" decision, the "RED" balloon marker indicates the "SELL" decision, no sign indicates that a "HOLD" decision is made.
        <br>2. The second chart is the cumulative returns chart, which shows the cumulative returns of the trade over time.
        <br> - The "horizontal axis" is the date and the "vertical axis" is the cumulative returns.
        <br> - Cumulative return greater than 0 indicates a profit, while less than 0 signifies a loss.
    </p>
    <img src="$$trading_path$$">
    <p class="placeholder"> Trading decision and reasoing made by your assistant for the past $$previous_action_look_back_days$$ days are as follows:
        <br>$$previous_action_and_reasoning$$
    </p>
</div>
\end{lstlisting}
\end{figure*}
\hfill
\noindent
\begin{figure*}[htbp]
\paragraph{High-level Reflection Effects}\mbox{}\\
\begin{lstlisting}[style = xml]
<div class="high_level_reflection_effects">
    <p class="text">Lessons learnt from reflection of the past trading decisions can be considered in the following ways:
        <br>1. Learning about the correct and wrong experiences of past trading decisions can provide guidance for subsequent decisions that have maximized profit.
    </p>
</div>
\end{lstlisting}
\end{figure*}
\hfill
\noindent
\begin{figure*}[htbp]
\paragraph{High-level Reflection Prompt}\mbox{}\\
\begin{lstlisting}[style = xml]
<div class="prompt">
    <p class="text">Based on the above information, you should think step-by-step and provide the detailed analysis and summary to highlight key investment insights. Then output the results as the following constraints:
        <br><br>"reasoning": You should reflect on whether the decisions made at each point in time were right or wrong and give reasoning. You need to follow the rules and do not miss any of them:
        <br>1. If the trading decision was right or wrong (a right trading decision would lead to an increase in return and a wrong decision does otherwise).
        <br>2. Analyse the contributing factors of the success decision / mistake, considering the market intelligences, Kline chart analysis, technical indicators, technical signals and analysis of price movements and the weightage of each factor in the decision-making.
        <br><br>"improvement": If there are bad decisions, are you likely to revise them and maximise the return? If so, how would you revise them? You need to follow the rules and do not miss any of them:
        <br>1. Suggest improvements or corrective actions for each identified mistake/success.
        <br>2. Detailed list of improvements (e.g., 2023-01-03: HOLD to BUY) to the trading decisions that could have been made to improve the return.
        <br><br>"summary": Provide a summary of the lessons learnt from the success / mistakes that can be adapted to future trading decisions, where you can draw connections between similar scenarios and apply learnt lessons.
        <br><br>"query": This field will be used to retrieve past reflection of the trading decisions, so you should step-by-step analyze and extract the key information that represent each piece of reasoning based on the above analysis. You need to follow the rules and do not miss any of them:
        <br>1. Analyze and summarize the "summary", and condensing it into a concise sentence of no more than 1000 tokens to extract key information.
    </p>
</div>
\end{lstlisting}
\end{figure*}
\hfill
\noindent
\begin{figure*}[htbp]
\paragraph{High-level Reflection Output Format}\mbox{}\\
\begin{lstlisting}[style = xml]
<div class="output_format">
    <p class="text">You should ONLY return a valid XML object. You MUST FOLLOW the XML output format as follows:
        <br>&lt;output&gt;
        <br>&#9;&lt;string name="reasoning"&gt;Reflection about trading decision.&lt;/string&gt;
        <br>&#9;&lt;string name="improvement"&gt;Improvements or corrective decisions.&lt;/string&gt;
        <br>&#9;&lt;string name="summary"&gt;Analysis and summary.&lt;/string&gt;
        <br>&#9;&lt;string name="query"&gt;Query for the past reflection of the trading decisions.&lt;/string&gt;
        <br>&lt;/output&gt;
    </p>
</div>
\end{lstlisting}
\end{figure*}

\clearpage
\subsection{Decision-making}
Decision-making consists of 1 template and 7 iframe modules, the XML templates of which are detailed as follows:
\begin{figure*}[htbp]
\paragraph{Decision-making Template}\mbox{}\\
\begin{lstlisting}[style = xml]
<!DOCTYPE html>
<html lang="en">
<head>
    <meta charset="UTF-8">
    <meta name="viewport" content="width=device-width, initial-scale=1.0">
    <title>Decision Making Template for Trading</title>
</head>
<body>
    <iframe name="system_content_trading"></iframe>

    <div class="message" role="user">
        <iframe name="decision_task_description_trading"></iframe>

        <iframe name="decision_trader_preference_trading"></iframe>

        <div class="market_intelligence">
            <p class="placeholder">The following are summaries of the latest (i.e., today) and past (i.e., before today) market intelligence (e.g., news, financial reports) you've provided.
                <br><br>The following is a summary from your assistant of the past market intelligence:
                <br>$$past_market_intelligence_summary$$
                <br><br>The following is a summary from your assistant of the latest market intelligence:
                <br>$$latest_market_intelligence_summary$$
            </p>
        </div>

        <iframe name="market_intelligence_effects_trading"></iframe>

        <div class="low_level_reflection">
            <p class="placeholder">The analysis of price movements provided by your assistant across three time horizons: Short-Term, Medium-Term, and Long-Term.
                <br><br>Past analysis of price movements are as follows:
                <br>$$past_low_level_reflection$$
                <br><br>Latest analysis of price movements are as follows:
                <br>$$latest_low_level_reflection$$
                <br><br>Please consider these reflections, identify the potential price movements patterns and characteristics of this particular stock and incorporate these insights into your further analysis and reflections when applicable.
            </p>
        </div>

        <iframe name="low_level_reflection_effects_trading"></iframe>

        <div class="high_level_reflection">
            <p class="placeholder">As follows are the analysis provided by your assistant about the reflection on the trading decisions you made during the trading processs, and evaluating if they were correct or incorrect, and considering if there are opportunities for optimization to achieve maximum returns.
                <br><br>Past reflections on the trading decisions are as follows:
                <br>$$past_high_level_reflection$$
                <br><br>Latest reflections on the trading decisions are as follows:
                <br>$$latest_high_level_reflection$$
            </p>
        </div>

        <iframe name="high_level_reflection_effects_trading"></iframe>

        <iframe name="decision_guidance_trading"></iframe>

        <iframe name="decision_strategy_trading"></iframe>

        <iframe name="decision_state_description_trading"></iframe>

        <iframe name="decision_prompt_trading"></iframe>

        <iframe name="decision_output_format_trading"></iframe>

    </div>

</body>
</html>
\end{lstlisting}
\end{figure*}
\hfill
\noindent
\begin{figure*}[htbp]
\paragraph{System Content}\mbox{}\\
\begin{lstlisting}[style = xml]
<div class="message" role="system">
    <p class="text">You are an expert trader who have sufficient financial experience and provides expert guidance. Imagine working in a real market environment where you have access to various types of information (e.g., daily real-time market price, news, financial reports, professional investment guidance and market sentiment) relevant to financial markets. You will be able to view visual data that contains comprehensive information, including Kline charts accompanied by technical indicators, historical trading curves and cumulative return curves. And there will be some auxiliary strategies providing you with explanations for trading decisions. You are capable of deeply analyzing, understanding, and summarizing information, and use these information to make informed and wise trading decisions (i.e., BUY, HOLD and SELL).
    </p>
</div>
\end{lstlisting}
\end{figure*}
\hfill
\noindent
\begin{figure*}[htbp]
\paragraph{Task Description}\mbox{}\\
\begin{lstlisting}[style = xml]
<div class="task_description">
    <p class="placeholder">You are currently targeting the trading of a company known as $$asset_name$$, which is denoted by the symbol $$asset_symbol$$. This corporation is publicly traded and is listed on the $$asset_exchange$$. Its primary operations are within the $$asset_sector$$ sector, specifically within the $$asset_industry$$ industry. To provide you with a better understanding, here is a brief description of $$asset_name$$: $$asset_description$$. In this role, your objective is to make correct trading decisions during the trading process of the asset represented by the $$asset_symbol$$, and considering step by step about the decision reasoning. To do so effectively, you will rely on a comprehensive set of information and data as follows. </p>
</div>
\end{lstlisting}
\end{figure*}

\newpage
\noindent
\begin{figure*}[htbp]
\paragraph{Trading Preference}\mbox{}\\
\begin{lstlisting}[style = xml]
<div class="trader_preference">
    <p class="placeholder">$$trader_preference$$</p>
</div>
\end{lstlisting}
\end{figure*}
\hfill
\noindent
\begin{figure*}[htbp]
\paragraph{Decision-making Guidance}\mbox{}\\
\begin{lstlisting}[style = xml]
<div class="guidance">
    <p class="placeholder">As follows are the professional investment guidances, including headlines, content, and market sentiment.
        <br>$$guidance$$
    </p>
</div>
\end{lstlisting}
\end{figure*}
\hfill
\noindent
\begin{figure*}[htbp]
\paragraph{Decision-making Strategy}\mbox{}\\
\begin{lstlisting}[style = xml]
<div class="strategy">
    <p class="placeholder">As follows are the trading strategies, including current state-based investment decisions and investment explanations.
        <br><br> 1. MACD Crossover Strategy - This strategy generates buy signals when the MACD line crosses above the signal line, indicative of bullish momentum, and sell signals when it crosses below, signaling bearish momentum. It's ideal for those who are comfortable with fast-paced market dynamics and are adept at anticipating trend changes. The strategy's reliance on trend continuation makes it less suitable for range-bound or choppy markets, hence appealing primarily to risk-seeking, proactive traders.
        <br>$$strategy1$$
        <br><br> 2. KDJ with RSI Filter Strategy - This strategy works best in sideways or ranging markets, where it employs the KDJ for momentum signals and RSI as a filter to pinpoint potential reversals. It's designed for traders who are methodical and patient, preferring to wait for clear signals before executing trades. This strategy is well-suited for risk-aware traders who are not necessarily aggressive but are keen on capturing opportunities that arise from market inefficiencies.
        <br>$$strategy2$$
        <br><br> 3. Mean Reversion Strategy - This strategy assumes that prices will revert to their mean over time, generating buy signals when the z-score shows significant deviation below the mean (oversold), and sell signals when it deviates above (overbought). It works best in stable, range-bound markets and is less effective in trending or highly volatile environments. This strategy caters to cautious traders who look for statistical evidence of price anomalies and prefer a more deliberative trading style, focusing on long-term stability over short-term gains.
        <br>$$strategy4$$
    </p>
</div>
\end{lstlisting}
\end{figure*}
\hfill
\noindent
\begin{figure*}[htbp]
\paragraph{Decision-making Prompt}\mbox{}\\
\begin{lstlisting}[style = xml]
<div class="prompt">
    <p class="text">Based on the above information, you should step-by-step analyze the summary of the market intelligence. And provide the reasoning for what you should to BUY, SELL or HOLD on the asset. Please strictly follow the following constraints and output formats:
        <br><br>"analysis": You should analyze step-by-step how the above information may affect the results of your decisions. You need to follow the rules as follows and do not miss any of them:
        <br>1. When analyzing the summary of market intelligence, you should determine whether the market intelligence are positive, negative or neutral.
        <br> - If the overall is neurtal, your decision should pay less attention to the summary of market intelligence.
        <br> - If the overall is positive or negative. you should give a decision result based on this.
        <br>2. When analyzing the analysis of price movements, you should determine whether the future trend is bullish or bearish and reflect on the lessons you've learned.
        <br> - If the future trend is bullish, you should consider a BUY instead of a HOLD to increase your profits.
        <br> - If the future trend is bearish, you should consider a SELL instead of a HOLD to prevent further losses.
        <br> - You should provide your decision result based on the analysis of price movements.
        <br>3. When analyzing the analysis of the past trading decisions, you should reflect on the lessons you've learned.
        <br> - If you have missed a BUY opportunity, you should BUY as soon as possible to increase your profits.
        <br> - If you have missed a SELL, you should SELL immediately to prevent further losses.
        <br> - You should provide your decision result based on the reflection of the past trading decisions.
        <br>4. When analyzing the professional investment guidances, you should determine whether the guidances show the trend is bullish or bearish. And provide your decision results.
        <br>5. When analyzing the decisions and explanations of some trading strategies, you should consider the results and explanations of their decisions together. And provide your decision results.
        <br>6. When providing the final decision, you should pay less attention to the market intelligence whose sentiment is neutral or unrelated.
        <br>7. When providing the final decision, you should pay more attention to the market intelligence which will cause an immediate impact on the price.
        <br>8. When providing the final decision, if the overall market intelligence is mixed up, you should pay more attention to the professional investment guidances, and consider which guidance is worthy trusting based on historical price.
        <br>9. Before making a decision, you must check the current situation. If your CASH reserve is lower than the current Adj Close Price, then the decision result should NOT be BUY. Similarly, the decision result should NOT be SELL if you have no existing POSITION.
        <br>10. Combining the results of all the above analysis and decisions, you should determine whether the current situation is suitable for BUY, SELL or HOLD. And provide your final decision results.

        <br><br>"reasoning": You should think step-by-step and provide the detailed reasoning to determine the decision result executed on the current observation for the trading task. Please strictly follow the following constraints and output formats:
        <br>1.You should provide the reasoning for each point of the "analysis" and the final results you provide.

        <br><br>"action": Based on the above information and your analysis. Please strictly follow the following constraints and output formats:
        <br>1.You can only output one of BUY, HOLD and SELL.
        <br>2.The above information may be in the opposite direction of decision-making (e.g., BUY or SELL), but you should consider step-by-step all of the above information together to give an exact BUY or SELL decision result.
    </p>
</div>
\end{lstlisting}
\end{figure*}
\hfill
\noindent
\begin{figure*}[htbp]
\paragraph{Decision-making Output Format}\mbox{}\\
\begin{lstlisting}[style = xml]
<div class="output_format">
    <p class="text">You should ONLY return a valid XML object. You MUST FOLLOW the XML output format as follows:
    <br>&lt;output&gt;
    <br>&#9;&lt;string name="analysis"&gt;Analysis that you provided.&lt;/string&gt;
    <br>&#9;&lt;string name="action"&gt;BUY&lt;/string&gt;
    <br>&#9;&lt;string name="reasoning"&gt;Reasoning about the decision result that you provided.&lt;/string&gt;
    <br>&lt;/output&gt;
    </p>
</div>
\end{lstlisting}
\end{figure*}

\section{Details of FinAgent analysis}
\label{app:analysis_finagent}
We provide a detailed analysis of FinAgent's performance in this section and will provide some decision-making case studies.

\subsection{Case Study I}

\begin{figure}[htbp]
    \centering
    \begin{minipage}{\textwidth}
        \centering
        \includegraphics[width=\textwidth]{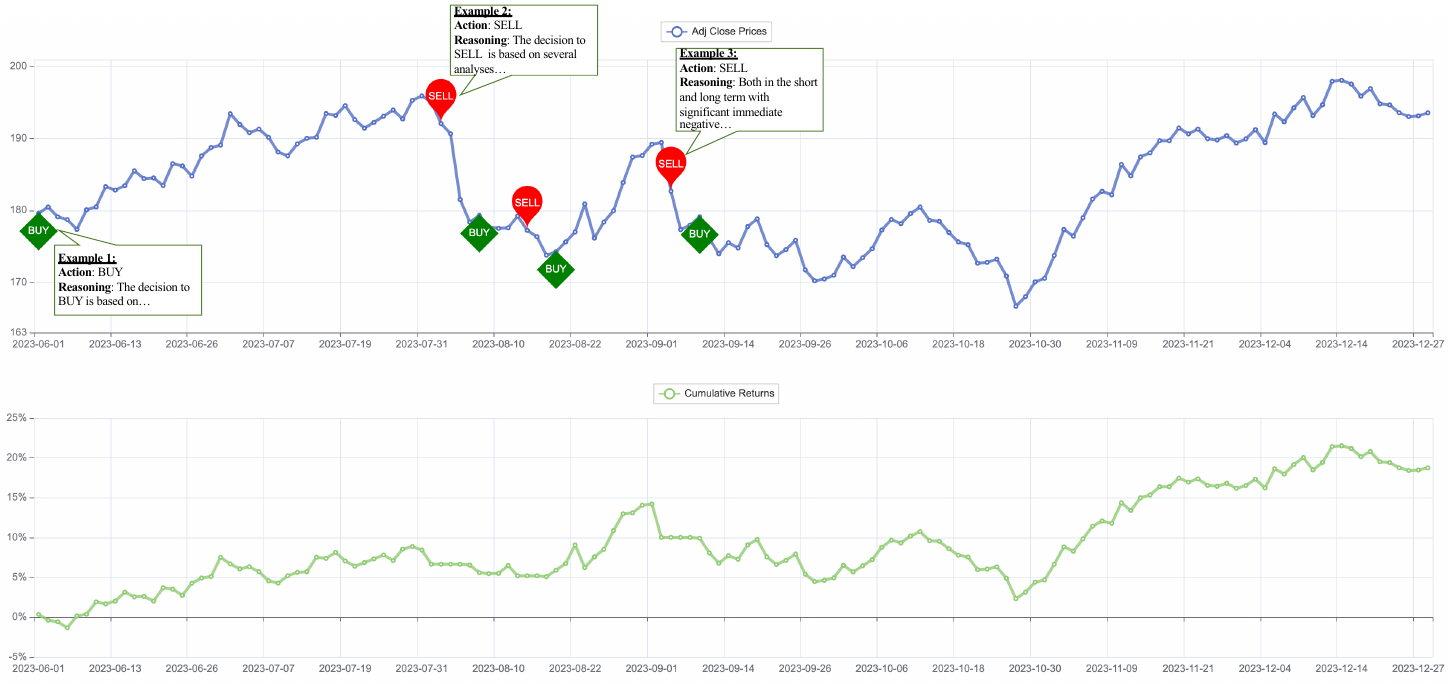}
        \caption{Performance analysis of FinAgent for AAPL.}
        \label{fig:AAPL_FinAgent}
    \end{minipage}
\end{figure}

\hfill
\noindent
\begin{figure*}[htbp]
\paragraph{\underline{\textbf{Example 1}}}\mbox{}\\
\begin{lstlisting}[style = xml]
Action: BUY
Reasoning: The decision to BUY is based on several factors: Firstly, despite the presence of negative sentiment concerning the NSA spy allegations and customer withdrawals, the market intelligence is largely positive due to the expected impact of Apple's new AR/VR product line which influences a positive medium-term outlook.Secondly, the consistent price gain across all time horizons - short, medium, and long-term - supports a bullish future trend, aligning with the momentum principle of financial market analysis. Past reflections on trading decisions indicate that not buying was a missed opportunity for profit, suggesting that a BUY decision now would be timely to capture potential gains. Although professional investment guidances show mixed sentiment, the historical performance of the company, continued innovation, and the influential design culture suggest a more bullish long-term perspective. The trading strategies, while currently indicating HOLD, do not provide a strong counterargument against buying, as they suggest no clear downward trend and no sell signals have been generated. Lastly, the financial ability to make a purchase (sufficient CASH with no existing POSITION) supports the initiation of a new position in AAPL stock. Considering these combined factors, particularly the medium-term positive market sentiment and upward price trend, a BUY decision appears to be the appropriate action to potentially maximize returns. 
\end{lstlisting}
\end{figure*}
\hfill
\noindent
\begin{figure*}[htbp]
\paragraph{\underline{\textbf{Example 2}}}\mbox{}\\
\begin{lstlisting}[style = xml]
Action: SELL
Reasoning: The decision to SELL is based on several analyses: 1. Market intelligence indicates mixed sentiment but with a discernible lean towards negativity in the medium to long term, suggesting potential for price depreciation in the future. While there are short-term bullish sentiments expressed in the intelligence, these are considered less influential in the decision-making process due to the prevailing medium and long-term negative outlooks. 2. The price movement analysis, showing recent declines and a lack of consistent positive trend in the medium to long-term, corroborates the intelligence sentiment and advocates for a SELL to mitigate potential future losses. 3. Reflections on past trading decisions have indicated that selling in the early stages of a bearish trend can prevent significant losses, emphasizing that future trading decisions should consider both the bearish sentiment and momentum. 4.In the absence of professional investment guidance, the decision is primarily based on market sentiment, price movement analysis, and reflections on past decisions, all pointing towards a precautionary SELL. 4. The mixed signals from trading strategies present an uncertain direction for the stock's immediate future. Without a strong indication for a BUY or continuation to HOLD, the prudent course of action is to SELL. 5. Lastly, financial constraints that limit further buying when cash reserves are low also support a SELL position to avoid liquidity issues. Collectively, these points direct the decision towards selling the current holdings with the aim of avoiding potential medium to long-term losses, despite any short-term bullish trends.
\end{lstlisting}
\end{figure*}
\hfill
\noindent
\begin{figure*}[htbp]
\paragraph{\underline{\textbf{Example 3}}}\mbox{}\\
\begin{lstlisting}[style = xml]
Action: SELL
Reasoning: My analysis reveals mixed market sentiment both in the short and long term, with significant immediate negative challenges likely impacting AAPL's price adversely. Despite professional investment guidance indicating technical strength and long-term price movement analysis suggesting a bullish trend, the immediate negative factors outweigh the potential for ongoing gains. A critical factor influencing the decision is the limited cash reserve, which not only prevents buying at the current stock price but also underscores the need for liquidity and risk mitigation. Past trading decisions highlight the importance of adjusting positions in response to prevailing market conditions, even when they suggest a departure from an overall bullish sentiment. The absence of a clear signal from the provided trading strategies further supports a shift towards a SELL decision. My reasoning is grounded in a comprehensive analysis of market intelligence, market trends, professional investment guidances, trading strategies, and financial constraints, all of which collectively point to SELL as the prudent action under current circumstances.
\end{lstlisting}
\end{figure*}

\texttt{FinAgent}'s decision-making prowess showcases a sophisticated blend of comprehensive analysis, adaptability, and rigorous risk management, each playing a pivotal role in navigating the intricate dynamics of financial markets. The methodology is rooted in a deep analysis that encompasses an array of factors such as market intelligence, price trends, and expert investment advice. This holistic approach enables \texttt{FinAgent} to distill actionable insights from complex data streams, as demonstrated in Example 1 where the BUY decision was informed by a confluence of positive market sentiment toward Apple's innovative AR/VR product line and a bullish price trajectory across different timelines.

The essence of \texttt{FinAgent}'s strategy lies in its remarkable adaptability and commitment to learning from historical trading patterns. This aspect is vividly illustrated in Example 2, where a SELL decision was guided by an analysis revealing a negative medium to long-term market sentiment, underscoring \texttt{FinAgent}'s capacity to preemptively adjust its positions in anticipation of future market movements.

Correcting the earlier oversight, Example 3 indeed underscores a SELL decision, driven by a nuanced evaluation of mixed market sentiments, immediate negative challenges potentially impacting AAPL's stock price, and the critical constraint of limited cash reserves. This example highlights \texttt{FinAgent}'s strategic foresight and prudence. Despite recognizing the technical strength and a bullish trend in the long-term price movement analysis, \texttt{FinAgent} opts to sell, prioritizing liquidity and risk mitigation over speculative gains in the face of significant immediate uncertainties and financial limitations.

In conclusion, \texttt{FinAgent}'s strategic approach to decision-making, illustrated through these examples, emphasizes the importance of a balanced, informed, and cautious investment strategy. By synthesizing comprehensive market analysis, drawing lessons from past trading experiences, and meticulously managing risks, \texttt{FinAgent} adeptly maneuvers through market volatility. This disciplined methodology not only optimizes potential returns but also diligently minimizes exposure to adverse market swings, demonstrating the criticality of adaptability and prudent risk assessment in achieving sustainable investment success.

\subsection{Case Study: II}

\begin{figure}[htbp]
    \centering
    \begin{minipage}{\textwidth}
        \centering
        \includegraphics[width=\textwidth]{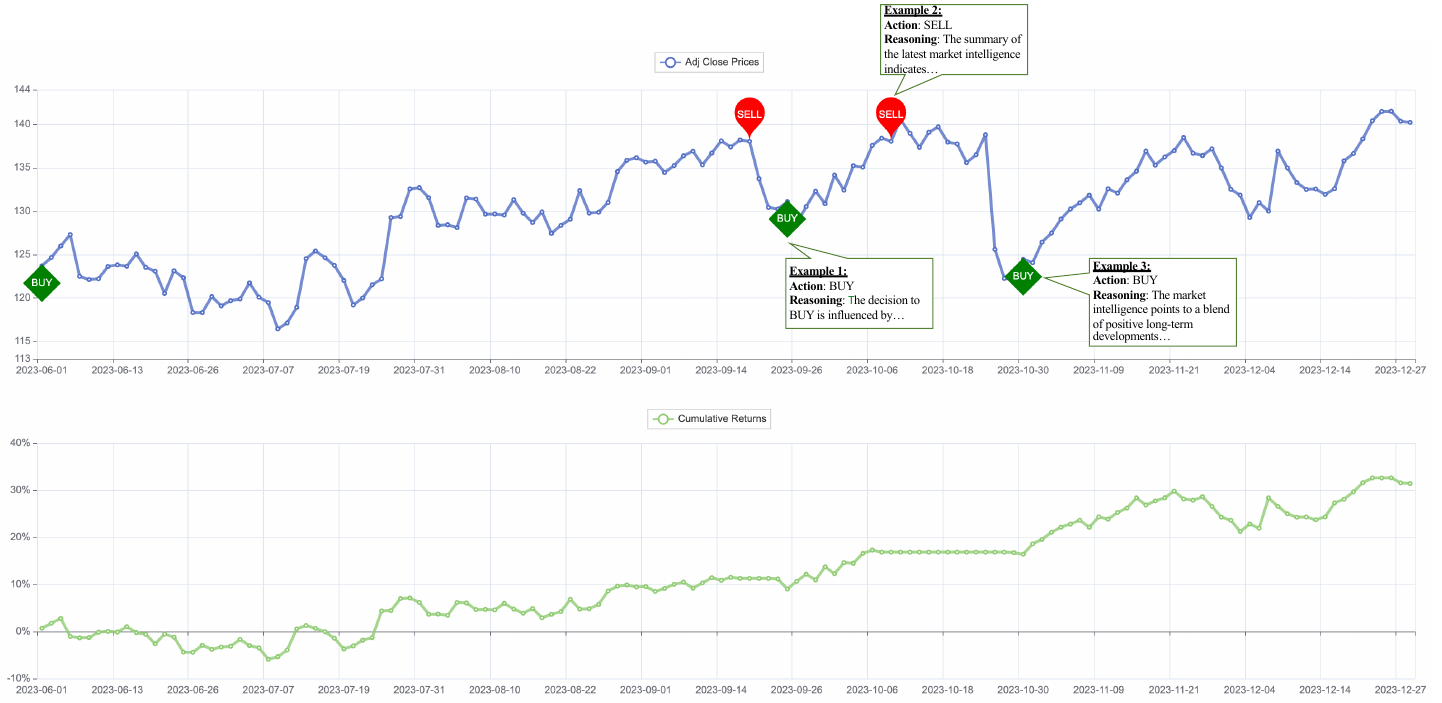}
        \caption{Performance analysis of FinAgent for GOOGL.}
        \label{fig:GOOGL_FinAgent}
    \end{minipage}
\end{figure}

\hfill
\noindent
\begin{figure*}[htbp]
\paragraph{\underline{\textbf{Example 1}}}\mbox{}\\
\begin{lstlisting}[style = xml]
Action: BUY
Reasoning: The decision to BUY is influenced by the summary of market intelligence, which leans toward a positive long-term outlook for Alphabet Inc., indicating strengthened areas like cloud and AI sectors, and persistent investor confidence, despite short-term concerns. Reflecting on past price movements and the upward trend in the latest short-term analysis confirms the bullish momentum, which aligns with the past lessons on the importance of acting on bullish signals and positive sentiment. Additionally, the reflection on past trading decisions advises us not to miss out on BUY opportunities when the overall trend and market sentiment are positive. Finally, with the absence of any current position and sufficient cash reserves, the constraints are satisfied for making a purchase decision.    Therefore, taking into account the bullish market outlook, positive price trend, and our capacity to act, the correct decision at this point is to BUY.
\end{lstlisting}
\end{figure*}
\hfill
\noindent
\begin{figure*}[htbp]
\paragraph{\underline{\textbf{Example 2}}}\mbox{}\\
\begin{lstlisting}[style = xml]
Action: SELL
Reasoning: Evaluation of the current market's risk of correction or downturn, a strategic SELL decision is considered. The decision to sell is further bolstered by the lack of liquidity to purchase additional shares and the desire to capitalize on the current stock value before any potential market adjustments that could erode gains. Additionally, reflections on past trading strategies underscore the risk of missing optimal selling points during bullish trends due to an overly conservative approach. Consequently, in light of the existing market optimism, the choice to SELL aims to preemptively secure profits and improve liquidity, providing a more flexible stance to navigate future market uncertainties and capitalize on buying opportunities at lower prices. This action respects financial limitations while strategically positioning for potential market recalibrations.
\end{lstlisting}
\end{figure*}
\hfill
\noindent
\begin{figure*}[htbp]
\paragraph{\underline{\textbf{Example 3}}}\mbox{}\\
\begin{lstlisting}[style = xml]
Action: BUY
Reasoning: The market intelligence presents a nuanced picture, highlighting short-term challenges alongside significant long-term opportunities, particularly with advancements in AI and a relaxation of regulatory pressures. This analysis points to a mild bearish trend in the short to medium term but underscores the positive long-term implications of these technological and regulatory developments. Past reflections on trading decisions emphasize the importance of seizing bullish momentum when it aligns with long-term growth prospects. Contrary to previous constraints, my current cash reserve is ample, exceeding the Adj Close Price, which positions me well to take advantage of the buying opportunity. Given the market's recent positive response, indicating early recognition of the long-term potential, there's a clear rationale for a BUY decision. This action is not only feasible due to the sufficient cash reserves but is also strategically sound, aiming to capitalize on the anticipated long-term appreciation. The decision to BUY, therefore, aligns perfectly with both the financial capability to make such a purchase and the optimistic long-term outlook. This strategic move is intended to pre-emptively lock in potential long-term gains before these prospects are fully priced into the market, leveraging the current financial flexibility to strengthen the portfolio's position for future growth.
\end{lstlisting}
\end{figure*}

The decision-making approach of \texttt{FinAgent}, as showcased through the examples, underscores a comprehensive and nuanced understanding of the market, demonstrating several core strengths that contribute to its effective trading strategy.

Firstly, in Example 1, \texttt{FinAgent}'s decision to BUY is informed by a detailed analysis of market intelligence that points to a positive long-term outlook for Alphabet Inc. This example highlights \texttt{FinAgent}'s ability to integrate diverse sources of market intelligence, including sector-specific developments in cloud and AI, and investor sentiment, into its decision-making process. It reflects a strategic consideration of both current market conditions and future growth prospects, showcasing the importance of acting on bullish signals when the overall market sentiment and trend analysis suggest potential for appreciation.

Example 2 presents a scenario where \texttt{FinAgent} opts for a SELL decision, emphasizing risk management and liquidity considerations. The reasoning behind the sell decision illustrates \texttt{FinAgent}'s capacity to preemptively navigate potential market downturns or corrections by capitalizing on current stock values. This decision is particularly noteworthy for its forward-looking nature, aiming to secure profits and improve liquidity in anticipation of future market uncertainties. It demonstrates \texttt{FinAgent}'s adaptability and its ability to incorporate reflections on past trading strategies to avoid missing optimal selling points, underscoring a proactive approach to risk management.

In Example 3, the decision to BUY despite short-term challenges, based on the strength of long-term opportunities, particularly advancements in AI and regulatory relaxations, showcases \texttt{FinAgent}'s strategic foresight. This example illustrates how \texttt{FinAgent} leverages comprehensive market analysis to identify and act on long-term growth prospects, even in the face of temporary market downturns. The ample cash reserve facilitates this decision, highlighting the importance of financial flexibility in capitalizing on strategic buying opportunities.

Collectively, these examples demonstrate \texttt{FinAgent}'s strengths in leveraging detailed market analysis, integrating short and long-term market trends, managing risks proactively, and maintaining financial flexibility to make informed trading decisions. \texttt{FinAgent} is characterized by a balance of strategic foresight, adaptability, and a disciplined consideration of past trading reflections, positioning it to capitalize on market opportunities while navigating potential risks effectively.

\subsection{Case Study III}

\begin{figure}[htbp]
    \centering
    \begin{minipage}{\textwidth}
        \centering
        \includegraphics[width=\textwidth]{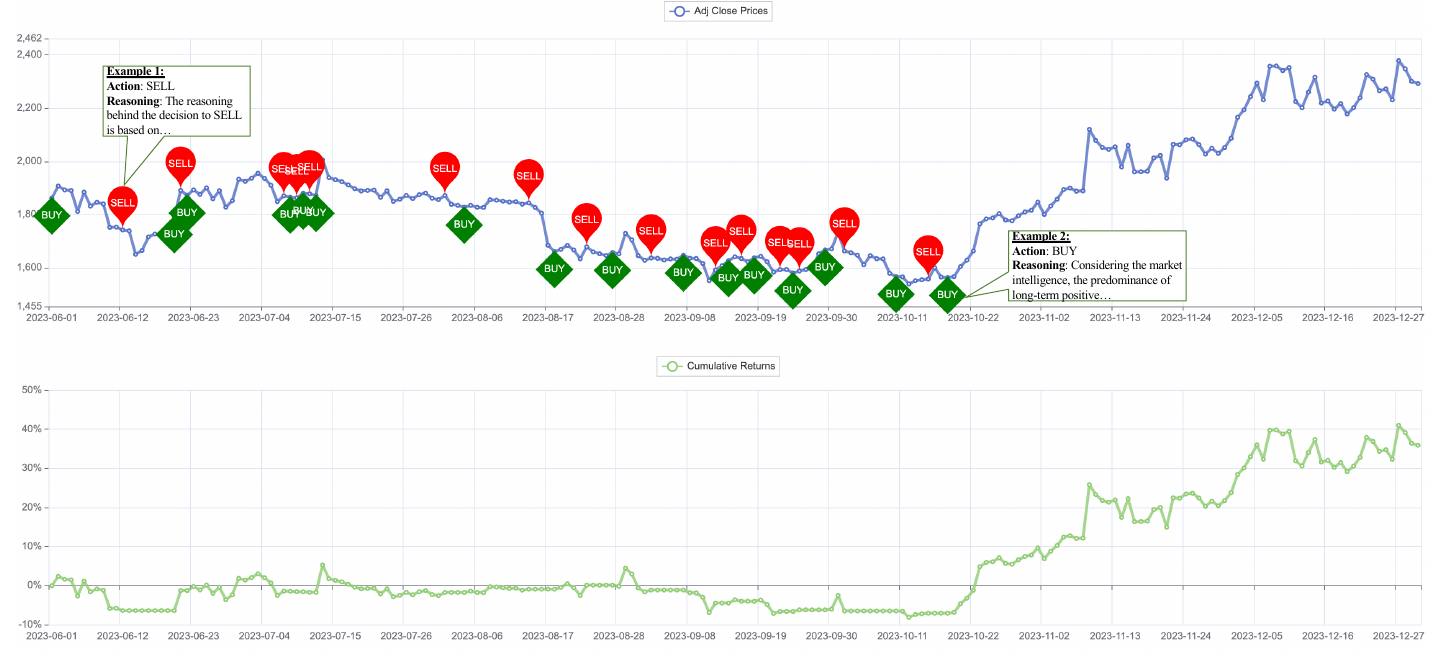}
        \caption{Performance analysis of FinAgent for ETHUSD.}
        \label{fig:GOOGL_FinAgent}
    \end{minipage}
\end{figure}

\hfill
\noindent
\begin{figure*}[htbp]
\paragraph{\underline{\textbf{Example 1}}}\mbox{}\\
\begin{lstlisting}[style = xml]
Action: SELL
Reasoning: The decision to SELL the ETHUSD pair is a strategic outcome of integrating market intelligence, historical trend analysis, technical indicators, and risk management principles, revealing a comprehensive picture of the market's downward trajectory. Negative sentiment prevails in the short to medium term, suggesting further depreciation, a view supported by consistent downtrends in historical price movements. Technical analyses, including the Moving Average (MA5) and Bollinger Bands, underscore the bearish momentum, with the ETHUSD pair's position near the lower Bollinger Band indicating not just an oversold state but the persistence of the bearish trend. Financial considerations, especially the current position's loss and insufficient cash reserves for additional purchases, necessitate a sell to prevent further losses. This decision, shaped by lessons from past trades and the observed continuity of the downtrend, aligns with a strategy focused on capital protection and risk management amidst negative market sentiment. Thus, selling is identified as the most prudent action to mitigate risks and capitalize on informed market analysis.
\end{lstlisting}
\end{figure*}
\hfill
\noindent
\begin{figure*}[htbp]
\paragraph{\underline{\textbf{Example 2}}}\mbox{}\\
\begin{lstlisting}[style = xml]
Action: BUY
Reasoning: The decision to BUY ETH/USD is strongly supported by a blend of market intelligence indicating long-term positive sentiment, technical analysis via MA5 and Bollinger Bands, and the advantage of having sufficient cash reserves. The MA5 confirms the asset's robust momentum by showcasing its price consistently above this short-term moving average, highlighting a sustained bullish trend. Concurrently, Bollinger Bands signal the asset's strength and potential for further growth as it trends towards or breaches the upper band, indicative of high market performance and volatility in favor of an upward trajectory. This comprehensive analysis, coupled with the financial readiness to invest, aligns with an aggressive strategy to capitalize on the anticipated bullish market dynamics. The convergence of positive long-term outlooks, technical indicators underscoring the continuation of upward trends, and liquidity positions BUY as a strategically sound decision aimed at leveraging expected market gains.
\end{lstlisting}
\end{figure*}

The decision-making capabilities of \texttt{FinAgent}, as illustrated through examples, reveal a sophisticated blend of strategic insight and analytical depth in navigating the cryptocurrency market. This approach leverages comprehensive market analysis, integrating market intelligence, historical trends, and technical indicators like the Moving Average (MA5) and Bollinger Bands, to inform its trading decisions.

Example 1 showcases \texttt{FinAgent}'s methodical use of data to make a SELL decision on the ETHUSD pair, reflecting a commitment to capital protection and risk management amidst negative market trends. This demonstrates \texttt{FinAgent}'s strength in crafting informed strategies that prioritize long-term capital security over short-term gains, illustrating a prudent approach to trading in volatile markets.

Example 2 illustrates the proactive side of \texttt{FinAgent}, where a BUY decision is made based on a positive market outlook. This decision, supported by solid technical analysis and sufficient liquidity, shows \texttt{FinAgent}'s ability to seize market opportunities, highlighting its adaptability and aggressive strategy to leverage potential market upswings for significant gains.

\texttt{FinAgent}'s strategic and analytical framework, as evidenced by these examples, effectively balances risk and reward, demonstrating a nuanced understanding of the cryptocurrency market's complexities. The methodical integration of diverse data points into coherent trading strategies underscores \texttt{FinAgent}'s capability to navigate market dynamics adeptly.

While the cryptocurrency market's inherent volatility poses challenges, \texttt{FinAgent}'s approach exhibits a level of resilience. The firm's strategic decisions, though cautious, are not overly conservative but are designed to adapt to market conditions, seeking to optimize outcomes within the realm of calculated risk. Moving forward, refining analytical models and strategies to further align with the unpredictable nature of cryptocurrencies remains an area for gradual enhancement. By continuing to evolve its decision-making framework, \texttt{FinAgent} aims to maintain its strategic edge while mitigating the impacts of market volatility, positioning it well for sustained success in the dynamic landscape of cryptocurrency trading.

\section{Details of Benchmark Analysis}
\label{app:analysis_benchmarks}

\subsection{Analysis of Rule-based Strategies for Trading }
\label{app:analysis_rule_based}
We examine four fundamental rule-based strategies using technical indicators, chosen for their broad coverage of market scenarios and strategic diversity in trading. Moving Average Convergence Divergence (MACD) and stochastic oscillators combined with Bollinger Bands represent trend-following methods fundamental to capturing market momentum. In contrast, KDJ with RSI Filter and Z-score Mean Reversion offer insights into potential market reversals, crucial for risk management and exploiting countertrend opportunities. Together, these strategies encompass a wide spectrum of trading situations, from following prevailing trends to identifying reversal points, thereby providing a comprehensive decision-making toolkit. To set a benchmark, these strategies use the same market state and trading environment as other RL benchmark methods. In the training stage, we applied OPTUNA for hyperparameter optimization to adapt them to the financial instrument traded. 

\subsubsection{Moving Average Convergence Divergence (MACD) Crossover Strategy.}

This strategy utilizes MACD indicators for trend-following, generating buy/sell signals based on bullish or bearish momentum shifts, making it suitable for moderate-risk traders favoring clear market trends.

\begin{figure}[htbp]
    \centering
    \begin{minipage}{0.45\textwidth}
        \centering
        \includegraphics[width=\textwidth]{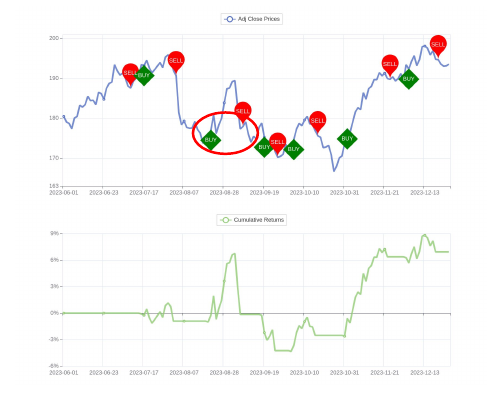}
        \caption{Performance analysis of MACD for AAPL.}
        \label{fig:MACD_AAPL}
    \end{minipage}
    \hfill
    \begin{minipage}{0.45\textwidth}
        \centering
        \includegraphics[width=\textwidth]{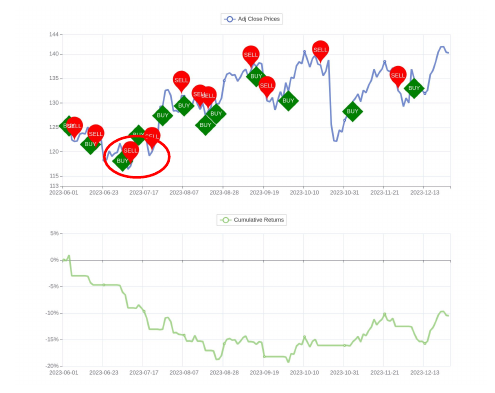}
        \caption{Performance analysis of MACD for GOOGL.}
        \label{fig:MACD_GOOGL}
    \end{minipage}
\end{figure}

\begin{figure}[htbp]
    \centering
     \begin{minipage}{0.45\textwidth}
        \centering
        \includegraphics[width=\textwidth]{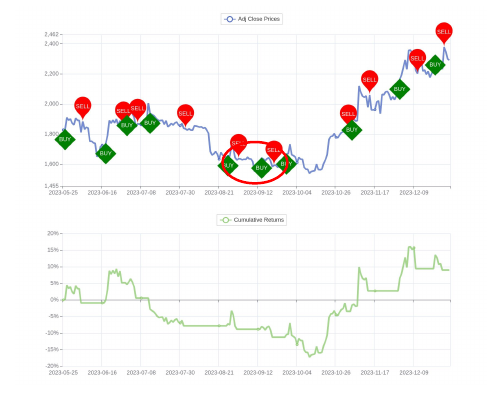}
      \caption{Performance analysis of MACD for ETHUSD.}
      \label{fig:MACD_ETHUSD}
    \end{minipage}
    \hfill
      \begin{minipage}{0.45\textwidth}
        \centering
        \includegraphics[width=\textwidth]{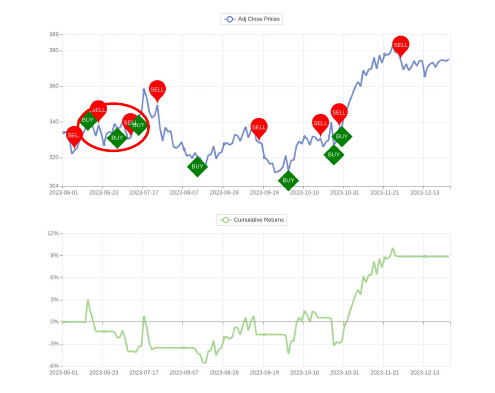}
      \caption{Performance analysis of MACD for MSFT.}
      \label{fig:MACD_MSFT}
    \end{minipage}
\end{figure}

\textbf{Limitation 1: Sub-optimal position-changing points.} 
The MACD strategy often exhibits delayed reactions to market changes, leading to sub-optimal position-changing points. This lag is particularly pronounced in fast-moving or volatile markets. For example, in the case of stocks like AAPL as marked in Fig. \ref{fig:MACD_AAPL}, where market dynamics can shift rapidly, the MACD might signal a buy or sell too late, thereby missing the optimal entry or exit points. This limitation is intrinsic to the MACD’s reliance on historical moving averages, making it less effective in markets that experience quick reversals or where the trend direction changes frequently.

\textbf{Limitation 2: Incapability of handling volatility.} 
The MACD strategy struggles in highly volatile markets due to its dependency on moving averages, which may not accurately reflect current market conditions under such circumstances. In volatile trading environments like the marked period in Fig.\ref{fig:MACD_GOOGL}, the MACD can generate misleading signals, leading to a series of unprofitable trades. The standard signal thresholds of MACD may not be adaptable enough for all market conditions, particularly in volatile settings. This limitation underscores the need for a more dynamic approach, possibly integrating adaptive thresholds that account for the prevailing market volatility.

\textbf{Limitation 3: High turnover rate results in high transaction fee.} 
A significant drawback of the MACD strategy is its tendency to generate a high turnover rate, as shown in the marked periods of Fig.\ref{fig:MACD_ETHUSD} and Fig.\ref{fig:MACD_MSFT}, leading to substantial transaction costs. In financial markets where transaction fees are considerable, like the stock market, frequent trading by MACD can significantly lower profit margins.

\subsubsection{KDJ\&RSI Strategy.} 

This strategy merges the KDJ stochastic oscillator and RSI to identify overbought or oversold conditions, issuing buy or sell signals based on momentum and RSI thresholds, ideal for risk-averse traders preferring extreme market conditions.

\begin{figure}[htbp]
    \centering
    \begin{minipage}{0.45\textwidth}
        \centering
          \includegraphics[width=\textwidth]{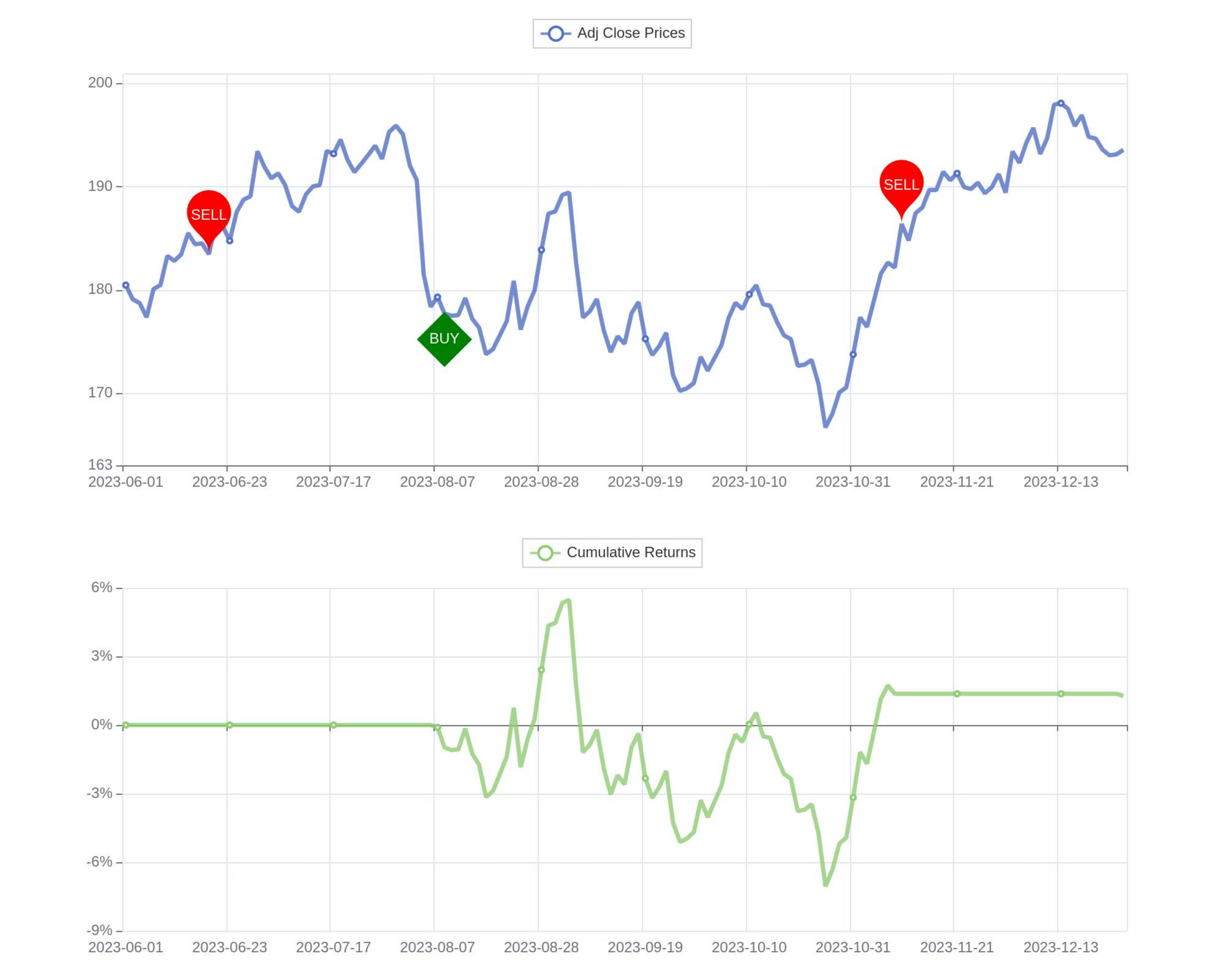}
          \caption{Performance analysis of KDJ\&RSI for AAPL.}
          \label{fig:KDJ&RSI_AAPL}
    \end{minipage}
    \hfill
    \begin{minipage}{0.45\textwidth}
        \centering
          \includegraphics[width=\textwidth]{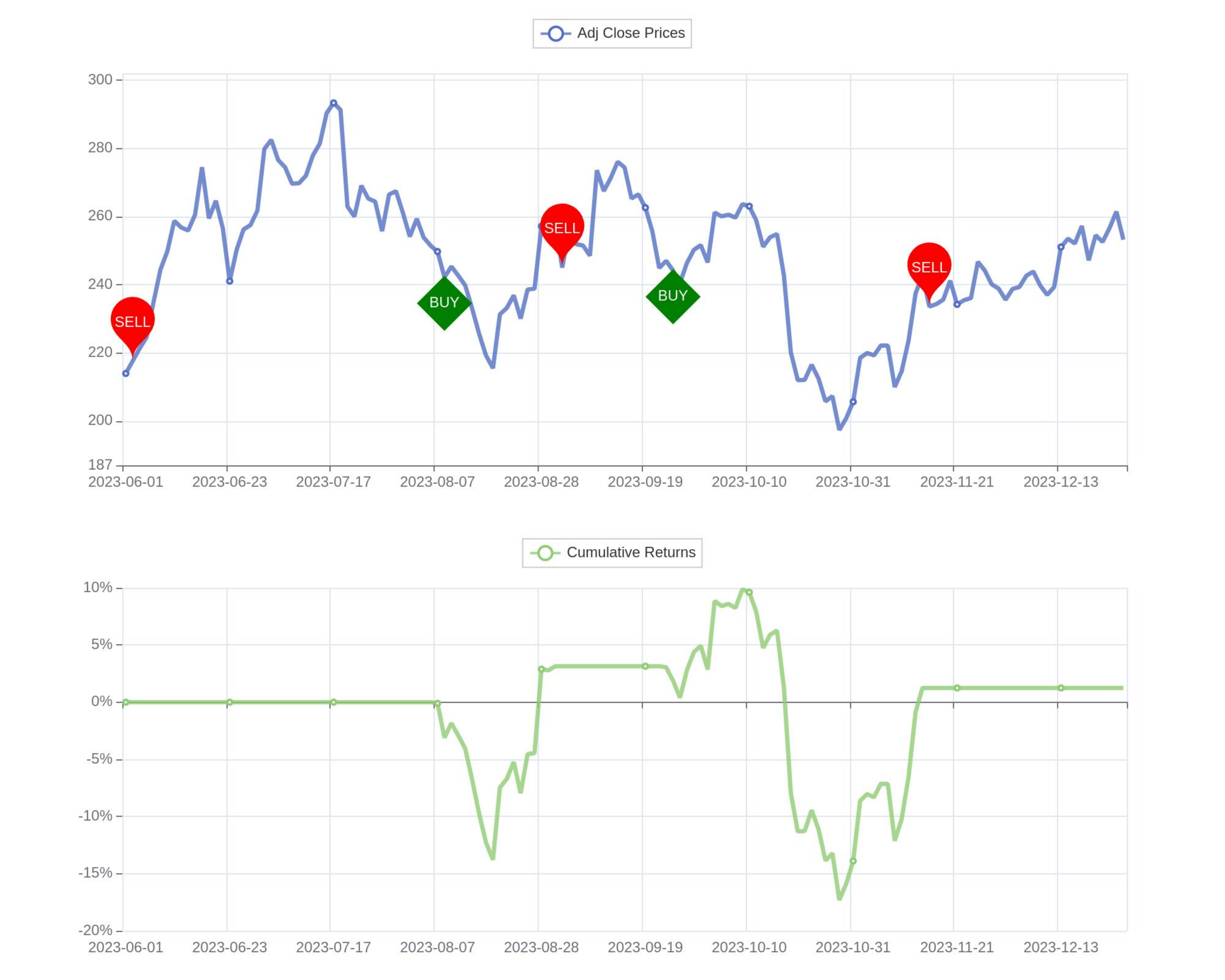}
          \caption{Performance analysis of KDJ\&RSI for TSLA.}
          \label{fig:KDJ&RSI_TSLA}
    \end{minipage}
\end{figure}

\textbf{Limitation 1: Lack of generalization ability.} 
The KDJ\&RSI strategy, while useful in certain market conditions, is not a universally applicable strategy. It lacks generalization across different market environments due to its sensitivity to specific parameters and market trends. This means that its performance can vary greatly between different stocks. Despite the hyper-partmeters are trained with historical data, it is still not able to get profitable results on some instruments as shown in Fig. \ref{fig:KDJ&RSI_AAPL}, \ref{fig:KDJ&RSI_TSLA}.

\subsubsection{Z-score Mean Reversion.} This strategy employs z-score mean reversion, generating buy or sell signals when prices deviate significantly from the mean, making it suitable for risk-averse traders who capitalize on price normalization.

\begin{figure}[htbp]
    \centering
    \begin{minipage}{0.45\textwidth}
        \centering
        \includegraphics[width=\textwidth]{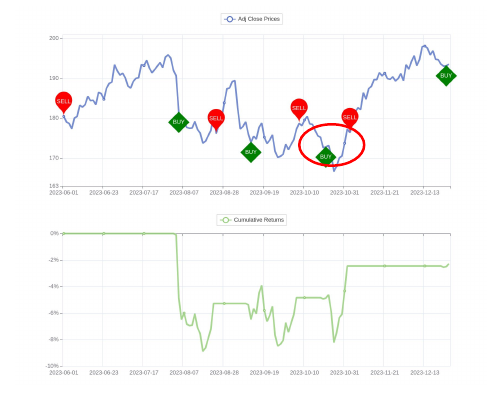}
        \caption{Performance analysis of ZMR for AAPL.}
        \label{fig:ZMR_AAPL}
    \end{minipage}
    \hfill
    \begin{minipage}{0.45\textwidth}
        \centering
        \includegraphics[width=\textwidth]{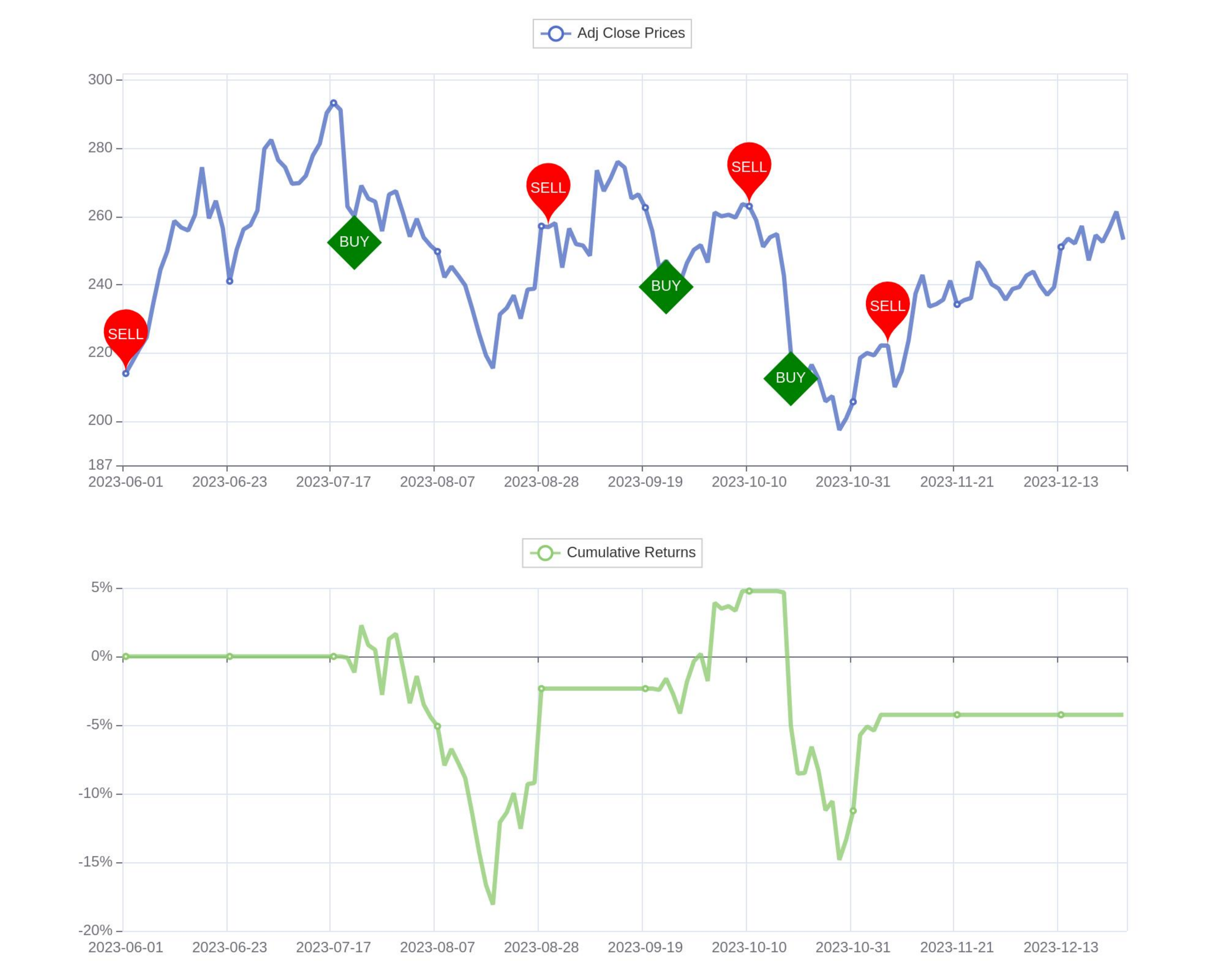}
        \caption{Performance analysis of ZMR for ETHUSD.}
        \label{fig:ZMR_TSLA}
    \end{minipage}
\end{figure}

\textbf{Limitation 1: Sub-optimal decision due to insensitivity to market trend.} The ZMR trading strategy, while adept at capturing dramatic shifts in market patterns, often exhibits a notable insensitivity to more stable, long-term market trends. This limitation becomes particularly apparent in its tendency to make sub-optimal trading decisions during periods of gradual market changes. A clear example of this can be observed in the trading behavior of AAPL, as depicted in the marked period of Fig. \ref{fig:ZMR_AAPL}. The strategy decided to sell after the price met the reverting point, missing the bullish trend afterward. The strategy's focus on mean reversion overlooks the significance of persistent trend movements, leading to sub-optimal trading outcomes.

\textbf{Limitation 2: Poor performance in non-mean-reverting Market conditions.} A significant limitation of the z-score mean-reversion trading strategy is its poor performance in market conditions that do not exhibit mean-reverting behavior. The strategy inherently assumes that stock prices will revert to their historical mean, which is often not the case in trending or momentum-driven markets. In such environments, where prices consistently move in one direction without reverting, the mean-reversion strategy can result in sustained losses. This is particularly evident in strongly bullish or bearish markets, where the strategy may continuously take counter-trend positions, leading to adverse trading outcomes. The inability of the mean-reversion strategy to adapt to non-mean-reverting market conditions highlights its lack of versatility and the potential risk of relying solely on historical mean-reversion patterns. This is evident by the trading decisions in TSLA as shown in Fig. \ref{fig:ZMR_TSLA} .
% \begin{figure}[H]
%   \centering
%   \includegraphics[width=0.45\textwidth]{assets/finagent_and_baselines/ZMR/ZMR_AMZN.pdf}
%   \caption{Performance analysis of ZMR for AMZN.}
%   \label{fig:ZMR_AMZN}
% \end{figure}

% \begin{figure}[H]
%   \centering
%   \includegraphics[width=0.45\textwidth]{assets/finagent_and_baselines/ZMR/ZMR_GOOGL.pdf}
%   \caption{Performance analysis of ZMR for GOOGL.}
%   \label{fig:ZMR_GOOGL}
% \end{figure}

% \begin{figure}[H]
%   \centering
%   \includegraphics[width=0.45\textwidth]{assets/finagent_and_baselines/ZMR/ZMR_TSLA.pdf}
%   \caption{Performance analysis of ZMR for TSLA.}
%   \label{fig:ZMR_TSLA}
% \end{figure}

\subsection{Analysis of Using FinAgent as a router for technical indicators-driven strategies for trading.}

We examined the potential of FinAgent to function as decision-making routers within a Mixture-of-Experts (MoE) framework of trading strategies. This experiment setting only utilized the decision module of FinAgent, intentionally excluding components such as market intelligence, memory, and reflection modules. This exclusion eliminates the necessity for a training phase for FinAgent. The strategies implemented in this research have undergone fine-tuning using Optuna on a designated training dataset. Each strategy's effectiveness is meticulously documented and provided to the LLM, enabling it to make informed decisions based on the detailed performance data.

According to the numerical data presented in Table \ref{tab:baselines}, the FinAgent router, while not always achieving the optimal strategy outcome, consistently delivers robust and positive trading results. This observation is further substantiated by the analysis of trading behavior across various assets, which exhibits less extreme behavior than using a single trading strategy. This suggests that the FinAgent router is capable of adapting its decision-making process to various market conditions. Although there is an improvement compared to employing single trading strategies, we have also identified certain limitations. These limitations are further addressed by other modules within the FinAgent system.

\begin{figure}[htbp]
    \centering
    \begin{minipage}{0.45\textwidth}
        \centering
        \includegraphics[width=\textwidth]{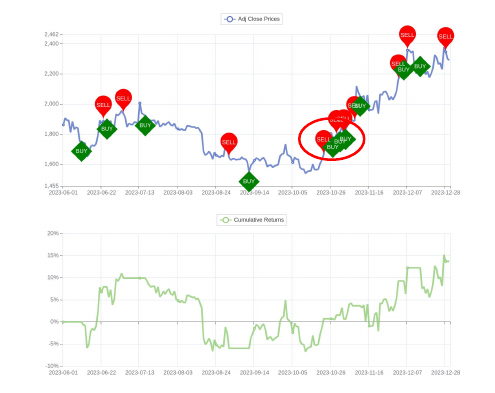}
        \caption{Performance analysis of using FinAgent as a strategy router for ETHUSD.}
        \label{fig:router_ETHUSD}
    \end{minipage}
    \hfill
    \begin{minipage}{0.45\textwidth}
        \centering
          \includegraphics[width=\textwidth]{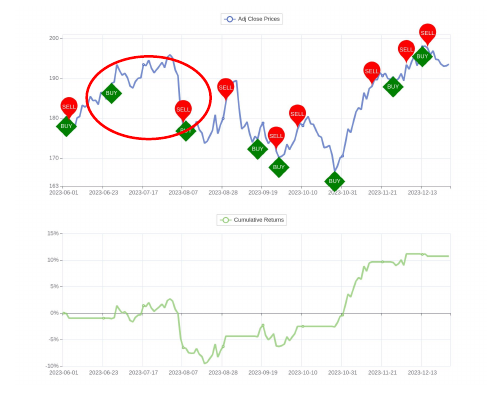}
          \caption{Performance analysis of using FinAgent as a strategy router for AAPL.}
          \label{fig:router_AAPL}
    \end{minipage}
\end{figure}

\textbf{Limitation 1: Gap in using multiple strategies to make a single decision}
In the role of a router, FinAgent demonstrates variable preferences for strategy decisions at each stage. This variability introduces a potential risk of unprofitable trading in such sequential decision-making scenarios, as each strategy operates based on unique assumptions. For instance, consider the period highlighted in Fig. \ref{fig:router_ETHUSD}. Within this short timeframe, multiple buy and sell decisions are executed. The buy decisions are mostly influenced by the KDJ\&RSI and SO\&BB strategies, which signal a buying opportunity. On the other hand, the sell decisions are primarily guided by the ZMR strategy, which indicates a strong sell signal during the same period.

\textbf{Limitation 2: Limited performance with weak strategies.}
The router's performance is constrained by the effectiveness of the underlying strategies. This limitation becomes evident in scenarios where all strategies demonstrate weak performance. An example of this can be observed during the period marked in Fig. \ref{fig:router_AAPL}. As illustrated in Fig. \ref{fig:MACD_AAPL}, Fig. \ref{fig:KDJ&RSI_AAPL}, and Fig. \ref{fig:ZMR_AAPL}, none of the strategies managed to generate a profit during this period. In the absence of other FinAgent modules, such as reflection, the decision-making capability of the router is limited.

\textbf{Limitation 3: Result is sensitive to prompt engineering and randomness.}
In our experiments, we observed a distinct variation in the trading decisions when modifying the decision prompt and executing the agent across different trials. Owing to the inherent nature of the API service provided by OpenAI, controlling the randomness in the responses is not feasible. This limitation adds to the risk factor when relying solely on the decision module in FinAgent for robustness.

\subsection{Analysis of RL for Trading}
\label{app:analysis_drl}
Reinforcement learning has been applied to trading with varying degrees of success. Traditional algorithms like PPO (Proximal Policy Optimization), DQN(Deep Q-learning), and SAC(Soft Actor-Critic) have shown potential but also limitations in stock prediction accuracy and practicality due to large data requirements. To set a benchmark, these algorithms are trained in uniform conditions and same dataset, using OPTUNA for hyperparameter optimization, to explore their effectiveness in trading and provide insights into their applicability and optimization for investment decision-making. 

\subsubsection{Analysis of PPO for trading} 
\begin{figure}[htbp]
    \centering
    \begin{minipage}{0.45\textwidth}
        \centering
          \includegraphics[width=\textwidth]{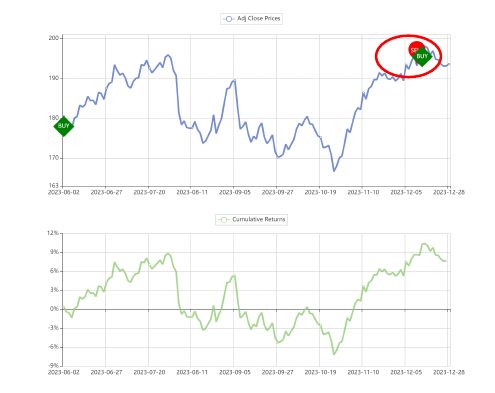}
          \caption{Performance analysis of PPO for AAPL.}
          \label{fig:PPO_AAPL}
    \end{minipage}
    \hfill
    \begin{minipage}{0.45\textwidth}
        \centering
            \includegraphics[width=\textwidth]{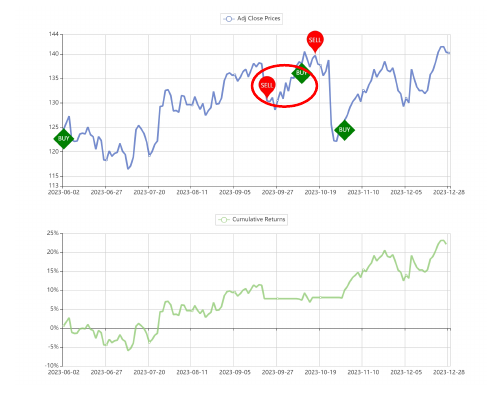}
            \caption{Performance analysis of PPO for GOOGL.}
            \label{fig:PPO_GOOGL}
    \end{minipage}
\end{figure}

\
\newline
\indent The Proximal Policy Optimization (PPO) model, when applied to stock trading, exhibits notable drawbacks that can impact its overall performance. One significant limitation is the tendency for the model to remain idle for extended periods. This proclivity for inactivity stems from the algorithm's cautious learning process, prioritizing risk avoidance over seeking potential gains. Consequently, during these prolonged idle phases, the model may miss lucrative trading opportunities, leading to suboptimal returns. This issue becomes particularly pronounced in dynamic and rapidly changing stock market conditions, where the hesitancy of the PPO model may hinder its ability to capitalize on favorable market movements. The model's conservative nature becomes particularly evident in Figure \ref{fig:PPO_AAPL}, where it remains idle even as the cumulative return drops below 0, highlighting its reluctance to capitalize on favourable market movements. This limitation underscores the need to refine the PPO algorithm to balance risk management better and seize profitable opportunities in dynamic market environments.

Another drawback associated with PPO in stock trading lies in its challenge to achieve optimal timing for buying and selling. The model may encounter difficulties in accurately predicting short-term price movements, resulting in purchasing stocks at relatively high prices or missing opportunities to buy at lower prices. This suboptimal timing can directly impact the model's overall performance, leading to lower returns and less effective utilization of available capital. The inherent uncertainty and volatility of stock markets pose challenges for PPO models to consistently make well-timed decisions, highlighting the complexity of accurately forecasting market movements. Figure \ref{fig:PPO_GOOGL} exemplifies this issue, where the model fails to capture the low-value on September 27, 2023, resulting in a decision to buy at a relatively high price and subsequently leading to suboptimal income. This highlights the need for improved timing mechanisms within the PPO algorithm to enhance its effectiveness in navigating the uncertainties of stock markets.

To address these drawbacks, it becomes imperative to fine-tune the model parameters and optimize the training process. Additionally, augmenting the PPO model with complementary techniques or incorporating more sophisticated features, such as sentiment analysis and macroeconomic indicators, can enhance its decision-making capabilities. Regular monitoring and adjustments are crucial to adapting the model to changing market conditions, mitigating the impact of extended idle periods, and improving its ability to time trades effectively. By embracing a holistic approach that combines machine learning with domain expertise, one can work towards overcoming these limitations and developing a more robust and adaptive stock trading algorithm.

\subsubsection{Analysis of DQN Algorithm for Stock Trading}
\
\newline
\indent Though the implementation of the Deep Q-Network (DQN) algorithm in stock trading can yield relatively positive results in the long term, its effectiveness in navigating market complexities has shown limitations.

\begin{figure}[htbp]
    \centering
    \begin{minipage}{0.45\textwidth}
        \centering
        \includegraphics[width=\textwidth]{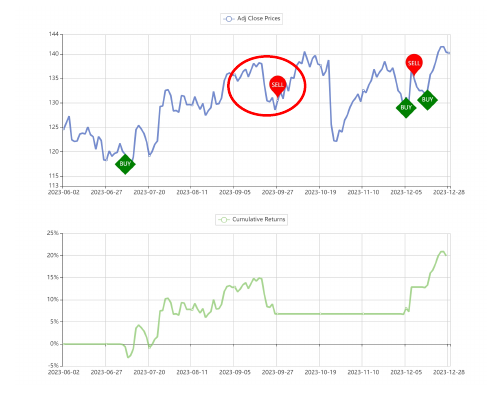}
        \caption{Performance analysis of DQN for GOOGL.}
        \label{fig:DQN_GOOGL}
    \end{minipage}
    \hfill
    \begin{minipage}{0.45\textwidth}
        \centering
            \includegraphics[width=\textwidth]{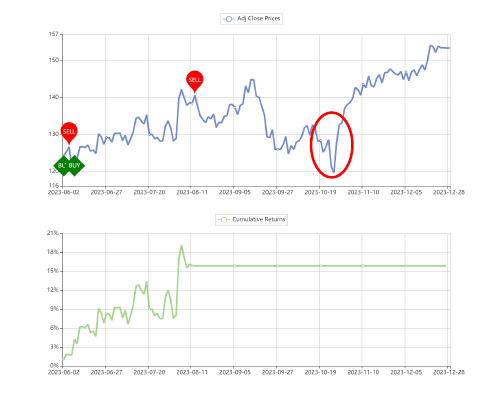}
            \caption{Performance analysis of DQN for AMZN.}
            \label{fig:DQN_AMZN}
    \end{minipage}
\end{figure}

One primary drawback of the DQN algorithm is its inability to balance risk and reward effectively. It often adopts a conservative stance, potentially leading to missed opportunities for significant gains during bullish market phases. Conversely, in bearish conditions, the model's failure to swiftly adjust its strategy can result in prolonged holding of depreciating assets, exacerbating losses. This is evident in Figure\ref{fig: DQN_GOOGL}, where the model fails to capture the high-value point, selling at a lower price on a later day, September 27, 2023, and missing the chance for better earnings. Additionally, it takes no action and remains in a zero position during the bullish phase from October 26, 2023, to November 23, 2023, missing this period of increase. A similar issue occurs in the implementation of AMZN, missing the bullish phase from October 23, 2023, to December 2023.

Another drawback of DQN in this context is its susceptibility to market volatility and unpredictable nature, which can lead to suboptimal decision-making. This issue is particularly noticeable in Figure \ref{fig:DQN_ETH}, where DQN's performance in managing the ETHUSD asset demonstrates a pattern of delayed reactions, often lagging behind rapid market movements, especially from July 18, 2023, to August 15, 2023, during periods of extreme price fluctuations.

To address these challenges, it is crucial to refine the DQN algorithm by incorporating advanced features, such as sophisticated market indicators, real-time data analysis, and diverse market data. Enhancing the algorithm's sensitivity to market dynamics and integrating effective risk management strategies can improve its decision-making quality. Regular updates of the model with current market data and trends are essential to maintain its relevance and accuracy in a rapidly changing financial landscape.

\subsubsection{Analysis of SAC Algorithm for Stock Trading}

\begin{figure}[htbp]
    \centering
    \begin{minipage}{0.45\textwidth}
        \centering
        \includegraphics[width=\textwidth]{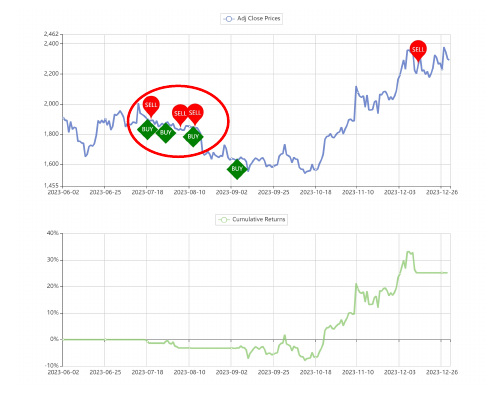}
        \caption{Performance analysis of DQN for ETHUSD.}
        \label{fig:DQN_ETH}
    \end{minipage}
    \hfill
    \begin{minipage}{0.45\textwidth}
        \centering
            \includegraphics[width=\textwidth]{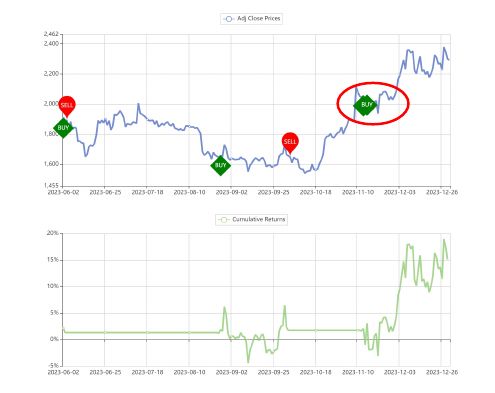}
            \caption{Performance analysis of SAC for ETHUSD.}
            \label{fig:SAC_ETH}
    \end{minipage}
    \hfill
    \begin{minipage}{0.45\textwidth}
        \centering
            \includegraphics[width=\textwidth]{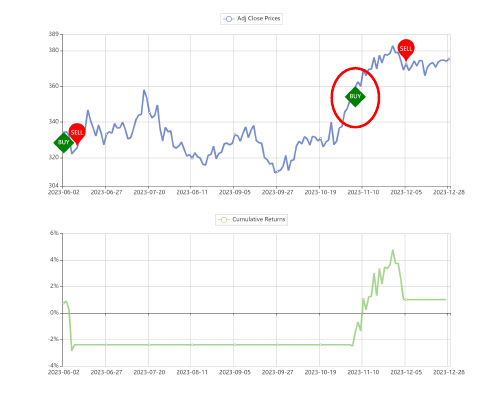}
            \caption{Performance analysis of SAC for MSFT.}
            \label{fig:SAC_MSFT}
    \end{minipage}
    \hfill
    \begin{minipage}{0.45\textwidth}
        \centering
        \includegraphics[width=\textwidth]{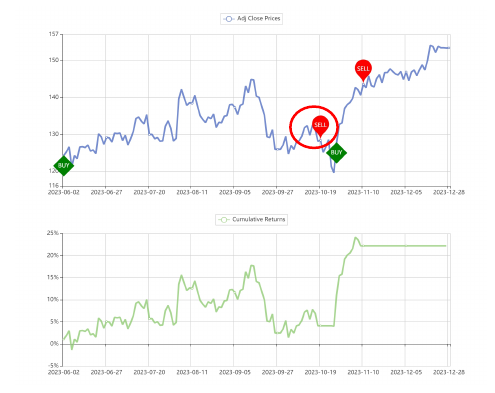}
        \caption{Performance analysis of SAC for AMZN.}
        \label{fig:SAC_AMZN}
    \end{minipage}
\end{figure}

\
\newline
\indent The Soft Actor-Critic (SAC) algorithm, when applied to stock trading, demonstrates certain strengths but also faces challenges that limit its performance. The introduction of entropy regularization enables the agent to better explore unknown states, learning to formulate strategies through interaction with the environment in order to maximize the expected cumulative reward. However, it still encounters difficulties when dealing with the intricacies of the stock market.

One limitation observed in the implementation of SAC is its handling of market ·volatility, especially in high-fluctuation environments like cryptocurrency trading.  For instance, when trade on Ethereum (ETHUSD), as depicted in Figure \ref{fig:SAC_ETH}, SAC tends to be overly cautious. It performed two buy actions around Nov 15, 2023, during a bullish phase. These decisions were made over a relatively long-term observation, which led to missed opportunities during significant market movements. This conservative approach, though avoiding potential losses, can result in the model not fully capitalizing on the profitable market trends.

Similarly, in the stock market, SAC's performance varies based on the specific characteristics of each stock. For example, with Microsoft (MSFT) and Amazon (AMZN), depicted in Figures \ref{fig:SAC_AMZN} and \ref{fig:SAC_MSFT}, SAC struggles to optimally time its entry and exit points. For AMAZ, it sold the holding stocks on October 19, 2023, missing several peaks and losing potential higher returns. For MSFT, it performed sold operations on November 10, 2023, when the bullish market trend still went on. Both late responses to market uptrends and premature exits from profitable positions affect the overall return on trading.

In summary, while SAC offers a robust framework for reinforcement learning tasks, its application in the dynamic and volatile stock requires careful tuning and enhancement to better leverage its capabilities. 

\subsection{Analysis of FinGPT for Trading}
\label{app:analysis_fingpt}
FinGPT's performance in predicting stock prices has exhibited several significant flaws, highlighting critical limitations in its ability to provide reliable insights for investment decisions.

\begin{figure}[htbp]
    \centering
    \begin{minipage}{0.45\textwidth}
        \centering
        \includegraphics[width=\textwidth]{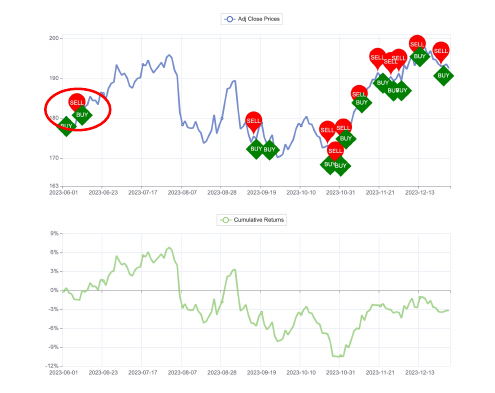}
        \caption{Performance analysis of FinGPT for AAPL.}
        \label{fig:chart1}
    \end{minipage}
    \hfill
    \begin{minipage}{0.45\textwidth}
        \centering
        \includegraphics[width=\textwidth]{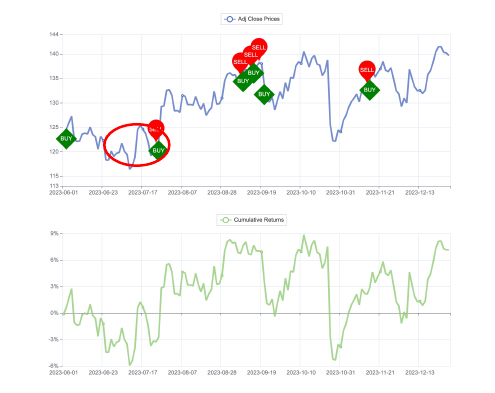}
        \caption{Performance analysis of FinGPT for GOOGL.}
        \label{fig:chart2}
    \end{minipage}
    \hfill
        \begin{minipage}{0.45\textwidth}
        \centering
        \includegraphics[width=\textwidth]{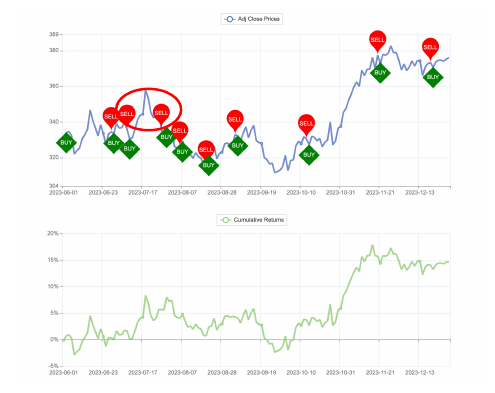}
        \caption{Performance analysis of FinGPT for MSFT.}
        \label{fig:chart3}
    \end{minipage}
    \hfill
    \begin{minipage}{0.45\textwidth}
        \centering
        \includegraphics[width=\textwidth]{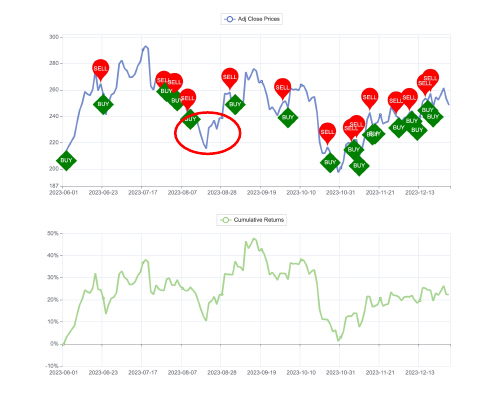}
        \caption{Performance analysis of FinGPT for TSLA.}
        \label{fig:chart4}
    \end{minipage}
\end{figure}

\textbf{Limitation 1: Failure to capitalize on peak price opportunities. } During a specific period, FinGPT missed the chance to sell the stock at its zenith, opting instead to hold onto it. Unfortunately, this strategy proved detrimental as the stock's value plummeted below the initial purchase price, leading to avoidable financial losses. This reveals a deficiency in FinGPT's adaptive decision-making, as it neglected to adjust its strategy despite the clear opportunity to maximize returns. For example, on July 10, 2023, FinGPT made a buy action for MSFT at a price of \$330.46. Despite reaching a peak price of \$358 on July 18, 2023, FinGPT continued to hold the stock, ultimately making a misprediction about a further increase. The reasoning behind this misprediction involved positive developments, market conditions, and the anticipation of strong earnings. However, the model failed to adjust its strategy and sold the stock at a lower price of \$336.37 on July 26, 2023, resulting in a significant loss as shown in Figure \ref{fig:chart3}.

\textbf{Limitation 2: Wrong prediction for a sharp price drop. } Another noted pattern of inaccurate predictions by FinGPT in stock market forecasting raises serious concerns about its reliability as a tool for investment guidance. The model's propensity to erroneously predict increases in stock prices, leading to unexpected downturns, suggests a fundamental flaw in its predictive capabilities, rendering it unsuitable for investors seeking profitable opportunities. 

A glaring example of FinGPT's failure in predictive accuracy occurred in its handling of TSLA (Tesla) stocks on October 8, 2023. The model recommended purchasing TSLA stocks at a specific price of \$245.34, operating under the assumption that the stock would subsequently experience an upward trajectory. However, this prediction proved grossly inaccurate, as the stock value deviated significantly from the forecasted trajectory, plummeting to \$215.49 by August 18, 2023. Figure \ref{fig:chart4} shows the details of this failure.

This significant misalignment between FinGPT's predictions and the actual market trajectory exposed its limitation in adapting to dynamic conditions and eroded the confidence of investors who relied on its recommendations. The incident highlights the imperative need for continuous improvement in FinGPT's algorithms to enhance its predictive accuracy, ensuring investors receive more reliable guidance in fluctuating financial landscapes.

The inability of FinGPT to accurately anticipate the dynamic nature of financial markets emphasizes the critical need for refining its predictive algorithms. This refinement is crucial to ensure a more nuanced and accurate understanding of market dynamics, ultimately enhancing the model's reliability as a tool for investment guidance. The incident with TSLA is a compelling example of the imperative nature of continuous improvement in FinGPT's capabilities to meet the demands of a rapidly changing and unpredictable financial landscape. Furthermore, this incident with TSLA serves as a poignant reminder of the evolving challenges within the financial landscape, necessitating ongoing efforts to fine-tune FinGPT's algorithms. 

\textbf{Limitation 3: Wrong prediction for a stock price surge. } Conversely, when FinGPT wrongly predicts a decrease in stock prices, it opts to sell, but the actual market behaviour contradicts its forecast by demonstrating a continued increase in stock value. This inconsistency in predicting market trends raises questions about the model's effectiveness in providing actionable insights aligned with real-world financial dynamics. For example, on June 7, 2023, the model recommended selling AAPL stocks, forecasting a 1-2\% decrease based on positive developments and potential concerns. The cited positive developments included the expected valuation of \$4 trillion by 2025 and the successful launch of the Vision Pro. However, the market response contradicted FinGPT's projection, as the stock price increased for three consecutive days, as shown in Figure \ref {fig:chart1}. This highlights a substantial misalignment between the model's analysis and the market dynamics, exposing a critical flaw in its ability to anticipate short-term stock movements accurately.

\textbf{Limitation 4: Mismatch between the action and reasoning.} The further drawback of FinGPT is its tendency to recommend continuous holding of stocks, even when successful price predictions are made. Despite accurately forecasting market movements, the model often fails to provide timely and proactive investment strategies. This inclination towards recommending a prolonged holding strategy may cause investors to miss out on valuable opportunities for profit maximization. For example, during the period spanning from June 6, 2023, to July 6, 2023, FinGPT showcased an impressive ability to predict all price changes for GOOGL accurately. However, a critical flaw emerged in its decision-making process, as the model consistently maintained the same investment action without adapting to the foreseen market dynamics. Despite believing that the stock was destined to experience a decline in the near future, as shown in Figure \ref{fig:chart2}, FinGPT failed to promptly translate this insight into a strategic move such as selling the stock. This lack of agility in responding to its predictions represents a significant shortcoming. The essence of successful trading lies not only in the accurate anticipation of market trends but also in the timely execution of appropriate actions to capitalize on those predictions or, conversely, to mitigate potential losses. FinGPT's inability to recalibrate its investment strategy when confronted with predicted downturns accentuates a key area where the model could significantly improve its decision-making capabilities. While the model excels in forecasting, integrating a more dynamic and proactive approach to trading would enhance its overall efficacy. Financial markets are renowned for their fluidity and susceptibility to rapid changes, making adaptability a crucial element for success. The failure to adjust to evolving conditions hampers the model's ability to navigate market complexities and leads to missed opportunities to leverage favourable price fluctuations. This limitation underscores the critical necessity of incorporating a more responsive decision-making approach into FinGPT's functionality. The dynamic nature of financial markets requires accurate predictions and the ability to interpret changing conditions swiftly and proactively, ensuring that FinGPT remains a valuable and reliable tool.

\begin{table}[htbp]
  \centering
  \caption{AMZN wrong prediction for 3 consecutive days }
  \small % Set the text size to \small for the table
  \label{tab:stock_info}
  \begin{threeparttable}
    \begin{tabular}{ccccc}
    \toprule
    \textbf{Symbol} & \textbf{Date} & \textbf{Price Change} & \textbf{Action} & \textbf{Prediction} \\
    \midrule
    AMZN & 11/10 & -0.97 & HOLD & ... increase by 0-1\% ... \\
    AMZN & 14/10 & 3.21 & HOLD &... decrease by 0.5-1\% ...  \\
    AMZN & 15/10 & -2.6 & HOLD & ... increase by 0-1\% ...  \\
    \bottomrule
  \end{tabular}
  \end{threeparttable}
\end{table}

\textbf{Limitation 5: Meaningless prediction after consecutive failures.} The last notable drawback of FinGPT is when FinGPT encounters three consecutive wrong predictions. In response, the model generates explanations that lack substance and offer little value to investors. These responses often involve generic statements, such as "it's difficult to predict the exact stock price movement," accompanied by vague positive and negative factors. This tendency to provide meaningless responses further diminishes FinGPT's credibility and utility in guiding investment decisions during challenging periods. For example, The prediction FinGPT made on November 16, 2023, for AMZN, after the wrong prediction from November 11, 2023, to November 15, 2023, as shown in Table \ref{tab:stock_info}, instead of providing a meaningful analysis or adjusting its approach, the model generated a generic response, acknowledging the difficulty of predicting stock movements and offering vague positive factors and potential concerns. This lack of nuanced reasoning and the provision of a seemingly irrelevant response indicates a weakness in the model's ability to learn from its mistakes and adjust its predictions coherently.

\subsection{Analysis of FinMem for Trading}
\label{app:analysis_finmem}
FinMem primarily relies on layered memory to retrieve historical news and financial reports, and depends on the reflection mechanism to contemplate the potential connections between this information and future price movements. In this section, we conduct a detailed analysis of the decision-making results and reasoning provided by FinMem during the decision-making process.

\begin{figure}[htbp]
    \centering
    \begin{minipage}{0.45\textwidth}
        \centering
        \includegraphics[width=\textwidth]{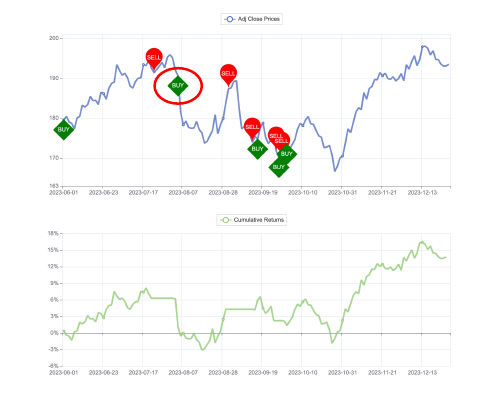}
        \caption{Performance analysis of FinMem for AAPL.}
        \label{fig:AAPL_FinMem}
    \end{minipage}
    \hfill
    \begin{minipage}{0.45\textwidth}
        \centering
        \includegraphics[width=\textwidth]{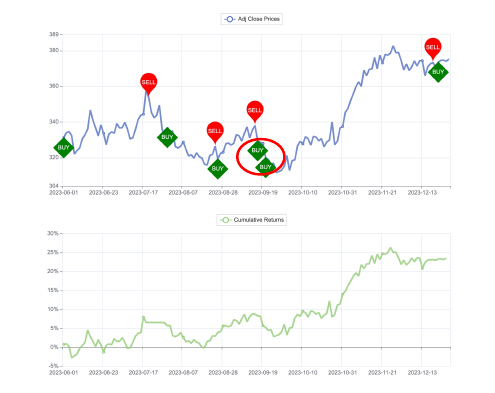}
        \caption{Performance analysis of FinMem for MSFT.}
        \label{fig:MSFT_FinMem}
    \end{minipage}
    \hfill
        \begin{minipage}{0.45\textwidth}
        \centering
        \includegraphics[width=\textwidth]{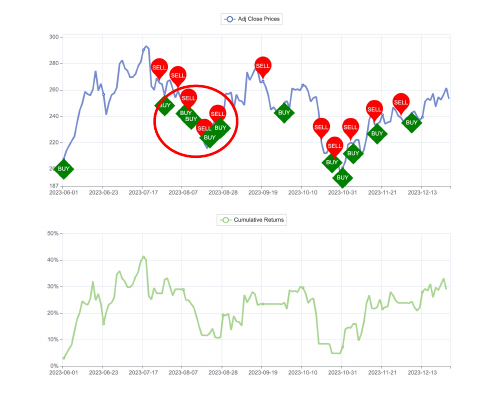}
        \caption{Performance analysis of FinMem for TSLA.}
        \label{fig:TSLA_FinMem}
    \end{minipage}
    \hfill
    \begin{minipage}{0.45\textwidth}
        \centering
        \includegraphics[width=\textwidth]{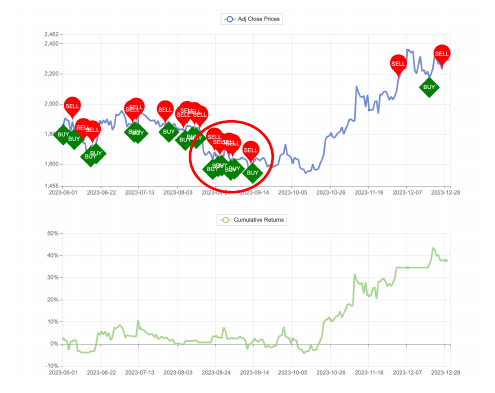}
        \caption{Performance analysis of FinMem for ETHUSD.}
        \label{fig:ETHUSD_FinMem}
    \end{minipage}
\end{figure}

\textbf{Limitation 1: Incorrect decision-making due to partially positive market news.} The wrong buying decision was generated on 4 August 2023 as shown in Figure \ref{fig:AAPL_FinMem}. The overall AAPL stock price from 1 August 2023 to 20 August 2023 was a continuous downward trend. On August 4, 2023, most of the news was negative for AAPL. For example, "Huawei's upgraded mobile operating system might intensify competition in the smartphone sector," and "Apple's loss of \$3T market value due to lower iPhone demand," etc. There were only two pieces of positive news: "Services revenue offsets iPhone sales drag in earnings report," and "Subscription-based business model indicates growth potential." However, FinMem ignored the long-term downward trend and provided a buying rationale based on the belief that the earnings report and subscription-based business model could generate positive signals for stock price appreciation. This erroneous buying decision led to a shift in returns from 6\% to -3\%.

\textbf{Limitation 2: The cash is insufficient for more shares but provides a buying decision result.} As shown in the Figure \ref{fig:MSFT_FinMem}, our setting is that buying means going all in, while selling means going all out. In the chart, it can be seen that there were two buying operations, and although the stock price declined, the remaining cash was not sufficient to support further buying. FinMem lacks a clear understanding of the environment and its own current situation, which could lead to decisions that contradict reality. In \texttt{FinAgent}, we have a specially designed state description iframe module to avoid this situation.

\textbf{Limitation 3: Unstable decision-making.}  FinMem's decisions are unstable, as shown in Figure \ref{fig:TSLA_FinMem} and \ref{fig:ETHUSD_FinMem}, where FinMem made numerous buy-then-sell and sell-then-buy operations. Although in a stable market with minor price fluctuations, this would not significantly impact the return rate, in a highly volatile market, a single unstable operation could lead to substantial losses. We analyze the main reasons for this phenomenon from two aspects. First, market information is complex and often contains both positive and negative news, making it difficult to clearly determine what causes stock price fluctuations. Second, FinMem is unable to perceive the historical record of past decisions. It cannot clearly identify the historical conditions of the current choice of decision, leading to unstable decision-making.

\end{document}